\documentclass[titlepage,12pt]{article}
\usepackage{amssymb,psfig,epsfig,pslatex}
\usepackage{cite}
\makeatletter
\def\@sect#1#2#3#4#5#6[#7]#8{\ifnum #2>\c@secnumdepth
  \def\@svsec{}\else
  \refstepcounter{#1}\edef\@svsec{\csname the#1\endcsname.\hskip0.5em}\fi
  \@tempskipa #5\relax
  \ifdim \@tempskipa>\z@
    \begingroup
      #6\relax
      \@hangfrom{\hskip #3\relax\@svsec}{\interlinepenalty \@M #8\par}%
    \endgroup
    \csname #1mark\endcsname{#7}\addcontentsline
      {toc}{#1}{\ifnum #2>\c@secnumdepth \else
        \protect\numberline{\csname the#1\endcsname}\fi #7}%
  \else
    \def\@svsechd{#6\hskip #3\@svsec #8\csname #1mark\endcsname
      {#7}\addcontentsline{toc}{#1}{\ifnum #2>\c@secnumdepth \else
        \protect\numberline{\csname the#1\endcsname}\fi #7}}%
  \fi \@xsect{#5}}
\@addtoreset{equation}{section}
\makeatother
\renewcommand\thesection{\Roman{section}}
\renewcommand\theequation{\ifnum \value{section}>0
 \thesection.\arabic{equation}%
\else
\arabic{equation}%
\fi}
\def\shat{\hat{s}}

\newcommand{\kh}{{{\bf \hat k}}}
\newcommand{\ph}{{\bf \hat p}}
\newcommand{\dhh}{{\bf \hat d}}
\newcommand{\one}{1\!\mbox{l}}

\newcommand{\nn}{\nonumber}
\def\hats{{\hat s}}
\def\Li2{{{\rm Li}_2}}
\def\yp{{y_+}}
\def\ym{{y_-}}
\def\yt{{y_t}}

\def\B{{\overline{B}_0}}
\def\C{{\overline{C}_0}}
\def\Dsixa{{D^6_0(p_1,p_2,-k_2,0,0,0,m)}}
\def\Dsixb{{D^6_0(p_1,p_2,-k_1,0,0,0,m)}}
\def\Dsixone{{D^6_0(-k_1,p_1,p_2,0,m,m,m)}}
\def\Dsixcone{{D^6_0(-k_2,p_1,p_2,0,m,m,m)}}
\def\Dsixthree{{D^6_0(p_1,-k_1,p_2,0,0,m,m)}}

\def\msbar{\ensuremath{\overline{\mbox{MS}}}}
\renewcommand{\thefootnote}{\small\fnsymbol{footnote}}

\begin{document}
\begin{titlepage}
  \begin{flushright}
    hep-ph/0403035\\
    CERN-PH-TH/2004-046 \\
    DESY-04-026  \\
    PITHA 04/06  \\
    TTP04-03  \\
  \end{flushright}
\vspace{0.01cm}
\begin{center}
{\LARGE {\bf Top quark pair production and decay \\ at hadron colliders
}  \\
\vspace{2cm}
\large{\bf W. Bernreuther\,$^{a,}$\footnote{Supported
in part  by BMBF contract 05 HT1 PAA 4 and by D.F.G., SFB/TR9.},
A. Brandenburg\,$^{a,b,}$\footnote{Work supported in part 
by a Heisenberg fellowship
of D.F.G.}, Z. G. Si\,$^{c,}$\footnote{Supported
in part by the National Natural Science Foundation of China.} 
and P. Uwer\,$^d$}}
\par\vspace{1cm}
$^a$Institut f\"ur Theoretische Physik, RWTH Aachen, 52056 Aachen, Germany\\
$^b$DESY-Theorie, 22603 Hamburg, Germany\\
$^c$Department of Physics, Shandong University, Jinan, Shandong
250100, China\\
$^d$Institut f\"ur Theoretische Teilchenphysik, Universit\"at Karlsruhe,
76128 Karlsruhe, Germany and Department of Physics, 
TH Division, CERN, CH-1211 Geneva 23, 
Switzerland
\par\vspace{1cm}
{\bf Abstract}\\
\parbox[t]{\textwidth}
{In ongoing and upcoming hadron collider experiments, top quark physics
will play an important r\^ole in testing the Standard Model and its
possible extensions. 
In this work we present analytic results for the differential
cross sections of top quark pair production
in hadronic collisions at next-to-leading order
in the QCD coupling, keeping the full dependence on the spins of the top
quarks.
These results are combined with the corresponding next-to-leading
order results for the 
decay of polarized top quarks into dilepton, lepton plus jets, and
all jets final states. 
As an application we predict double differential angular
distributions which are due to the QCD-induced top quark spin
correlations in the intermediate state. In addition to the analytic
results, we give numerical 
results in terms of fit functions that can easily be 
used  in an experimental analysis.}
\end{center}
\vspace*{2cm}

PACS number(s): 12.38.Bx, 13.88.+e, 14.65.Ha
\end{titlepage}
%
%
\setcounter{footnote}{0}
\renewcommand{\thefootnote}{\arabic{footnote}}
\setcounter{page}{1}

\section{Introduction} 
\label{introduction}
A large number of top
quarks will be produced at the Fermilab Tevatron and at the 
CERN Large Hadron Collider (LHC). This 
makes the exploration of the interactions of these quarks 
one of the main physics issues  at these  facilities. 
Top quark spin phenomena are expected to
play an important r\^ole in these  efforts:
the spin-polarization and
spin-correlation phenomena reflect in detail the interactions
involved in top quark production and decay, thus give an opportunity
for precise tests of these interactions.
In contrast to light quarks the top quark polarization/correlation effects 
are not washed out by hadronization.
This is because these quarks  are  extremely short-lived and thus 
decay weakly before hadronization can take place \cite{Bigi:1986jk}. 

As far as theoretical predictions on top quark pair 
production are concerned, the cross sections for
spin-averaged top quark pair production have been known for 
quite some time  to next-to-leading order (NLO) in QCD 
\cite{Nason:1987xz,Nason:1989zy,Beenakker:1988bq,Beenakker:1990ma}.
The NLO results were refined later by resummation of soft gluon
and threshold logarithms;
see 
Refs.~\cite{Laenen:1993xr,Kidonakis:1997gm,Bonciani:1998vc,Kidonakis:2001nj}, 
and the review Ref.~\cite{Chakraborty:2003iw} 
and references therein.
As to top quark spin phenomena at hadron colliders  
there exists an extensive literature on theoretical investigations
within  the standard model (SM)
\cite{Kuhn:1983ix,Barger:1988jj,Arens:1992wh,Arens:1992fg,Mahlon:1995zn,Stelzer:1995gc,Chang:1995ay,Brandenburg:1996df,Mahlon:1997uc} 
and beyond \cite{Atwood:1992vj,Kane:1991bg,Schmidt:1992et,Brandenburg:be,Bernreuther:1993df,Bernreuther:1993hq,Haberl:1995ek,Cheung:1996kc,Grzadkowski:1997yi,Bernreuther:1997gs,Bernreuther:1998qv,Atwood:2000tu,Khater:2003wq}. 
Yet the precision of these analyses --- in particular
within the SM  --- must be
increased for the data samples that will
be recorded at the Tevatron and at the LHC to
be fully explored.
\par
In this paper we study 
hadronic top quark pair production and decay at NLO QCD by taking
the spin of the top and antitop quarks  into account.
In particular we present the differential cross sections 
to order $\alpha_s^3$, describing the $t{\bar t}+X$ production in a
general spin configuration.
Short reports on parts of this work were given in
Refs.~\cite{Bernreuther:2000yn,Bernreuther:2001bx,Bernreuther:2001rq}.

The theoretical description of top quark pair production in proton--proton and 
proton--antipro\-ton collisions
at next-to-leading order in the QCD coupling involves  the
following
parton reactions:  quark--antiquark annihilation
\begin{equation}
q + {\bar q} \rightarrow t + {\bar t},
\label{eq:qq}
\end{equation}
\begin{equation}
q + {\bar q} \rightarrow t + {\bar t} + g,
\label{eq:qqgluon}
\end{equation}
which is the dominant production mechanism at the Tevatron, gluon--gluon fusion
\begin{equation}
g + g \rightarrow t + {\bar t},
\label{eq:gg}
\end{equation}
\begin{equation}
g + g  \rightarrow t + {\bar t}  + g,
\label{eq:ggg}
\end{equation}
which dominates at the LHC, and
\begin{equation}
g + q  \rightarrow t + {\bar t}  + q,
\label{eq:qg}
\end{equation}
\begin{equation}
g + {\bar q}  \rightarrow t + {\bar t}  + {\bar q},
\label{eq:qbarg}
\end{equation}
which gives, in general, only a tiny correction.

As mentioned earlier the top quarks decay before they can form
hadronic bound states. To construct
realistic observables, the decays of the top and antitop quarks 
have to be taken into account.
We consider the SM decays of polarized (anti)top quarks both into
semileptonic and non-leptonic final states, taking  into account the
order $\alpha_s$ QCD corrections \cite{Czarnecki:1990pe,Brandenburg:2002xr}.
The main SM top decay modes at that order are: 
\begin{equation}
  t\to b W  \to  b \ell \nu_{\ell} (g),  b q {\bar q}' (g) 
  \label{eq:topdecay}
\end{equation}
where $q {\bar q}'= u {\bar d}, c {\bar s}$.
A complete next-to-leading order
QCD analysis of  top quark production {\it and} decay
thus involves the parton reactions
\begin{equation}
  gg, q{\bar q} \ {\buildrel
    t{\bar t}\over \longrightarrow} \  b {\bar b} + 4 f,
  \label{eq:ttrec1}
\end{equation}
\begin{equation}
  gg, q{\bar q} \  {\buildrel
    t{\bar t}\over \longrightarrow} \  b {\bar b} + 4f  + g,
  \label{eq:ttrec2}
\end{equation}
\begin{equation}
  g + q ({\bar q})\  {\buildrel
    t{\bar t}\over \longrightarrow}\   b {\bar b} + 4f  + q ({\bar q}),
  \label{eq:ttrec3}
\end{equation}
where $f=q,\ell,\nu_{\ell}$. 
In view of the fact that the total width
$\Gamma_t$
of the top quark is much smaller than its mass, $\Gamma_t/m_t ={\cal
  O}(1\%)$, one may expand 
the amplitudes  of the reactions Eqs.~(\ref{eq:ttrec1})--(\ref{eq:ttrec3})
around the poles of the unstable top and
antitop quarks, which corresponds to
an expansion in powers of $\Gamma_t/m_t$. Only the leading term of this
expansion, i.e.
the residue of the double poles, is  considered here. In this framework the
radiative 
corrections to Eq.~(\ref{eq:ttrec1})  can be classified into so-called
factorizable and
non-factorizable
corrections --- and, likewise, the contributions
to the squared matrix elements of Eq.~(\ref{eq:ttrec2}). 
We compute the factorizable corrections in
the narrow width approximation $\Gamma_t/m_t\to 0$. In this
approximation the squared matrix element $|{\cal M}|^2$ of
the respective reaction is
of the form
\begin{equation}
  \vert{\cal M}{\vert}^2 \propto {\rm Tr}\;[\rho
  R{\bar{\rho}}]
  = \rho_{\alpha'\alpha}
  R_{\alpha\alpha',\beta\beta'}{\bar{\rho}}_{\beta'\beta} .
  \label{eq:trace}
\end{equation}
Here $R$ denotes the  density matrix that describes
 the production of on-shell
top quark pairs in a specific spin configuration by one of the six
reactions Eqs.~(\ref{eq:qq})--(\ref{eq:qbarg}). The matrices
$\rho,{\bar{\rho}}$ are the density matrices describing the decay
of polarized top and antitop quarks into specific final states.
The subscripts in  Eq.~(\ref{eq:trace}) denote the  top and antitop
spin indices.
Both the production and decay density matrices are gauge invariant.
The production density matrices are determined from the NLO results
for the reactions Eq.~(\ref{eq:qq})--Eq.~(\ref{eq:qbarg}). The decay
density matrices are derived from the results for the reactions
shown in Eq.~(\ref{eq:topdecay}).

With these building
blocks the factorizable NLO QCD corrections to top quark pair production
and decay can be computed,  keeping the full
information on the spin configuration  of the intermediate $t\bar{t}$ state.
As far as applications of these
results are concerned, our primary aim in this paper is the study of 
final-state angular correlations of the top quark and antiquark decay products,
which reflect the spin properties  of the top and antitop quarks. 
For this purpose we consider the following channels: 
\begin{eqnarray}
p {\bar p}, p p &\rightarrow& t{\bar t}  + X 
\rightarrow \ell^+ \ell\,'^- + X, \label{eq:ttll} \\
p {\bar p}, p p &\rightarrow& t{\bar t}  + X \rightarrow \ell^+ j_2  +
X, 
\label{eq:ttlj1}\\
p {\bar p}, p p &\rightarrow& t{\bar t}  + X \rightarrow j_1
\: \ell\,'^-  + X\, ,\label{eq:ttlj2}\\
p {\bar p}, p p &\rightarrow& t{\bar t}  + X \rightarrow j_1 \, j_2  + X,
\label{eq:ttjj} 
\end{eqnarray}
where
$\ell = e,\mu,\tau,$ and $j_1,$ $j_2$ denote jets originating from
top and antitop decays.
For these final states we study the following  distributions 
at NLO in the coupling
$\alpha_s$:
\renewcommand{\labelenumi}{\roman{enumi}.}
\begin{enumerate}
\item The double distributions
  \begin{eqnarray}
    {1\over \sigma}{d\sigma\over d\cos\theta_1 d\cos\theta_2}=
    {1\over 4} (1 +
    {\rm B}_1 \cos\theta_1
    + {\rm B}_2 \cos\theta_2 
    - {\rm C} \cos\theta_1 \cos\theta_2)\,\, ,
    \label{eq:ddist1}
  \end{eqnarray}
  where $\sigma$ denotes the cross section for the channel under consideration.
  Here  $\theta_1$ ($\theta_2$) describes the angle between the
  direction of flight of the lepton $\ell^+$ or jet $j_1$ ($\ell\,'^-$ or
  $j_2$) in the $t$ ($\bar{t}$)
  rest frame  and a reference 
  direction ${\bf \hat a}$ (${\bf \hat b}$). In particular we will
  discuss three specific choices for the reference directions 
  ${\bf \hat a}, {\bf \hat b}$ which allow a simple physical interpretation.
\item The opening angle distributions 
  \begin{eqnarray}
    {1\over \sigma}{d\sigma \over d\cos\varphi}=
    {1\over 2} (1 - {\rm D} \cos\varphi)\,\, ,
    \label{eq:ddist2}
  \end{eqnarray}
  where $\varphi$ denotes the angle 
  between the
  direction of flight of the lepton $\ell^+$ (or jet $j_1$)
  and  of $\ell\,'^-$ (or $j_2$), defined  in the $t$ or $\bar{t}$
  rest frames, respectively.
\end{enumerate}
The functional forms of the r.h.s. of Eqs. 
(\ref{eq:ddist1}) and  (\ref{eq:ddist2}) hold if no
kinematic cuts are applied and will be derived in Section
\ref{angular}. 
\par
The results presented in this work are definite predictions of
QCD. Specifically, comparison of our results for
the above distributions with future measurements will allow for detailed
investigations of top quark production and decay dynamics.
\vspace*{0.3cm}

The paper is organized as follows. In Section \ref{calc} we present the
differential cross sections for the $2\to 2$ processes
Eq.~(\ref{eq:qq}) and Eq.~(\ref{eq:gg}) to order $\alpha_s^3$.
In Section \ref{real} we treat the real gluon radiation Eq.~(\ref{eq:qqgluon}),
Eq.~(\ref{eq:ggg}) 
and the $qg$ ($\bar{q}g$) fusion processes.
In particular, we compute the differential cross sections in the soft
and collinear limits and perform the mass factorization.
In Section \ref{partonsection} we discuss the 
QCD-induced top and antitop spin effects
at NLO, and we treat the  issue
of constructing infrared and collinear safe
$t, \bar t$ spin observables at the parton level.
 Specifically, we compute, for all relevant parton reactions
  $i\to t\bar{t}X$, the expectation values of
four different spin observables. These observables correspond to
different choices of the directions ${\bf\hat{a}}$, ${\bf\hat{b}}$
used to define the angles in Eq.~(\ref{eq:ddist1}). We also show that
the spin observables have a simple interpretation as double spin
asymmetries with respect to a given quantization axis.
The spin observables serve as one building block for the calculation
of the  distributions  Eq.~(\ref{eq:ddist1}), Eq.~(\ref{eq:ddist2}). 
The other building blocks are given in Section \ref{decay}: these are the 
one-particle inclusive angular distributions
of the semileptonic \cite{Czarnecki:1990pe} and non-leptonic
\cite{Brandenburg:2002xr} decays of 
polarized top quarks and antiquarks to order $\alpha_s$.  
In Section \ref{angular} we show that the functional form of the
distributions (\ref{eq:ddist1}), (\ref{eq:ddist2}) is indeed 
as given in (\ref{eq:ddist1}), (\ref{eq:ddist2}).
In addition we give 
NLO formulae for the coefficients ${\rm C}$ and ${\rm D}$ 
defined in Eq.~(\ref{eq:ddist1}) and Eq.~(\ref{eq:ddist2}). 
We further investigate 
the important question of whether the distributions Eqs.~(\ref{eq:ddist1}),
(\ref{eq:ddist2}) are affected by non-factorizable QCD corrections. 
In particular we show, using the results of 
Refs.~\cite{Fadin:1993dz,Fadin:1993kt,Melnikov:np,Beenakker:1997ir,Beenakker:1999ya}, 
that the non-factorizable corrections
at order $\alpha_s^3$ do not contribute to 
Eqs.~(\ref{eq:ddist1}), (\ref{eq:ddist2}).
In Section \ref{numres} 
we present our NLO predictions for these distributions
for the dilepton, for the lepton+jet, and for the jet+jet decay channels,
both at the Tevatron and the LHC.    
Section \ref{conclusions} contains our conclusions.
Appendix \ref{sec:LoopIntegrals} contains a collection of one-loop 
integrals that appear in
the virtual corrections to the squared matrix elements of
$q\bar{q}\to t\bar{t}$ and  $gg\to t\bar{t}$, which are given in
Appendices \ref{sec:VirtResultsqq} and \ref{sec:VirtResultsgg}, 
respectively. In Appendix \ref{sec:HelAmps} we give the 
amplitudes for the processes Eqs.~(\ref{eq:qqgluon}), (\ref{eq:ggg}), 
(\ref{eq:qg}), and (\ref{eq:qbarg}), for arbitrary spins of the
$t$ and $\bar{t}$ and arbitrary helicities of the massless partons.
Appendix \ref{sec:FitFunctions} contains fit functions
for the NLO results for 
the expectation values
of the four spin observables computed in Section \ref{partonsection}.

\newpage
\section{One-loop QCD corrections to $\mathbf{q\bar q}$ annihilation
  and to $\mathbf{g g}$ fusion}
\label{calc}
In this section we present the differential cross sections
of the parton processes
\begin{equation}
q(p_1)+\bar{q}(p_2)\rightarrow t(k_1, s_t)+\bar{t}(k_2, s_{\bar t})\, ,
\label{qqs12}
\end{equation}
\begin{equation}
g(p_1)+g(p_2)\rightarrow t(k_1, s_t)+\bar{t}(k_2, s_{\bar t}) \, .
\label{ggs12}
\end{equation}
at NLO in $\alpha_s$. 
Here $p_1$, $p_2$, $k_1$, and $k_2$ denote the parton momenta,
and the vectors $s_t$,  $s_{\bar t}$,
with   
\begin{equation}
  s^2_t = s^2_{\bar t} = -1\quad 
  \mbox{and}\quad k_1\cdot s_t = k_2\cdot  s_{\bar t} = 0,
\end{equation}
describe the
spin of the top and antitop quarks.
All quarks but the top quark are taken to be massless. The top
quark mass is denoted by $m$. 
In the calculation of the radiative
corrections, ultraviolet as well as soft/collinear singularities 
are encountered.
The singularities are regulated by
using dimensional regularization. We keep the spin vectors
in 4 dimensions. Note that no scheme dependence is introduced
in this way, in particular no `$\gamma_5$ problem' arises. 
This is because
the cancellation of the UV singularities is independent
of the external spin state and the soft/collinear singularities
are cancelled in a universal way.
\par 
In the (anti)top 
rest frame, the spin of the (anti)top  is described by a unit vector
${\bf\hat s}_t$ (${\bf\hat s}_{\bar{t}}$).
We define the $t$ ($\bar{t}$) rest frame by a rotation-free Lorentz
boost from the zero momentum frame of the $t\bar{t}$ quarks
($t\bar{t}$-ZMF). In this frame
\begin{eqnarray}
(k_1+k_2)^{\mu} =\bigg(\sqrt{(k_1+k_2)^2},0,0,0\bigg).
\end{eqnarray}
As we will see in Section \ref{partonsection}, 
choosing this frame simplifies the  definition of infrared and collinear safe
observables at the parton level.
For the $2\to2$ processes Eqs.~(\ref{qqs12}) and (\ref{ggs12}),
the  $t\bar{t}$-ZMF coincides with the centre-of-mass
frame of the initial partons.
Note that in principle the  $t\bar{t}$-ZMF is only defined up to an 
arbitrary rotation.
One can resolve this, for instance, as done in an experiment 
where the direction of one
of the initial hadron beams is chosen as $z$-axis and one orthogonal 
direction as $x$-axis. After having measured the 4-momenta of the
$t$ and $\bar{t}$ in this laboratory frame, the $t\bar{t}$-ZMF can be
defined unambiguously by a rotation-free boost.     
The observables that we consider in Sections \ref{partonsection} and 
\ref{angular} are actually
the same for any choice of the $t\bar{t}$-ZMF.\par
In the  $t\bar{t}$-ZMF we have
\begin{eqnarray} 
  s_t^{\mu} &=& \left ({{\bf k}\cdot{\bf\hat s}_t \over m}\, ,
    {\bf\hat s}_t + {{\bf k}({\bf k}\cdot{\bf\hat
        s}_t)\over m(m+E)}  \right ), \\
  s_{\bar t}^{\mu} &=& \left (-{{\bf k}\cdot{\bf\hat s}_{\bar t}\over 
      m}\, ,
    {\bf\hat s}_{\bar t} + {{\bf k}({\bf k}\cdot{\bf\hat
        s}_{\bar t})\over m(m+E)}  \right),
\end{eqnarray}
where $E$ and ${\bf k}$ 
are the energy and the 3-momentum of the top quark in the $t\bar{t}$-ZMF.
\par
In $d = 4 - 2\epsilon$ dimensions 
the Born+virtual parts of the cross sections 
of the processes Eqs.~(\ref{qqs12}), (\ref{ggs12})
are, to order $\alpha_s^3$,  of the following form:
\begin{equation}
d\sigma^i(s_t, s_{\bar t})=d\sigma_B^i + d\sigma_V^i =
{1 \over 2\hat{s}}\Phi_i  d\Gamma_2 \left [ |{\cal M}_{B}^i|^2
+  {\cal M}_{B}^{*i} {\cal M}_{V}^i+
{\cal M}_{V}^{*i} {\cal M}_{B}^i \right ] \, ,
\label{dsigbv}
\end{equation}
where $i=q\bar{q}, gg$, and $\hat{s}=(p_1+p_2)^2$. The factors
\begin{equation} 
\Phi_{q\bar{q}}={1\over 4N^2} \, , \qquad 
\Phi_{gg}={1\over (N^2-1)^2 (d-2)^2}
\end{equation}
arise from  averaging over the colours  ($N=3$) and spins of the initial
partons.  The 2-particle phase-space measure is denoted by
$d \Gamma_2$,  ${\cal M}_{B}^{i}$ are the Born amplitudes and
${\cal M}_{V}^i$ those of the virtual corrections. 
As indicated earlier, Eq.~(\ref{dsigbv}) still contains soft and collinear 
singularities regulated in the framework of conventional 
dimensional regularization. 
The production spin
density matrices $R^i$ that enter the formula (\ref{eq:trace})
are obtained from the identity
\begin{equation} 
d\sigma^i(s_t, s_{\bar t}) = {1 \over 2\hat{s}}\Phi_i d\Gamma_2
{1\over 4}{\rm Tr} \left [ R^i (\one + {\bf\hat s}_t\cdot\mathbf{\tau})
\otimes (\one + {\bf\hat s}_{\bar t}\cdot\mathbf{\tau})  \right ] \, ,
\label{spinid1}
\end{equation}
where  $\tau_i$ are the Pauli matrices.

\subsection{Born matrix elements}
\label{sec:BornResults}
\subsubsection{$\mathbf{q\bar{q}}$ initial states}
\label{sec:BornResults_qq}
For the sake of fixing our notation, we present the squared Born
matrix element for the reaction (\ref{qqs12}) 
in two ways, which prove useful in the presentation of the
virtual, soft, and collinear contributions to the cross sections.

The squared Born  matrix element reads 
\begin{equation}
|{\cal M}_{B}^{q\bar{q}}|^2=4\pi^2\alpha_s^2 (N^2-1)
\Big\{A_0^{q\bar{q}}+{\cal
  C}^{q\bar{q}}_{\mu\nu}s_t^{\mu}s_{\bar{t}}^{\nu}\Big\} \, ,
\label{mqqborn}
\end{equation}
where 
\begin{equation}
A_0^{q\bar{q}}  
= 2 -\beta^2\* \left( 1 -y^2 \right) - 2 \epsilon  \, .
\end{equation} 
Here 
\begin{equation}
\beta=\sqrt{1-{4m^2\over \hat{s}}} \, , 
\end{equation}
and $y$ is the cosine of the scattering angle in the $t\bar{t}$-ZMF,
i.e.   $y={\bf \hat p}\cdot{\bf \hat k}$, where ${\bf \hat p}$
is the direction of the initial quark
(or one of the
gluons in the case of $ gg$ fusion) in that frame. 
The spin-dependent
term can  be decomposed as follows:
\begin{eqnarray}
{\cal
  C}^{q\bar{q}}_{\mu\nu}s_t^{\mu}s_{\bar{t}}^{\nu}&=&C_0^{q\bar{q}} 
(s_t\cdot s_{\bar{t}})
+ D_0^{q\bar{q}} \Big[(p_1\cdot s_t)(p_1\cdot s_{\bar{t}})+
 (p_2\cdot s_t)(p_2\cdot s_{\bar{t}})\Big]   \nonumber \\
&+&E_0^{q\bar{q}} (p_1\cdot s_t)(p_2\cdot s_{\bar{t}})
+F_0^{q\bar{q}}(p_2\cdot s_t)(p_1\cdot s_{\bar{t}}),
\end{eqnarray}
where
\begin{eqnarray}
C_0^{q\bar{q}}&=&
   \beta^2 \*(1-y^2)+2 \epsilon \, , \nonumber \\
D_0^{q\bar{q}}&=&
-{4 \over \hat{s}}\epsilon, \nonumber \\
E_0^{q\bar{q}}&=& 
 -{4\over \hat{s}}\*\Big\{ 1+\beta \*y+\epsilon \Big\}, \nonumber \\
F_0^{q\bar{q}} &=& 
 -{4\over \hat{s}}\*\Big\{ 1-\beta \*y+\epsilon \Big\}\, .
\end{eqnarray}
In view of a compact presentation of the virtual corrections
$d\sigma_V^{q\bar q}$ (see  Section \ref{sec:qqvir}), it is  useful
to write down  the spin-dependent  term in an alternative way. Defining the
two unit vectors ${{\bf \hat u}_1}, {{\bf \hat u}_2}$ which are
orthogonal to each other:
\begin{eqnarray}
  {{\bf \hat u}_1}&=&{1\over \sqrt{1-y^2+\gamma^2 y^2}}
  \left [-\sqrt{1-y^2} \: {\bf \hat q}-\gamma y\:  
    {\bf \hat k} \right ], \nonumber\\
  {{\bf \hat u}_2}&=&{1\over \sqrt{1-y^2+\gamma^2 y^2}}
  \left [\sqrt{1-y^2} \: {\bf \hat k}-\gamma y \: {\bf \hat q} \right ],
  \label{dd12}
\end{eqnarray}
where
${\bf\hat{q}}=({{\bf \hat p}-y \, {\bf \hat k}})/ {\sqrt{1-y^2}}$ 
and $\gamma=E/m$,
we obtain
\begin{equation}
{\cal C}_{\mu\nu}^{q\bar{q}} s_t^{\mu} s_{\bar{t}}^{\nu}=
\tilde{\cal C}_{ij}^{q\bar{q}} \hat{s}_t^{i} \hat{s}_{\bar t}^j,
\label{decij}
\end{equation}
with
\begin{equation}
\tilde{\cal C}_{ij}^{q\bar{q}}={1\over 3} e_0 \delta_{ij}+
\left(\hat{u}_{1i}\hat{u}_{1j}-{1\over 3}\delta_{ij}\right) e_1 +
\left(\hat{u}_{2i}\hat{u}_{2j}-{1 \over 3}\delta_{ij}\right) e_2 +
(\hat{u}_{1i}\hat{u}_{2j}+\hat{u}_{2i}\hat{u}_{1j}) e_3 \, ,
\label{ctilij}
\end{equation}
and
\begin{eqnarray}
e_0&=&A_0^{q\bar{q}}-4\epsilon, \nonumber \\
e_1&=&2 ,\nonumber\\
e_2&=&2 \*\beta^2 \*(1-y^2), \nonumber \\
e_3 &=& 0 .
\end{eqnarray}

\subsubsection{$\mathbf{gg}$ initial state}
\label{sec:BornResults_gg}
The squared  Born matrix element is given by 
\begin{eqnarray}
|{\cal M}_{B}^{gg}|^2
 = 2\*\Big[N^2 \*(1+\beta^2 \* y^2)-2\Big]\*
|\tilde{\cal M}_{B}^{gg}|^2,
\end{eqnarray}

where
\begin{eqnarray}
|\tilde{\cal M}_{B}^{gg}|^2&=& {4\*\pi^2 \* \alpha_s^2\* (N^2-1)\over N
\*(1-\beta^2\*y^2)^2}
\* \Big\{ A_0^{gg}+C_0^{gg} \* (s_t\cdot s_{\bar{t}})+D_0^{gg}
\* \Big[(p_1\cdot s_t)\* (p_1\cdot s_{\bar{t}})+ 
(p_2\cdot s_t)\* (p_2\cdot s_{\bar{t}})\Big] \nonumber \\
&+&E_0^{gg}\*(p_1\cdot s_t)\*(p_2\cdot s_{\bar{t}})
+E_0^{gg}|_{y\rightarrow -y} \*(p_2\cdot s_t)\*
(p_1\cdot s_{\bar{t}})
\Big\} \, ,
\end{eqnarray}
and 
\begin{eqnarray}
A_0^{gg}&=&
1+2\* \beta^2 \*(1-y^2)-\beta^4 \*\Big[1+(1-y^2)^2\Big]+
\epsilon \*\Big[(1+\beta^2 \*y^2)^2-4\Big]+2 \*\epsilon^2 \* 
(1-\beta^2 \*y^2),
 \nonumber \\
C_0^{gg}
&=&1-2\* \beta^2+\beta^4 \*\Big[1+(1-y^2)^2\Big]
-\epsilon \*(1-\beta^2\* y^2)^2
-2\*\epsilon^2\* (1-\beta^2 \*y^2),  \nonumber \\
D_0^{gg}
&=&{32 \*\epsilon \*m^2 \over \hat{s}^2},
 \nonumber \\
E_0^{gg}
&=&{4\* (1+\beta \*y)\over \hat{s}}\*\Big\{
-\beta^2 \*(1-y^2)+\epsilon \*(1-\beta^2\* y^2)
+2\*\epsilon^2
\Big\} \, .
\end{eqnarray}
The virtual corrections $d\sigma_V^{gg}$ 
do not simplify when using a  decomposition similar to Eq.~(\ref{decij}).

\subsection{Virtual corrections}
\label{sec:VirtualCorrections}
We now give our results for the order $\alpha_s^3$ virtual
corrections
to the Born cross  sections of Section \ref{sec:BornResults}. 
The ultraviolet singularities are removed by
using the $\overline{\rm{MS}}$ prescription for the QCD coupling 
and the on-shell definition of the top mass. 
In the following,  $\alpha_s$ denotes the QCD coupling
defined in the  $\overline{\rm{MS}}$ scheme of $N_f=6$ flavour QCD
(5 massless and one massive quark), $m$ denotes the mass of the
top quark defined in the on-shell scheme, and $\mu_R$ is the
renormalization scale. 
The renormalized differential cross sections $d\sigma^i_V$ 
still contain single and double poles in $\epsilon = (4-d)/2$ 
due to soft and collinear divergences, which we collectively
call infrared (IR) singularities in the following.
These poles are removed  after including the  contributions from
soft gluon radiation 
 and  mass factorization. This will be done in the next section. 
\par
The absorptive parts of the one-loop amplitudes induce small
polarizations of the $t$ and $\bar t$ quarks orthogonal to 
the $2\to 2$ scattering plane, i.e. the 
$d\sigma^i_V$ contain also terms proportional to
$\epsilon_{ijl}{\hat s_t}^i {\hat p}^j {\hat k}^l$ and 
$\epsilon_{ijl}{\hat s_{\bar t}}^i {\hat p}^j {\hat k}^l$,
 which are ultraviolet- 
and infrared-finite at this order in the
perturbative expansion.  These terms were computed in 
Refs.~\cite{Bernreuther:1995cx,Dharmaratna:xd}. 
They do not contribute to the observables, which we
investigate in Sections \ref{partonsection} 
and \ref{angular}, so that we omit them 
in the following. 
\subsubsection{$\mathbf{q\bar{q}}$ initial states}
\label{sec:qqvir}
The NLO virtual corrections as defined in
Eq.~(\ref{dsigbv}) can  be presented as follows:
\begin{eqnarray}
d\sigma_{V}^{q\bar{q}}&=&{\alpha_s \over \pi}\* 
C_{\epsilon} \*d\sigma_{B}^{q\bar{q}}\*
\Bigg\{-{C_F\over \epsilon^2}+{1\over \epsilon}\*\Bigg[
-{5 \* C_F\over 2}+{1+\beta^2\over 4\*\beta \*N}\*\ln(x)+
{2\over N}\* \ln\left({1+\beta\* y\over 1-\beta \*y}\right)  \nonumber \\
&+&C_F\*\ln\left ({\hat{s}\over m^2}\right )
+{N\over 2}\*\ln\left({(1-\beta \*y)^2\over 1-\beta^2}\right)\Bigg]\Bigg\}
+{1\over 2\* \hat{s}}\*\Phi_{q\bar{q}} \*{\cal F}_{q\bar{q}} \*d\Gamma_2 \, ,
\end{eqnarray}
where 
\begin{equation}
C_{\epsilon}=\Big[{4\*\pi\* \mu^2\over m^2}\Big]^{\epsilon} 
e^{-\epsilon \*\gamma_E} \, ,
\end{equation}
$C_F=(N^2-1)/(2\*N)$, $\gamma_E$ is the Euler constant, $\mu$
is an arbitrary mass scale, and 
\begin{eqnarray}
x={1-\beta\over 1+\beta}.
\end{eqnarray} 
The IR-finite part ${\cal F}_{q\bar{q}}$ is of the form  
\begin{equation}
{\cal F}_{q\bar{q}}={{\cal F}}_1 \*|{\cal M}_{B}^{q\bar{q}}|^2+
4\*\pi^2\*\alpha_s^2 \*(N^2-1)\* {\cal F}_2,
\end{equation}
with
\begin{eqnarray}
{{\cal F}}_1&=&{\alpha_s\over 2\* \pi}\*\Big\{
2 \*G+{\hat{s}\over N}\*\C(p_1,p_2,0,0,0)+
{N^2-2\over N}\*\hat{s}\*(1-\beta \*y)\*\C(-p_1,k_1,0,0,m) \nonumber
\\
&+&{2\*\hat{s}\*(1+\beta\* y)\over N} \*\C(-p_1,k_2,0,0,m)
+{\hat{s}\*(1+\beta^2)\over 2\*N}\*\C(k_1,k_2,m,0,m)\Big\}.
\label{resf1}
\end{eqnarray}
The contribution $G$ is the finite part of the
sum of the gluon self-energy diagram
(computed in the Feynman gauge) and the counterterms:  
\begin{eqnarray}\label{G}
G &=&
{11\*N - 2\*N_f \over 6}\*
     \ln\left({\mu_R^2\over m^2}\right) +
{N\over 6}\* \left (5\*\B(k_1+k_2,0,0)+{1\over 3}\right) \nonumber \\
&-&{1\over 3}\*\left[(N_f-1)\* \B(k_1+k_2,0,0)+\left(
1+{2\*m^2\over \hat{s}}\right)\B(k_1+k_2,m,m)-{1\over 3}\*N_f\right] 
\, .
\end{eqnarray}
The functions  $\C,\ \B$,
which appear in Eqs.~(\ref{resf1}), (\ref{G}), are defined in 
Appendix \ref{sec:LoopIntegrals}.
Omitting the contributions from the absorptive parts, 
the term ${\cal F}_2$ can be decomposed as follows:
\begin{equation}
{\cal F}_2={\alpha_s\over \pi}\left[A_V^{q\bar{q}}+\tilde{\cal C}_{ij}^{V} 
\hat{s}_t^i \hat{s}_{\bar{t}}^j\right] \,. 
\label{calf2}
\end{equation}
The $3\times 3$ matrix
 $\tilde{\cal C}_{ij}^{V}$ has the same structure as Eq.~(\ref{ctilij})
with the coefficients $e_a$ being replaced by $e_a^{V}$. 
These coefficients and  $A_V^{q\bar{q}}$ are listed in 
Appendix \ref{sec:VirtResultsqq}.

\subsubsection{$\mathbf{gg}$ initial state}

The one-loop contributions to the differential cross section 
of the gluon fusion reaction  (\ref{ggs12}) are
of the following form:
\begin{equation}
d\sigma_{V}^{gg}=
\frac{1}{2\hat{s}}\Phi_{gg}\Big\{
C_{\epsilon} {\cal G}_{gg}
+{\cal H}_{gg}
\Big\} d \Gamma_2,
\end{equation}
where the IR-singular term  ${\cal G}_{gg}$ is given by 
\begin{eqnarray}
{\cal G}_{gg}&=&{\alpha_s\over \pi}\Big\{
{\cal V}_A |{\cal M}_{B}^{gg}|^2
+|\tilde{\cal M}_{B}^{gg}|^2\Big[{\cal V}_B+{\cal V}_C+
{\cal V}_C|_{y\rightarrow -y}\Big]\Big\} \, ,
\end{eqnarray}
with
\begin{eqnarray}
{\cal V}_A&=&-{N\over \epsilon^2} 
-{1\over \epsilon}\*{11\*N-2\*(N_f-1)\over 6}
-{1\over \epsilon}\*\left[C_F-N\*\ln\left({\hat{s}\over m^2}\right)\right],  \nonumber \\
{\cal V}_B&=&-{1+\beta^2\over 2\*N\*\beta}\*\left[2+N^2\*(1-\beta^2 \*y^2)
\right]\*
{\ln(x)\over \epsilon},  \nonumber \\
{\cal V}_C&=&{N\over 2\*\epsilon}\*\left[N^2 \*(1+\beta \*y)^2-
4\right]\*
\ln\left({(1-\beta \*y)^2\over 1-\beta^2}\right).
\end{eqnarray}
The infrared-finite part reads
\begin{eqnarray}
{\cal H}_{gg}={\alpha_s \over \pi}\* 
\*\ln\left({\mu_R^2\over m^2}\right) {11N-2 N_f\over 6}\*|{\cal M}_{B}^{gg}|^2
+{\cal H}_{1}|\tilde{\cal M}_{B}^{gg}|^2 + 4\*\pi^2\*\alpha_s^2\*(N^2-1)
\*{\cal H}_{2},
\end{eqnarray}
with
\begin{eqnarray}
{\cal H}_{1}&=&-{\alpha_s\over \pi}\*\hat{s} \*\Bigg\{
\C(k_1,k_2,m,0,m)\*{(1+\beta^2) [2+N^2\*(1-\beta^2 \*y^2)]\over
2\*N} \nn \\ 
&+& \C(p_1,p_2,0,0,0)\* N^3 
\left(1+\beta^2\* y^2\right) \nonumber \\
&+&\C(-p_1,k_1,0,0,m) \* {N\over 2} \*
\left[4-N^2\* (1+\beta \*y)^2\right]\*(1-\beta\* y) \nn \\
&+&\C(-p_1,k_2,0,0,m)\* {N\over 2} \left[4-N^2\* (1-\beta \*y)^2\right]\*
(1+\beta\*y)
\Bigg\}.
\end{eqnarray}

The term  ${\cal H}_{2}$ has the structure (here too,
contributions from absorptive parts are omitted)

\begin{eqnarray}
{\cal H}_{2}&=&{\alpha_s\over \pi}\*\Big\{
A_{V}^{gg}+C_V^{gg}\*(s_t\cdot s_{\bar{t}})+
D_V^{gg}\Big[  (p_1\cdot s_t) (p_1\cdot s_{\bar{t}})+
 (p_2 \cdot s_t) (p_2\cdot s_{\bar{t}}) \Big]  \nonumber \\
&+& E_V^{gg}\*(p_1\cdot s_t) (p_2\cdot s_{\bar{t}})
+E_V^{gg}|_{y\rightarrow -y} (p_2\cdot s_t) (p_1\cdot s_{\bar{t}})
\Big\} \, .
\label{fgg}
\end{eqnarray}
The coefficients $A_V^{gg},\ldots, E_V^{gg}$
are given in Appendix \ref{sec:VirtResultsgg}.

\section{Real gluon radiation and (anti)quark--gluon fusion}
\label{real}
In this section we consider the reactions
\begin{equation}
q(p_1)+\bar{q}(p_2)\rightarrow t(k_1, s_t)+ \bar{t}(k_2, s_{\bar t})+g(p_3) \, ,
\label{ttg}
\end{equation}
\begin{equation}
g(p_1)+g(p_2)\rightarrow t(k_1, s_t)+ \bar{t}(k_2, s_{\bar t})+g(p_3) \, ,
\label{ggg}
\end{equation}
\begin{equation}
q(p_1)+g(p_2)\rightarrow t(k_1, s_t)+ \bar{t}(k_2, s_{\bar t})+q(p_3) \, ,
\label{qgq}
\end{equation}
\begin{equation}
{\bar q}(p_1) + g(p_2) \rightarrow t(k_1, s_t)+ 
\bar{t}(k_2, s_{\bar t})+{\bar q}(p_3) \, . 
\label{qbarg}
\end{equation}
In order to deal with the IR divergences of the corresponding
lowest order QCD cross sections, we employ the phase-space slicing
method: 
in the case of the reactions Eq.~(\ref{ttg}), 
Eq.~(\ref{ggg}), we divide the phase-space
into four regions, namely the region where the gluon is soft, the two 
regions where the gluon is  
collinear (but not soft) to one of the initial-state massless
partons,   and the complement of these three regions, where
all partons are `resolved'.
The Born cross sections  of the  reactions in Eqs.~(\ref{qgq}) and 
(\ref{qbarg}) develop collinear divergences but no soft ones. Thus,
in this case, we split the phase-space  into the two regions
where the outgoing $q$ $({\bar q})$ becomes collinear
to  the incoming (anti)quark or gluon, and the complement,
which is the resolved region.
For convenience we parametrize in this section the 3-parton phase
space in terms of angles and momenta defined in the c.m.s. of the
initial partons. 

\subsection{Soft gluon cross sections}
For the processes  of Eqs.~(\ref{ttg}), 
and (\ref{ggg}), we define the soft region by the requirement
that the scaled gluon energy in the c.m. frame of the initial partons,
$x_g={2 E_g}/{\sqrt{\hat{s}}}$, be smaller than some
cut parameter $x_{\rm min}$, where $x_{\rm min}\ll 1$. That is, 
`hard' and `soft' gluon refers to the 
decomposition
\begin{equation}
1=\Theta ( x_g-x_{\rm min})+\Theta ( x_{\rm min}-x_g) \, .
\label{soft}
\end{equation}
In the soft region we use the eikonal approximation for  the matrix
elements ${\cal M}^{q\bar{q},g}$ and  ${\cal M}^{gg,g}$ 
of Eqs.~(\ref{ttg}) and (\ref{ggg}), and the soft limit of the
$d$-dimensional phase-space measure. \\
For quark--antiquark annihilation, we obtain
\begin{eqnarray}
|{\cal M}_{\rm soft}^{q\bar{q},g}|^2&=&4\pi \alpha_s \mu^{2\epsilon}
\Bigg\{-C_F \Big[ {m^2\over (k_1\cdot p_3)^2}+
{m^2\over (k_2\cdot p_3)^2}\Big ] 
-{1\over  N}\Big[{k_1\cdot k_2\over (k_1\cdot p_3)(k_2\cdot p_3)}
 \nonumber \\
 &+& {p_1\cdot p_2\over (p_1 \cdot p_3) (p_2\cdot p_3)}
   -{2 k_1\cdot p_2\over (k_1\cdot p_3) (p_2\cdot p_3)}
 -{2 k_2\cdot p_1\over (k_2\cdot p_3) (p_1\cdot p_3)}\Big ]
 \nonumber \\
 &+& {N^2-2\over N} \Big[ 
{k_1\cdot p_1\over (k_1\cdot p_3)(p_1\cdot p_3)}
+{k_2\cdot p_2\over (k_2\cdot p_3)(p_2\cdot p_3)}\Big]
\Bigg\} ~ |{\cal M}_{B}^{q\bar{q}}|^2 \, .
\end{eqnarray}
Here  $|{\cal M}_{B}^{q\bar{q}}|^2$
is the  $d$-dimensional squared Born matrix element (\ref{mqqborn}). 
Integrating the gluon  momentum $p_3$ over  the soft region 
 we obtain
\begin{eqnarray}
d\sigma_{\rm soft}^{q\bar{q},g} &= &
{1\over 2\hat s} \Phi_{q{\bar q}} \, d\Gamma_2
\int |M_{\rm soft}^{q{\bar q},g}|^2 \, \Theta(x_{\rm min}-x_g)
{d^{d-1}p_3\over (2\pi)^{d-1}2 E_g} \nonumber \\
 &=&  d\sigma_{B}^{q\bar{q}} \: {\alpha_s\over \pi}
 \tilde{C}_{\epsilon} \, 
\Bigg\{ 
{C_F\over \epsilon^2}+{1\over \epsilon}\*
\Bigg[
C_F-{1+\beta^2\over 4\*\beta \*N}\*\ln(x) -
{2\over N}\* \ln\left({1+\beta\* y\over 1-\beta \*y}\right)
-{N\over 2}\*\ln\left({(1-\beta \*y)^2\over 1-\beta^2}\right)\Bigg]
\nonumber \\
&-&C_F\*{\pi^2\over 6}+{1+\beta^2\over 2\*N\* \beta}\*\Big\{
{\rm Li}_2(1-x)+{1\over 4}\*\ln^2(x)\Big\}-{C_F\over \beta} \* \ln(x)
-{N\over 4}\*\ln^2(x)\nonumber \\
&+&{N^2-2\over 2\*N}\*{\cal S}(y)+{1\over N}\*{\cal S}(-y)
\Bigg\},
\end{eqnarray}
where
\begin{eqnarray}
\tilde{C}_{\epsilon}&=&
\left({4\*\pi\*\mu^2\over \hat{s}}\right)^{\epsilon} 
\*{x_{\rm min}^{-2\*\epsilon} 
\over \Gamma(1-\epsilon)} \, \nonumber \\ 
\end{eqnarray}
and
\begin{eqnarray}
{\cal S}(y)=2 \* {\rm Li}_2\left (-{\beta\* (1-y)\over 1-\beta}\right ) -
2 \* {\rm Li}_2\left (-{\beta \* (1+y)\over 1-\beta\*  y}\right )+
\ln^2\Big( {1-\beta \* y\over 1-\beta}\Big).
\end{eqnarray}

Using again the eikonal approximation the $d$-dimensional soft 
$gg$ fusion matrix element squared takes the form

\begin{eqnarray}
|{\cal M}_{\rm soft}^{gg,g}|^2 &=& 4\pi\alpha_s N \mu^{2\epsilon}\*
\Bigg\{\Big[(N^2-1)M_2-M_1 
-M_{12} \Big]\*
\Big[{p_2\cdot k_1\over (p_2\cdot p_3)(k_1\cdot p_3)}+
{p_1\cdot k_2\over (p_1\cdot p_3)(k_2\cdot p_3)}\Big]
\nonumber \\ &+&\Big[(N^2-1) M_1-M_2-M_{12}\Big]\*
\Big[{p_2\cdot k_2\over (p_2\cdot p_3)(k_2\cdot p_3)}
+{p_1\cdot k_1\over (p_1\cdot p_3)(k_1\cdot p_3)}\Big]   \nonumber \\
&+&N^2 \Big[M_1+M_2\Big] {p_1\cdot p_2\over (p_1\cdot p_3)(p_2\cdot
p_3)}\nonumber \\
&+&{N^2-1\over 2N^2}\* \Big[M_{12} - 
 (N^2-1) (M_1+M_2)\Big] \* \Big[
{m^2\over (k_1\cdot p_3)^2}+{m^2\over (k_2\cdot p_3)^2}\Big]
\nonumber \\
&+&{1\over N^2}\Big[ M_1+M_2+(N^2+1)M_{12}\Big]
{k_1\cdot k_2\over (k_1\cdot p_3)(k_2\cdot p_3)}
\Bigg\} \, ,
\end{eqnarray}
where 
\begin{eqnarray}
M_1&=&(1+\beta y)^2\* |\tilde{\cal M}_{B}^{gg}|^2 \, , 
\nonumber \\
M_2&=&(1-\beta y)^2\* |\tilde{\cal M}_{B}^{gg}|^2 \, ,
\nonumber \\
M_{12}&=&2\*(1-\beta^2 y^2)\* |\tilde{\cal M}_{B}^{gg}|^2 \, ,
\end{eqnarray}
and $|\tilde{\cal M}_{B}^{gg}|^2$ is given in Section 
\ref{sec:BornResults_gg}. 
Integrating  $p_3$ over the  soft region we obtain
\begin{eqnarray}
d\sigma_{\rm soft}^{gg,g}&=&{1\over 2\hat{s}}\Phi_{gg}~
{\alpha_s\over \pi}\tilde{C}_{\epsilon} 
\*
\Bigg\{
{\cal S}_A |{\cal M}_{B}^{gg}|^2
+|\tilde{\cal M}_{B}^{gg}|^2\Big[{\cal S}_B+{\cal S}_C+
{\cal S}_C|_{y\rightarrow -y}\Big]
\Bigg\} d\Gamma_2 \, ,
\end{eqnarray}
where
\begin{eqnarray}
{\cal S}_A&=&{N\over \epsilon^2}+{C_F\over \epsilon}-{N\over 6}\*\pi^2
-{C_F\over\beta}\*\ln(x),\nonumber \\
{\cal S}_B&=&{1+\beta^2\over 2\*\beta\*N}\*
\left[N^2\*(1-\beta^2\*y^2)+2\right]\*
\left[{\ln(x)\over \epsilon}-2\*\Li2(1-x)-{1\over 2}\*\ln^2(x)\right]
\nn \\
&+&{N\over 2}\*\left[4-N^2\*(1+\beta^2\*y^2)\right]\*\ln^2(x),
\nonumber \\
{\cal S}_C&=&{N\over 2}\*\left[4-N^2\*(1+\beta\*y)^2\right]\*
\left[{1\over \epsilon}\*
\ln\left({(1-\beta\*y)^2\over 1-\beta^2}\right)-{\cal S}(y)\right].
\end{eqnarray}

Adding  the virtual and soft contributions, we obtain the following 
results for  
the remaining singular contribution:
\begin{eqnarray}\label{IRdiv}
  \left[d\sigma_V^{q\bar{q}}+d\sigma_{\rm soft}^{q\bar{q},g}\right]_{\rm singular}
  &=&-{\alpha_s\over 2\*\pi}\*
  C_F\*{1\over \epsilon}\*\left[3+4\*\ln(x_{\rm min})\right]
  \*d\sigma_{B}^{q\bar{q}},\nn \\
  \left[d\sigma_V^{gg}+d\sigma_{\rm soft}^{gg,g}\right]_{\rm singular}
  &=&-{\alpha_s\over 2\*\pi}\*
  {1\over \epsilon}\*\left[{11\*N-2\*(N_f-1)\over 3}+
    4\*N\*\ln(x_{\rm min})\right]\*d\sigma_B^{gg}.
\end{eqnarray}
\subsection{Collinear contributions  and mass factorization}
The order $\alpha_s^3$ 
cross sections of the reactions Eqs.~(\ref{ttg})--(\ref{qbarg}) develop
collinear singularities when the momentum of the outgoing massless parton
becomes parallel to the momentum  of the incoming massless quark
or gluon. 
We use, for convenience, the parameter $x_{\rm min}$ 
for characterizing the collinear regions in phase-space, too. 
For Eqs.~(\ref{ttg}) and (\ref{ggg}),
we define the two collinear regions  by 
$\{\cos\theta >(1-x_{\rm min})$ and $ x_g>x_{\rm min}\}$
and  $\{\cos\theta <(-1+x_{\rm min})$ and $ x_g>x_{\rm min}\}$, 
where
$\theta$ is the angle between the gluon and one of the initial
partons in the c.m. frame of the initial partons.
For Eqs.~(\ref{qgq}) and (\ref{qbarg}) the two collinear regions
are defined by $\cos\theta >(1-x_{\rm min})$ and
$\cos\theta <(-1+x_{\rm min})$. Here 
$\theta$ denotes the angle between the outgoing massless (anti)quark
and one of the initial partons.
\par
In these regions we use the collinear 
approximations for both the squared matrix element of the
respective reaction and 
the phase-space measure in $d$ dimensions. We  parametrize 
$p_3=(1-z)p_i+ p_{\bot}+{\cal O}( p_{\bot}^2)$, $p'=zp_i
- p_{\bot}+{\cal O}( p_{\bot}^2)$, $p_3\cdot p_{\bot}=0$,
where $p_i=p_1, p_2$ is the momentum of one
of the initial partons, $p_3$ is the momentum of
the unobserved collinear outgoing parton,  and $p'$ is the 
momentum of parton $i$ after collinear emission. The momentum fraction $z$
varies in the interval 
$z\in [{4m^2}/{\hat{s}},1 - \delta]$, where $\delta=x_{\rm min}$ 
for $q\bar{q}$ 
and $gg$ initial states, and  $\delta=0$ for the $gq (\bar{q})$
initial state. Integrating over the angles of the  parton with
momentum $p_3$ in the collinear region, using the angular
measure in $d-1$ spatial dimensions, 
$d\Omega_{d-1}=(1-\cos^2\theta)^{-\epsilon}d\cos\theta d\Omega_{d-2}$,
we obtain for the collinear contribution to the differential cross sections:
\begin{equation}
  d\sigma_{\rm coll}^{q\bar{q},g}(z)=
  -{\alpha_s C_F\over 2\pi} F_{\epsilon}
  {1\over \epsilon}
  {1+z^2-\epsilon (1-z)^2\over 1-z}
  \Big\{ d\sigma_{B}^{q\bar{q}}(z p_1,p_2)+
  d\sigma_{B}^{q\bar{q}}(p_1,z p_2)\Big\}dz,
  \label{qqcol}
\end{equation}
\begin{equation}
  d\sigma_{\rm coll}^{gg,g}(z)=-{N\alpha_s\over \pi} F_{\epsilon} 
  {1\over \epsilon}\Big\{
  {1\over 1-z} 
  + {1\over z} -z^2 + z -2
  \Big\}\Big\{ d\sigma_{B}^{gg}(z p_1,p_2)+d\sigma_{B}^{gg}(p_1, z p_2)
  \Big\}dz,
  \label{ggcol}
\end{equation}
\begin{eqnarray}
  d\sigma_{\rm coll}^{gq}(z)
  =d\sigma_{\rm coll}^{g\bar{q}}(z) 
  &=& - {\alpha_s\over 2\*\pi} F_{\epsilon} {1\over \epsilon}
  \*\Big\{{1\over 2}\*{z^2+(1-z)^2-\epsilon\over 1-\epsilon}
  d\sigma_{B}^{q\bar{q}}(p_1, z p_2) \nonumber \\
  &+&   C_F\* {1+(1-z)^2-\epsilon z^2\over z}d\sigma_{B}^{gg}(z p_1,p_2)
  \Big\}dz,
  \label{qgcol}
\end{eqnarray}
where
\begin{equation}
  F_{\epsilon}=\Big[{8\pi\mu^2\over \hat{s}}\Big]^{\epsilon}
  {x_{\rm min}^{-\epsilon} (1-z)^{-2\epsilon} \over \Gamma(1-\epsilon)} \, .
\end{equation}
The Born cross sections  in Eqs.~(\ref{qqcol})--(\ref{qgcol}) 
are those given in Section 
\ref{sec:BornResults}.
The $z$-dependent terms in
front of the $d\sigma^i_B$,  which appear in 
these equations are, as expected, proportional to
the $d$-dimensional Altarelli--Parisi splitting functions 
$P^d_{qq}(z)$, 
$P^d_{gg}(z)$,  $P^d_{gq}(z)$, 
and $P^d_{qg}(1-z)$. 
In  Eqs.~(\ref{qqcol}) and (\ref{ggcol}), 
the integration over the momentum fraction $z$ can be performed
using the relation
\begin{eqnarray} \label{plusint}
\int\limits_{{4m^2\over \hat{s}}}^{1-x_{\rm min}} 
d z \, {g(z)\over (1-z)^{1+\epsilon}}
 = \int\limits_{{4m^2\over \hat{s}}}^1 dz \, g(z) \, 
\Big [ {1\over (1-z)_+}-  \delta(1-z) \, \ln(x_{\rm min}) \nonumber \\
- \epsilon \, \Big[{\ln(1-z)\over 1-z}\Big]_+ 
+{1\over 2}\epsilon \, \delta(1-z)\ln^2(x_{\rm min}) +
{\cal O}(\epsilon^2) \Big ].
\end{eqnarray}
The plus prescription defines distributions via
\begin{equation}
\left[F(z)\right]_+
= \lim_{\eta\rightarrow 0}
  \left\{ \Theta(1-z-\eta) F(z) - \delta(1-z-\eta) \int_0^{1-\eta} F(y) dy
   \right\},
\end{equation}
so that if $g(z)$ is well-behaved at $z=1$, then
\begin{eqnarray}
\int_x^1 dz\; {g(z)\over (1-z)_+} &=& \int_x^1 {g(z)-g(1)\over 1-z} + 
g(1) \ln(1-x),\\
\int_x^1 dz\; g(z)\Bigl[{\ln(1-z)\over 1-z}\Bigr]_+ &=&
   \int_x^1 {(g(z)-g(1))\ln (1-z)\over 1-z} + {g(1)\over2} \ln^2(1-x).
 \label{PlusPrescriptionExample}
\end{eqnarray}
Adding, in the case of  $i=q{\bar q}, gg$, the soft
and collinear contributions, the singularities proportional
to  $\ln(x_{\rm min})$ cancel for infrared- and collinear-safe observables. 
The singularities, which remain in
$d\sigma^i_V + d\sigma^{i,g}_{\rm soft}$  and $d\sigma^{i,g}_{\rm coll}$,
are removed by absorbing them
into the unphysical bare parton density functions,
i.e. by  renormalizing these  functions at 
the factorization scale $\mu_F$. At the level of the 
differential cross sections, this amounts to adding  
counterterms $d\sigma_c^i$ being composed of 
mass factorization counterfunctions. Finite Born cross sections
for $gq$ and $g\bar q$ scattering 
are obtained in an analogous fashion. Performing the mass factorization  
in the  $\overline{\rm{MS}}$ scheme, the counterterms read
\begin{equation}
  d\sigma^{q\bar q}_c(z) =
  {1\over \Gamma(1-\epsilon)} \left(
    {4\pi\mu^2\over \mu_F^2}\right)^{\epsilon}
  \,  {1\over \epsilon}{\alpha_s\over 2\pi} 
  \Big [ P_{qq}(z)
  d\sigma_{B}^{q\bar q}(z p_1,p_2)
  +P_{{\bar q} {\bar q}}(z)d\sigma_{B}^{q\bar q}(p_1,zp_2)
  \Big ]dz
\, ,
\end{equation}
\begin{equation}
  d\sigma^{gg}_c(z)=
  {1\over \Gamma(1-\epsilon)} \left(
    {4\pi\mu^2\over \mu_F^2}\right)^{\epsilon}
  \,  {1\over \epsilon} {\alpha_s\over 2\pi} 
  \Big [ P_{gg}(z)
  d\sigma_{B}^{g g}(z p_1,p_2)
  + P_{gg}(z) d\sigma_{B}^{gg}(p_1, zp_2)\Big ]dz
  \, ,
\end{equation}
\begin{equation}
  d\sigma^{gq}_c(z)=
  {1\over \Gamma(1-\epsilon)} \left(
    {4\pi\mu^2\over \mu_F^2}\right)^{\epsilon}
  \,  {1\over \epsilon} {\alpha_s\over 2\pi} 
  \Big [ P_{gq}(z)
  d\sigma_{B}^{g g}(z p_1,p_2)
  +P_{qg}(z)d\sigma_{B}^{q\bar q}(p_1, zp_2)\Big ]dz
  \, ,
\end{equation}
and $d\sigma^{g\bar{q}}_c(z)=d\sigma^{gq}_c(z)$.
The evolution kernels are given by \cite{Gribov:rt,Gribov:ri,Lipatov:qm,
Altarelli:1977zs,Dokshitzer:sg}
\begin{eqnarray}
  \label{eq:pgg}
  P_{gg}(z)&=&
  \delta(1-z) {11N-2 (N_f-1)\over 6}
  +2\*N\*\Big[ {z\over (1-z)_+}+{1-z\over z}+z (1-z)\Big], \\
  \label{eq:pqq}
  P_{qq}(z)&=&P_{{\bar q} {\bar q}}(z)
  = C_F\Big[
  {3\over 2}\delta (1-z)+{1+z^2\over (1-z)_+}\Big],
  \\
  \label{eq:pgq}
  P_{gq}(z)&=&P_{g\bar{q}}(z)= C_F
  {1+(1-z)^2\over z}, \\
  \label{eq:pqg}
  P_{qg}(z)&=&P_{\bar{q}g}(z)={1\over 2}
  \Big[ z^2+(1-z)^2\Big].
\end{eqnarray}

\def\Sp{{\mathbf S}_t}
\newcommand{\Sm}{{\mathbf S_{\bar t}}}

\subsection{Finite contributions}
In the resolved regions the four differential
cross sections $d\sigma^i_{\rm res}$, respectively the corresponding
production density matrices $R^i_{\rm res}$ for the
final states $t{\bar t}+g$, $t{\bar t}+q$, $t{\bar t}+\bar q$, are
obtained from the  helicity amplitudes given in Appendix
\ref{sec:HelAmps} according to
formula (\ref{density}). In the resolved regions no
singularities
arise and it is thus possible to work in $d=4$ dimensions.

\section{Parton level results for the final states $\mathbf{t{\bar t}X}$}
\label{partonsection}
Based on the results given in the previous sections 
we consider in this the inclusive
parton reactions 
\begin{equation}
i\to t{\bar t} \; X \, , \quad i=q{\bar q}, gg, gq, g{\bar q}
\label{ttink}
\end{equation}
at NLO. From the different contributions 
$d\sigma_B^i, d{\sigma}^i_V,
 d{\sigma}^{i,g}_{\rm soft}$,  given in Sections \ref{calc} and \ref{real}, 
we can extract 
the corresponding density matrices using
Eq.~(\ref{spinid1}). For $d{\sigma}^{i,g}_{\rm coll}$ and
$d{\sigma}^{i}_{c}$ an analogous formula holds.
These and the above-mentioned matrices
$R^i_{\rm res}$ have the following structure in the spin spaces
of the top and antitop quarks:
\begin{eqnarray}
(R^{i})_{\alpha\alpha',\beta\beta'} = 
A^i \delta_{\alpha\alpha'}\delta_{\beta\beta'}
+ B^i_{ta} (\tau^a)_{\alpha\alpha'}\delta_{\beta\beta'}
+ B^i_{{\bar t}a} \delta_{\alpha\alpha'}(\tau^a)_{\beta\beta'}
+\, C^{i}_{ab}(\tau^a)_{\alpha\alpha'}
(\tau^b)_{\beta\beta'}  \,\, ,
\label{eq:Rstruct}
\end{eqnarray}
where $\tau^a$ are the Pauli matrices. 
The functions $A^i$
determine the spin-averaged production cross section.  The
functions $B^i_{ta}$ and $B^i_{{\bar t}a}$ are associated with a 
polarization of
the top quarks and antiquarks. As mentioned earlier, parity invariance of QCD
only allows top and antitop polarizations that are induced by
absorptive parts of the scattering amplitudes. To NLO QCD, 
the corresponding polarization is orthogonal to the  $2\to 2$ scattering 
plane, i.e. the ${\bf B}^i_{t}$ 
and ${\bf B}^i_{{\bar t}}$ have no components in the
scattering plane.  The functions 
$C^{i}_{ab}$ encode the top--antitop spin-spin correlations.
For a general decomposition of the $2 \to 2$ 
density matrices and their classification with respect to
discrete symmetries, see  Ref.~\cite{Bernreuther:1993hq}.

In the following we consider a general observable ${\cal O}$ that 
may depend on the spins 
and momenta of the top and antitop quarks and is thus sensitive
to the different structures shown in  Eq.~(\ref{eq:Rstruct}). 
In particular we study  observables of the following form 
\begin{equation}
{\cal O} = 4 \,(\Sp\cdot {\bf{\hat a}})
(\Sm\cdot {\bf{\hat b}})  \,  , 
\label{genob}
\end{equation}
which are sensitive to the correlation $C^{i}_{ab}$. 
The top and antitop spin operators are given by 
\begin{equation}
  \Sp={1\over 2}(\mathbf{\tau}\otimes \one)
\end{equation}
and 
\begin{equation}
  \Sm={1\over 2}(\one\otimes\mathbf{\tau}).
\end{equation}
The unit vectors ${\bf{\hat a}}$ and ${\bf{\hat b}}$ are 
arbitrary reference directions. The factor 4 in Eq.~(\ref{genob}) is
introduced to have a simple relation between the expectation
values of the observables defined in 
Eq.~(\ref{genob}) and double spin asymmetries. At the parton level
the following relation holds:
\begin{equation}
\label{double}
  4 \langle (\Sp\cdot  {\bf \hat a})(\Sm \cdot  {\bf \hat b}) \rangle_i
  = {\sigma^i (\uparrow \uparrow)+\sigma^i(\downarrow \downarrow)
  - \sigma^i(\uparrow \downarrow)- \sigma^i(\downarrow 
\uparrow)\over
  \sigma^i(\uparrow \uparrow)+\sigma^i(\downarrow \downarrow)
  + \sigma^i(\uparrow \downarrow)+ \sigma^i(\downarrow \uparrow)
  }.
\end{equation}
The arrows on the right-hand side refer to the spin state of the top 
and antitop quarks  with respect to the  quantization axes 
$ {\bf \hat a}$ and $ {\bf \hat b}$. 
A prescription of how
to construct the correlated top and antitop rest frames in a unique way has 
to be given. At that point it is 
important to make sure that the specific prescription is soft and
collinear-safe. To illustrate the problem, consider
the top--antitop helicity correlation, in which case $ {\bf \hat a}$ and $
{\bf \hat b}$ are the $t$ and $\bar{t}$ directions of flight. An obvious
frame in which these directions may be defined is the c.m. frame of
the initial partons. In fact, this is how the helicity correlation
at parton level
is usually defined in Born level calculations for hadron colliders.
However, this frame can only be constructed by a measurement
of the 4-momenta of all final state particles/jets.
In particular, at NLO QCD, one needs to know, apart from the top and
antitop momenta, also the momentum of the hard gluon emitted in 
real radiation. Obviously, this information cannot be obtained
if the gluon is collinear to one of the initial partons. 
Thus the above definition of the helicity correlation is not
collinear-safe
and cannot be applied beyond the leading order. 
This applies also to other observables involving the momenta
of the top and antitop defined in the c.m. frame of the initial-state partons. 
In the situation at hand, a suitable frame is the
$t\bar{t}$-ZMF defined in Section \ref{calc}.   
As discussed in Section \ref{calc} the top (anti)quark rest frame is
then defined through a rotation-free Lorentz boost.
We study the following spin bases, which are
relevant to applications to the Tevatron and the LHC 
(see  Sections \ref{angular} and \ref{numres}):
\begin{eqnarray}
     {\bf\hat a} = -  {\bf\hat b} = \kh,&&
    \mbox{(helicity basis)}
\label{helbasis}, \\
     {\bf\hat a} = {\bf\hat b} = {\bf\hat p}, &&\mbox{(beam\
      basis)}
\label{beambasis},\\
     {\bf\hat a} =  {\bf\hat b} = \dhh,&&
    \mbox{(off-diagonal\ basis)},
\label{offbasis}
\end{eqnarray}
where $\kh $ denotes the direction of
flight of the top quark in the  $t\bar{t}$-ZMF
and ${\bf\hat p}$ is the direction of flight of one of the colliding
hadrons in that frame. The direction of the hadron beam can be identified to 
a very good approximation with the direction of flight of one of the
initial partons. Thus,  at Born level,  the beam basis  
$ {\bf\hat a} = {\bf\hat b} = {\bf\hat p}$ coincides with the
direction of flight ${\bf\hat p}^*$ 
of one of the initial partons in the c.m. frame
of the initial partons, and  this vector  is equal to 
the unit vector of one of the hadron beams
in the laboratory frame.
The term `off-diagonal basis' refers to axes 
with respect to which the spins
of tops and antitops produced by $q {\bar q}$ annihilation are 100\%
(anti)correlated \cite{Mahlon:1997uc} to leading order in $\alpha_s$.
(For $gg\to t\bar t$ one can show that no spin basis
with this property exists.) We use the definition:
\begin{equation}
\dhh = \frac{-\ph+(1-\gamma)(\ph\cdot\kh)\kh}
{\sqrt{1-(\ph\cdot\kh)^2(1-\gamma^2)}} \, , 
\label{dopt}
\end{equation}
where  $\gamma=E/m$.
 The spin bases defined
in Eqs.~(\ref{helbasis})--(\ref{offbasis}) lead to spin observables
(\ref{genob}) that are invariant under spatial rotations of the 
$t\bar{t}$-ZMF. Thus predictions involving these observables can be
unambiguously made without an explicit prescription on how to obtain this
frame. 
If in the case of the beam axis one would use 
$ {\bf\hat a} =  {\bf\hat b}={\bf\hat p}^*$ instead,
one would still obtain a collinear-safe observable,
which is however not invariant under rotations  of the $t\bar{t}$-ZMF.
Therefore we do not consider this choice any further. 

In addition we find that the observable
\begin{equation}
{\cal O}'= 4\, \Sp \cdot \Sm\ ,
\label{eq:sbasis}
\end{equation}
which is also infrared- and collinear-safe,
is also sensitive to $t \bar t$ spin correlations at both  the
Tevatron and the LHC. This observable can be 
expressed in terms of observables of the form of Eq.~(\ref{genob}),
because
\begin{equation}
  {\cal O}'=4 \sum_{i=1}^3 (\Sp\cdot {\bf{\hat e}}_i)
  ({\bf{\hat e}}_i\cdot \Sm)\, ,
\end{equation}
where ${\bf{\hat e}}_{i=1,2,3}$ forms an orthonormal basis.

As mentioned earlier, parity invariance of QCD tells us that the $t$ and 
$\bar t$ ensembles  have no polarizations with respect to the reference axes
of Eqs.~(\ref{helbasis})--(\ref{offbasis}):
\begin{equation}
  \label{zero}
  \langle \Sp\cdot {\bf{v}} \rangle_i = \langle \Sm \cdot {\bf{v}}
  \rangle_i
  = 0 \, , \quad\mbox{for}\quad
  {\bf{v}}=  {\bf\hat k },  {\bf\hat p },  {\bf\hat d } .
  \label{sppol}
\end{equation}
Equation (\ref{zero}) holds for any polar vector $\bf v$. 
These equations justify the term {\it spin correlations}
for the double spin asymmetries Eq.~(\ref{double}), since a simple
calculation shows that
\begin{equation}
  4 \langle (\Sp\cdot  {\bf\hat a})(\Sm \cdot  {\bf\hat b}) \rangle_i
  ={\rm corr}(\Sp\cdot  {\bf\hat a},\Sm \cdot  {\bf\hat b})_i,
\end{equation}
where for two observables ${\cal O}_1,{\cal O}_2$ the correlation
is defined in the standard way by
\begin{equation}
  {\rm corr}({\cal O}_1,{\cal O}_2)={\langle {\cal O}_1 {\cal O}_2\rangle
    -\langle {\cal O}_1 \rangle \langle {\cal O}_2\rangle\over
    \delta {\cal O}_1 \delta {\cal O}_2},
\end{equation}
where 
\begin{equation}
  \delta {\cal O}_k=\sqrt{\langle {\cal O}_k^2\rangle
    -\langle{\cal O}_k \rangle^2}.
\end{equation}
Using the results presented in Sections \ref{calc} and  \ref{real},
let us now discuss the observables defined through Eq.~(\ref{genob}),
Eqs.~(\ref{helbasis})--(\ref{offbasis}) and Eq.~(\ref{eq:sbasis}).
The unnormalized expectation value of the observable
 ${\cal O}$ at NLO in the QCD coupling --- and, as a special case,
the NLO cross sections ${\sigma^i}$ for the parton 
reactions in Eq.~(\ref{ttink}) ---  is obtained as follows:
\begin{eqnarray}
{\sigma^i}\langle{\cal O}\rangle_i &=& 
\frac{\Phi_i}{8 \hat s} \Bigg\{ \int d \Gamma_2 \, {\rm Tr} [(R_B^i + R^i_V
+ R^{i,g}_{\rm soft}) {\cal O}] \nn \\ &+& \,  
\int d \Gamma_2 dz \, {\rm Tr} [(R^{i,g}_{\rm coll} + R^{i}_{c} ) {\cal O}]
\, + \, \int d \Gamma_3 \, {\rm Tr} [R^{i}_{\rm res} {\cal O}] \Bigg\} \, 
\label{ggobs}
\end{eqnarray}
for $i = q{\bar q}, gg.$ Here $d\Gamma_3$ is the
3-particle phase-space measure.
The contributions to the left-hand side of Eq.~(\ref{ggobs}) from the soft 
and collinear regions and from the resolved region depend individually
on the slicing parameter $x_{\rm min}$. 
For  $x_{\rm min}\ll 1$ the leading
dependence is, by construction, logarithmic, 
but in the
sum only a residual linear dependence on $x_{\rm min}$ remains, which is due
to the approximations made in the soft and collinear regions.
By varying $x_{\rm min}$ between
$10^{-3}$ and $10^{-8}$, we have checked for the cross sections and for
the expectation values of the observables  given below 
that for $x_{\rm min}\le 10^{-4}$ this residual dependence
is negligible. 

For $i=gq, g{\bar q}$ we have 
\begin{equation}
{\sigma^i}\langle{\cal O}\rangle_i =  \frac{\Phi_i}{8 \hat s} \left\{
\int d \Gamma_2dz \, {\rm Tr}[(R^{i,g}_{\rm coll} + R^{i}_{c} ) {\cal O}]
\, + \, \int d \Gamma_3 \, {\rm Tr}[R^{i}_{\rm res} {\cal O}]  \right\} \, ,
\label{qgobs}
\end{equation}
where $\Phi_{gq}= 1/[2(d-2)N(N^2-1)]$. 
 Of course, $ {\sigma}^{gq}= {\sigma}^{g\bar q}$
and $\langle{\cal O}\rangle_{gq} = \langle{\cal O}\rangle_{g\bar q}$   within QCD. 
The statements made below Eq.~(\ref{ggobs})  apply also here.

The unnormalized expectation values
Eq.~(\ref{ggobs}) and Eq.~(\ref{qgobs}) are  building blocks for
computing the distributions  (\ref{eq:ddist1})
and  (\ref{eq:ddist2}) (see Section \ref{angular}). 
In computing the cross sections and  these 
expectation values  we have performed 
the phase-space integration of the resolved parts
numerically.  If one identifies the renormalization and mass 
factorization scales, and puts $\mu_R=\mu_F=\mu$, then the 
NLO parton cross sections and the above expectation values are of the
form
\begin{equation}
\sigma^{i}(\shat,m^2)={\alpha_s^2\over m^2}[
f^{(0)}_{i}(\rho) + 4\pi\alpha_s(f^{(1)}_{i}(\rho) +
{\tilde f}^{(1)}_{i}(\rho) \ln(\mu^2/m^2))] \, , \quad i=q{\bar q}, gg, gq
\label{eq:xsection}
\end{equation}
\begin{equation}
\sigma^{i} \langle {\cal O}_a \rangle_{i}  = {\alpha_s^2\over m^2}
[ g^{(0)}_{i,a}(\rho) + 4\pi\alpha_s(g^{(1)}_{i,a}(\rho) +
{\tilde g}^{(1)}_{i,a}(\rho) \ln(\mu^2/m^2))] \, , \quad i=q{\bar q}, gg, gq
\label{eq:expval}
\end{equation}
and $a=1,\ldots,4.$ The label $a=1$ refers to the observable
Eq.~(\ref{eq:sbasis}) and 
 $a=2,3,4$  to the 
helicity, beam, and off-diagonal basis. The  variable $\rho$ 
is  defined by
\begin{equation}
\rho =  {4m^2\over \hat{s}}\,  .
\label{defeta}
\end{equation}
For the beam
and off-diagonal basis, one has to compute {\it two} 
contributions for the $gq$ initial
state: either the quark or the gluon direction of 
flight in the $t\bar{t}$-ZMF can correspond
to the direction used to define the axes ${\bf\hat a}$, ${\bf\hat b}$, and
the contribution from hard gluon emission is different for these two 
cases.  
We have 
\begin{equation}
 {f}^{(0)}_{gq} =  g^{(0)}_{gq,a} =0 \, . 
\label{relgq}
\end{equation}
The lowest order scaling functions $f^{(0)}_{i}$ and $g^{(0)}_{i,a}$
 $(i=q{\bar q},gg)$
can be computed analytically.
In order to write these functions in a compact form 
it is advantageous to introduce the following functions: 
\begin{eqnarray}
\ell_1&=&{1\over \beta}\*\left[\ln(x)+2\*\beta
\right], \nn \\
\ell_2&=&{1\over \beta^3}\*
\left[\ln(x)+2\*\beta+{2\over 3}\*\beta^3
\right], \nn \\
\ell_3&=&{1\over \beta^5}\*
\left[\ln(x)+2\*\beta+{2\over 3}\*\beta^3
+{2\over 5}\*\beta^5\right], \nn \\
r_1&=&{1\over \beta}\*
\left[\arctan\left({\beta\over \sqrt{\rho}}\right)-\beta\right],
\end{eqnarray}
where $\beta = \sqrt{1-\rho}$ and $x=(1-\beta)/(1+\beta)$.

We then get:
\begin{eqnarray}
  f^{(0)}_{q\bar{q}}(\rho)&=&{\pi\*\beta\*\rho\over 27}\*(2+\rho), \\
  f^{(0)}_{gg}(\rho)&=&{\pi\*\beta\*\rho\over 192}\*\left[4+\rho+2\*\rho^2-
    (16+16\*\rho+\rho^2)\*\ell_1\right]\\
  g^{(0)}_{q\bar{q},1}(\rho)&=&f^{(0)}_{q\bar{q}}(\rho), \\
  g^{(0)}_{q\bar{q},2}(\rho)&=&{\pi\*\beta\*\rho \over 27}\*(-2+\rho),
  \\
  g^{(0)}_{q\bar{q},3}(\rho)&=&{\pi\*\beta\*\rho\over 27}\*
  {8+3\*\rho+4\*\sqrt{\rho}\over 5},  \\
  g^{(0)}_{q\bar{q},4}(\rho)&=&f^{(0)}_{q\bar{q}}(\rho)\\
  g^{(0)}_{gg,1}(\rho)&=&{\pi\*\beta\*\rho\over 192}\*\left[
    -28+5\*\rho+2\*\rho^2-(16+18\*\rho+\rho^2)\*\ell_1\right], \\
  g^{(0)}_{gg,2}(\rho)&=&{\pi\*\beta\*\rho\over 192}\*
  \left[{52-7\*\rho-22\*\rho^2-2\*\rho^3\over 3}+
    (16+2\*\rho+14\*\rho^2+\rho^3)\*\ell_2
  \right], \\
  g^{(0)}_{gg,3}(\rho)&=&{\pi\*\beta\*\rho\over 192}\*
  \Bigg[-{96\over 5}+{248\over 15}\*\sqrt{\rho}+25\*\rho
  -{44\over 15}\*\rho^{3/2}-{482\over 15}\*\rho^2
  -{68\over 5}\*\rho^{5/2}+{284\over 15}\*\rho^3
  +{2\over 5}\*\rho^4 \nn \\
  &-&
  (16+34\*\rho-64\*\rho^{3/2}-2\*\rho^2-34\*\rho^{5/2}
  +49\*\rho^3+\rho^4)\*\ell_3
  \Bigg], \\
  g^{(0)}_{gg,4}(\rho)&=&{\pi\*\beta\*\rho\over 192}\*{1\over (1+\rho)^2}
  \*\Bigg[
  -28-35\*\rho+14\*\rho^{3/2}+30\*\rho^2
  -18\*\rho^{5/2}+7\*\rho^3+2\*\rho^4
  \nn \\
  &-&(16+50\*\rho+68\*\rho^2+19\*\rho^3+\rho^4)\*\ell_1
  +2\*\rho^{3/2}\*(7-9\*\rho)\*r_1
  \Bigg].
\end{eqnarray}
The functions ${\tilde f}^{(1)}_{i}$ and 
${\tilde g^{(1)}}_{i,a}$ that determine the scale dependence are obtained
from the next formula, which follows from
the renormalization group equations:
\begin{eqnarray}\label{RGE}
  \tilde{f}_{ij}^{(1)}(\rho)=
  {1\over 8\*\pi^2}\*
  \left[
    \beta_0\*f_{ij}^{(0)}(\rho)
    -\int_\rho^1 dz f_{ik}^{(0)}\left({\rho\over z}\right)\*P_{kj}(z)
    -\int_\rho^1 dz f_{jk}^{(0)}\left({\rho\over z}\right)\*P_{ki}(z)
  \right],
\end{eqnarray}
and likewise for the functions $\tilde{g}^{(1)}_{i,a}(\rho)$. Here,
\begin{eqnarray}
\beta_0={1\over 3}(11 N-2\*N_f),
\end{eqnarray}
and the evolution kernels $P_{kj}$ are given 
in Eqs.~(\ref{eq:pgg})--(\ref{eq:pqg}).

For the $q\bar{q}$ initial state we obtain the following results for
the $\mu$-dependence:
\begin{eqnarray}
\tilde{f}^{(1)}_{q\bar{q}}&=&
{\beta\*\rho\over 648\*\pi}\*\left\{
\kappa_1\*(2+\rho)-16\*
{\ln(x)\over \beta}\right\},\nn\\
\tilde{g}^{(1)}_{q\bar{q},1}&=&\tilde{f}^{(1)}_{q\bar{q}}, \\
\tilde{g}^{(1)}_{q\bar{q},2}&=& 
{\beta\*\rho\over 648\*\pi}\*\left\{\kappa_1\*(-2+\rho)
+16\*\beta\*\ln(x)
-{32\over 3}\right\}           
, \\
\tilde{g}^{(1)}_{q\bar{q},3}&=& 
{\beta\*\rho\over 3240\*\pi}\*\Bigg\{
\kappa_1\*(8+3\*\rho+4\*\sqrt{\rho})
+{8\*(-8+\rho)\*\ln(x)\over \beta}
+{32\*(2\*\rho-1)\over \beta}\* 
\arctan\left({\beta\over \sqrt{\rho}}\right)        
\nonumber \\
&-&64\*\sqrt{\rho}\*\ln(\rho)+{32\over 3}\*\sqrt{\rho}+
{16\over 3}\Bigg\}\ , \\  
\tilde{g}^{(1)}_{q\bar{q},4}&=&\tilde{f}^{(1)}_{q\bar{q}},
\end{eqnarray}
where
\begin{eqnarray}
\kappa_1=16\*\ln\left({\rho\over 4\*\beta^2}\right)
+{127\over 3}-2\*N_f.
\end{eqnarray}

For the $gg$ initial state, we only give  analytic results
for the simple cases. With 
\begin{eqnarray}
\kappa_2&=&\ln(x)\*\ln({\rho\over 4})-2\*\Li2\left({1+\beta\over 2}\right)
+2\*\Li2\left({1-\beta\over 2}\right),\nonumber \\
\kappa_3&=&-3\*\ln(x)^2+2\*\pi^2-12\*\Li2(x),
\end{eqnarray}
we have
\begin{eqnarray}
\tilde{f}^{(1)}_{gg}&=&{\rho\over 768\*\pi}\*
\Big\{-6\*\ln\left({\rho\over 4\*\beta^2}\right)
\*(28+31\*\rho)\*\beta+3\*\kappa_2\*(32-16\*\rho+\rho^2)
+\kappa_3\*(16+16\*\rho+\rho^2) \nonumber \\
&-&{2\*\beta\over 15\*\rho}\*(724-3328\*\rho+7449\*\rho^2)
-(198\*\rho+59\*\rho^2-288)\*\ln(x)\Big\}-{f^{(0)}_{gg}\over 12\*\pi^2},
\\
\tilde{g}^{(1)}_{gg,1}&=&
{\rho\over 768\*\pi}\*
\Big\{-6\*\ln\left({\rho\over 4\*\beta^2}\right)\*(60+31\*\rho)\*\beta
+3\*\kappa_2\*(-2+\rho)\*(-16+\rho)
+\kappa_3\*(16+18\*\rho+\rho^2)
\nonumber \\
&-&{2\*\beta\over 15\*\rho}\*(-176-6118\*\rho+7179\*\rho^2)
-(108\*\rho+77\*\rho^2-672)\*\ln(x)\Big\}
-{g^{(0)}_{gg,1}\over 12\*\pi^2},
\\
\tilde{g}^{(1)}_{gg,2}&=&
{\rho\over 768\*\pi}\*
\Big\{{6\over\beta}\*\ln\left({\rho\over 4\*\beta^2}\right)
\*(60-25\*\rho+31\*\rho^2)
-6\*\kappa_2\*(16-9\*\rho+16\*\rho^2)
\nonumber \\ &-&
(16+2\*\rho+14\*\rho^2+\rho^3)\*{\kappa_3\over \beta^2}
-{2\*\beta\over 15\*\rho}\*(7384+1127\*\rho+3399\*\rho^2)
\nonumber \\ &-&(-231\*\rho+329\*\rho^2+870)\*\ln(x)
- {396\*(\rho^2-\rho+1)^2
\over \rho\*\beta^2}
\*\left[\Li2(-\beta)-\Li2(\beta)\right]\Big\}
-{g^{(0)}_{gg,2}\over 12\*\pi^2}.
\end{eqnarray}
For the remaining two functions $\tilde{g}^{(1)}_{gg,3,4}$ we give simple fits
in Appendix \ref{sec:FitFunctions}. 

For the $gq$ initial state,
we obtain:
\begin{eqnarray}
\tilde{f}^{(1)}_{gq}&=&{1\over 8\*\pi^2}\*{\pi\over 192}\*
\Big\{{32\over 3}\*\rho\*(2-\rho)\*\kappa_2
-{4\over 9}\*\rho\*
(14\*\rho^2+27\*\rho-136)\*\ln(x) \nonumber \\
&-&{8\*\beta\over 135}
\*(1319\*\rho^2-3468\*\rho+724)\Big\},
\\
\tilde{g}^{(1)}_{gq,1}&=&{1\over 8\*\pi^2}\*{\pi\over 192}\*
\Big\{{4\over 3}\*\rho\*(16-9\*\rho)\*\kappa_2
-{4\over 9}\*\rho\*(14\*\rho^2-27\*\rho-328)\*\ln(x)
\nonumber \\
&-&{8\*\beta\over 135}\*
(1319\*\rho^2-6258\*\rho-176)\Big\}
,
\\
\tilde{g}^{(1)}_{gq,2}&=&{1\over 8\*\pi^2}\*{\pi\over 192}\*
\Big\{{4\over 3}\*\rho\*(-16+9\*\rho)\*\kappa_2
-{8\over 9}\*\rho\*(3\*\rho^2-26\*\rho+263)\*\ln(x)\nonumber \\
&+&88\*(\rho^2-2\*\rho+2)\*
[\Li2(\beta)-\Li2(-\beta)]
-{8\*\beta\over 135}\*(1439\*\rho^2+1347\*\rho+7384)\Big\}.
\end{eqnarray}
Again we give simple fits
in Appendix \ref{sec:FitFunctions} for the more complicated functions 
$\tilde{g}^{(1)}_{gq,3,4}$.   
\begin{figure}[ht]
  \unitlength1.0cm
  \begin{center}
    \begin{picture}(7.5,7.5)
      \put(-5.85,-3.2){\psfig{figure=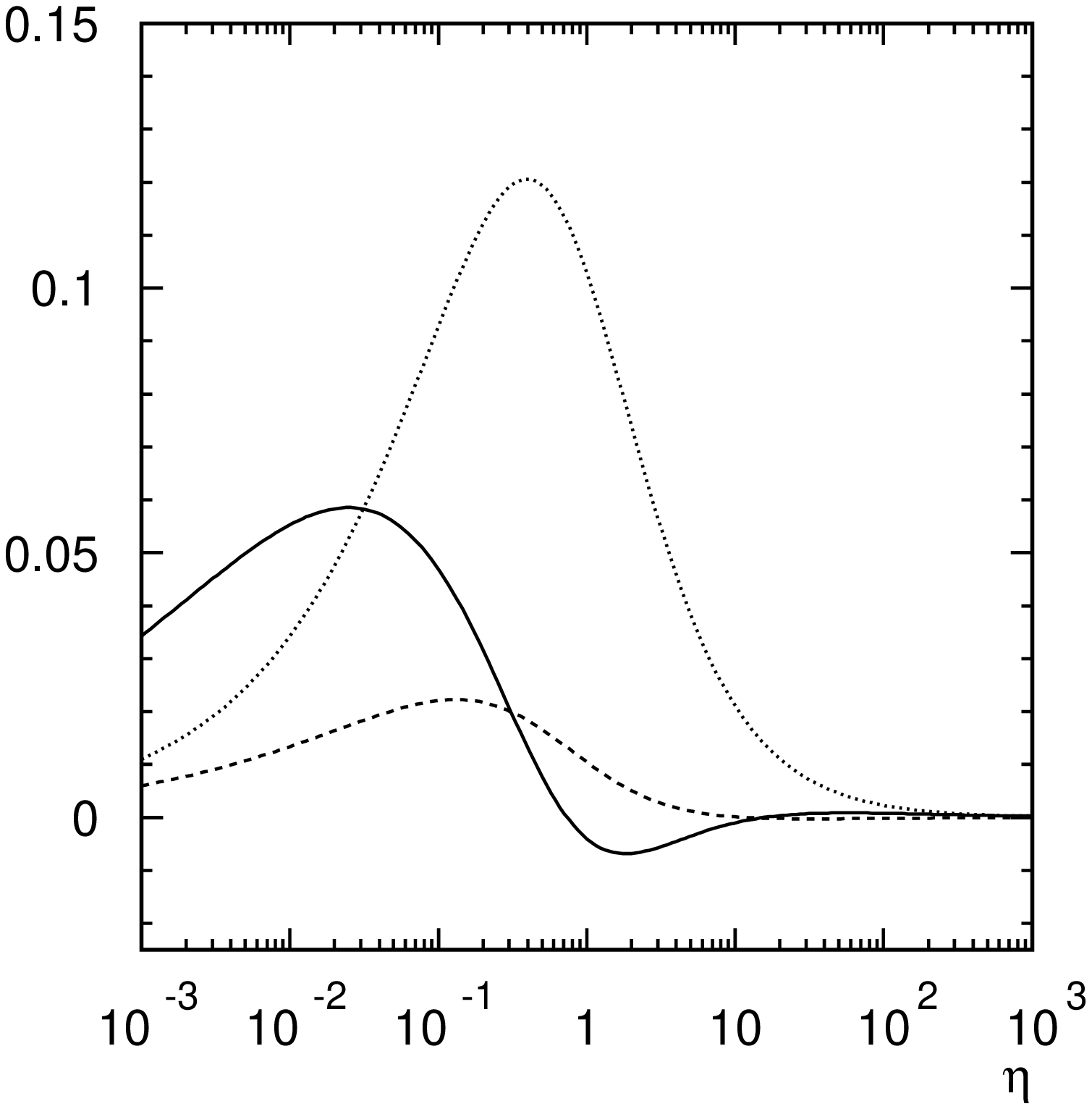,width=9cm}}
      \put(3.15,-3.2){\psfig{figure=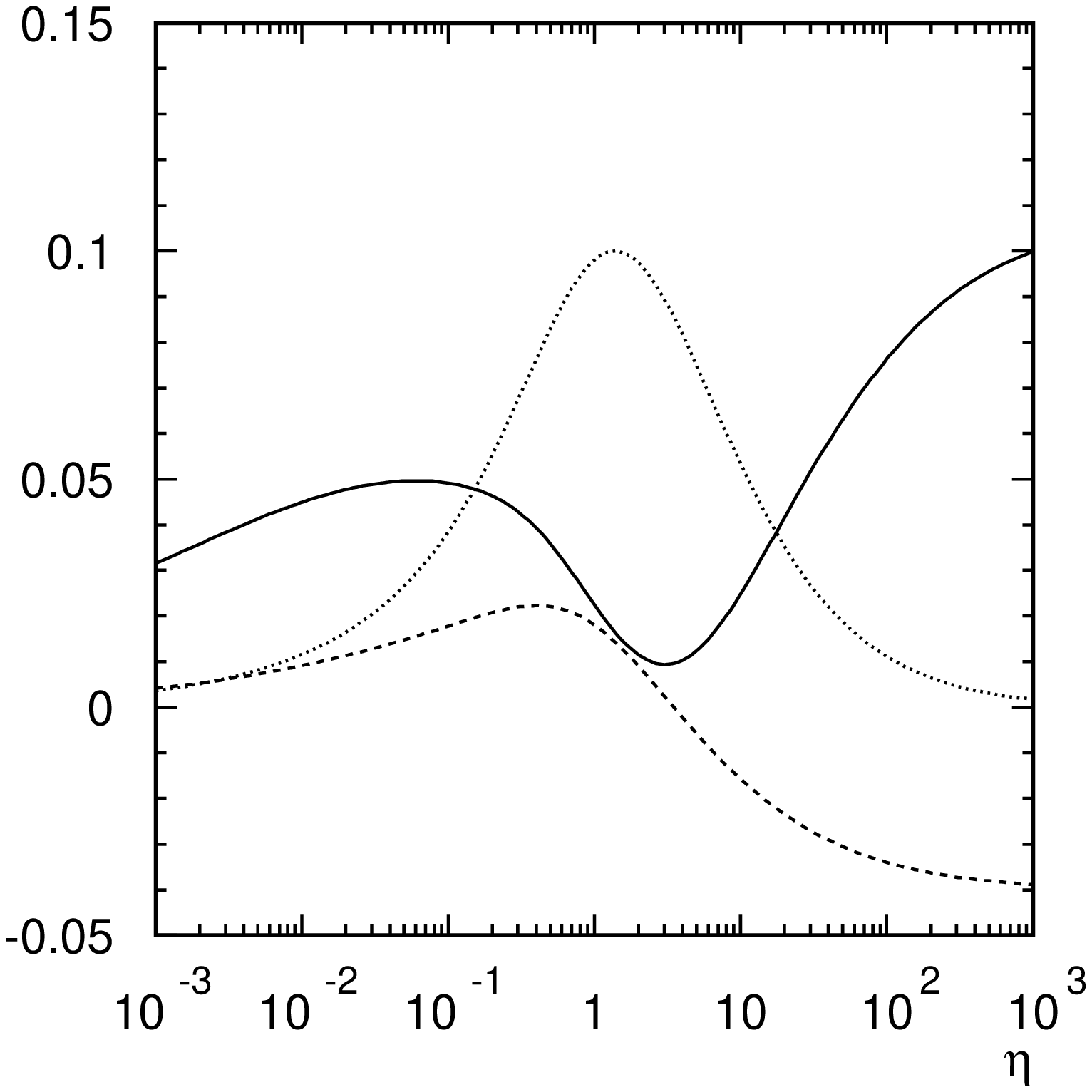,width=9cm}}
    \end{picture}
    \caption{\it Left: Scaling functions $f^{(0)}_{q\bar{q}}(\eta)$ (dotted), 
      $f^{(1)}_{q\bar{q}}(\eta)$ (full), 
      and $\tilde{f}^{(1)}_{q\bar{q}}(\eta)$ (dashed). 
      Right: The same 
      for the process $gg\to t\bar{t}(g)$.}
    \label{fig:o0}
  \end{center}
\end{figure}
\begin{figure}[ht!]
  \unitlength1.0cm
  \begin{center}
    \begin{picture}(7.5,7.5)
      \put(-5.85,-3.2){\psfig{figure=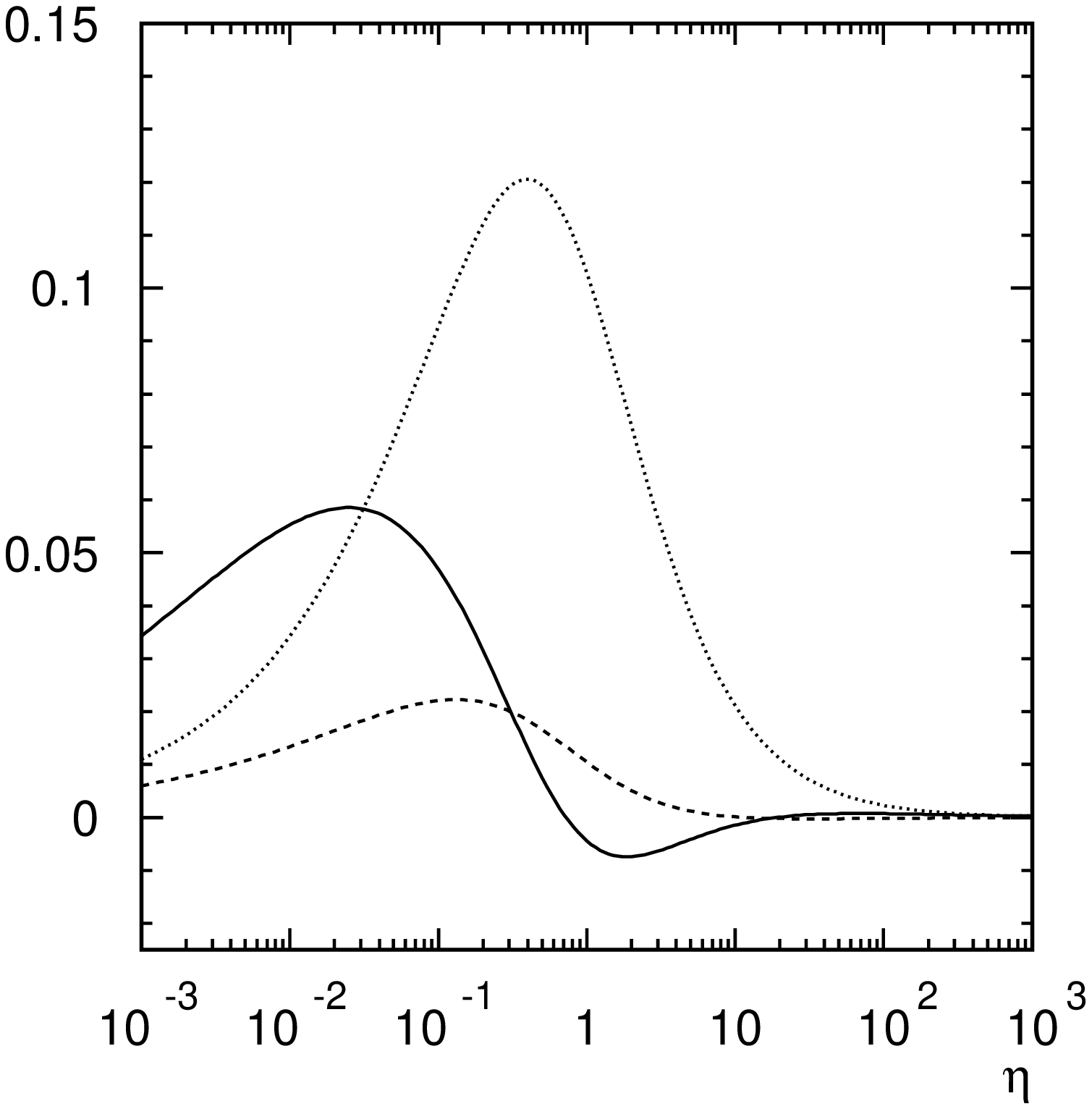,width=9cm}}
      \put(3.15,-3.2){\psfig{figure=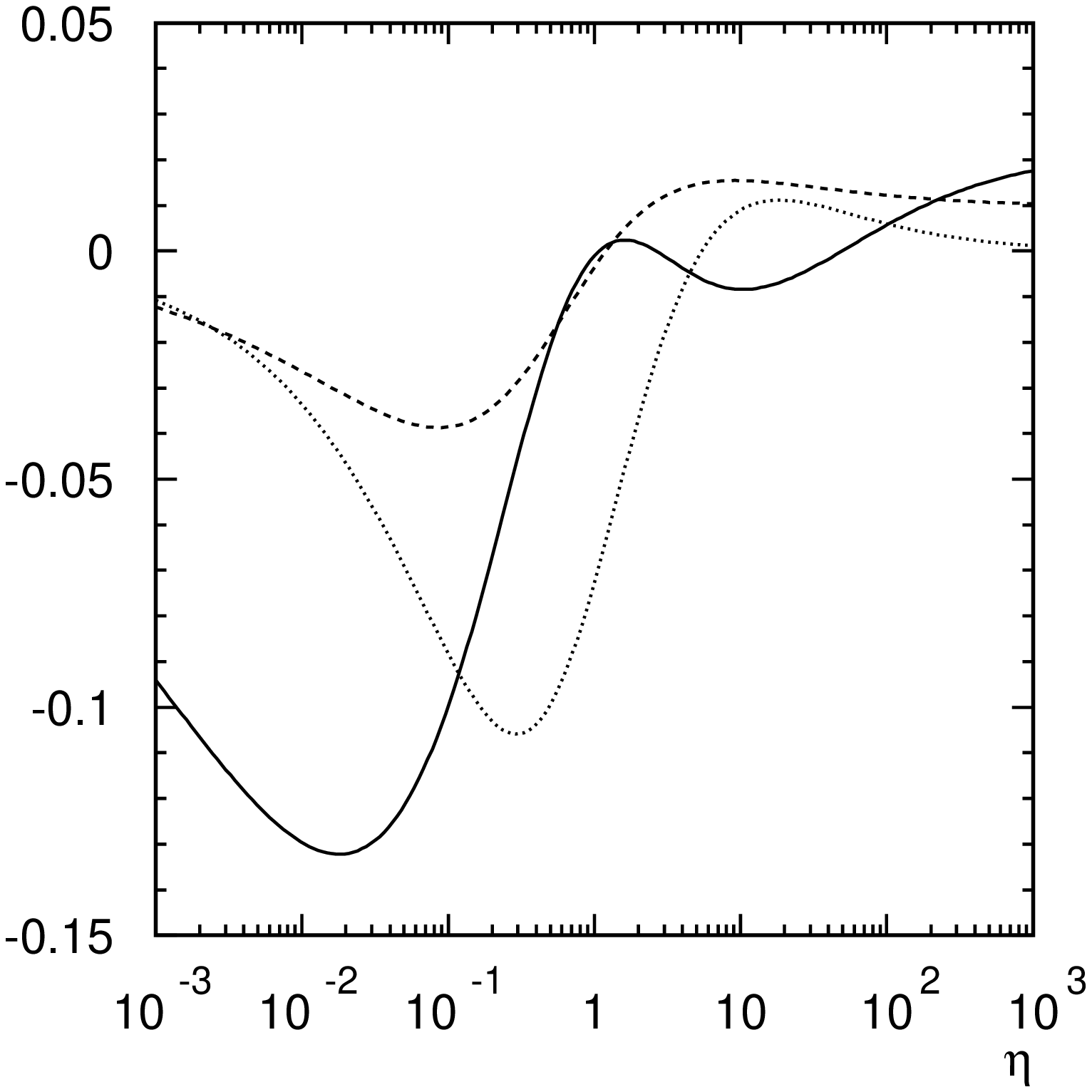,width=9cm}}
    \end{picture}
    \caption{\it Left: Scaling functions $g^{(0)}_{q\bar{q},1}
      (\eta)$ (dotted), 
      $g^{(1)}_{q\bar{q},1}(\eta)$ (full), 
      and $\tilde{g}^{(1)}_{q\bar{q},1}(\eta)$ (dashed).
      Right: The same 
      for the process $gg\to t\bar{t}(g)$.}
    \label{fig:o1}
  \end{center}
\end{figure}
\begin{figure}[ht!]
  \unitlength1.0cm
  \begin{center}
    \begin{picture}(7.5,7.5)
      \put(-5.85,-3.2){\psfig{figure=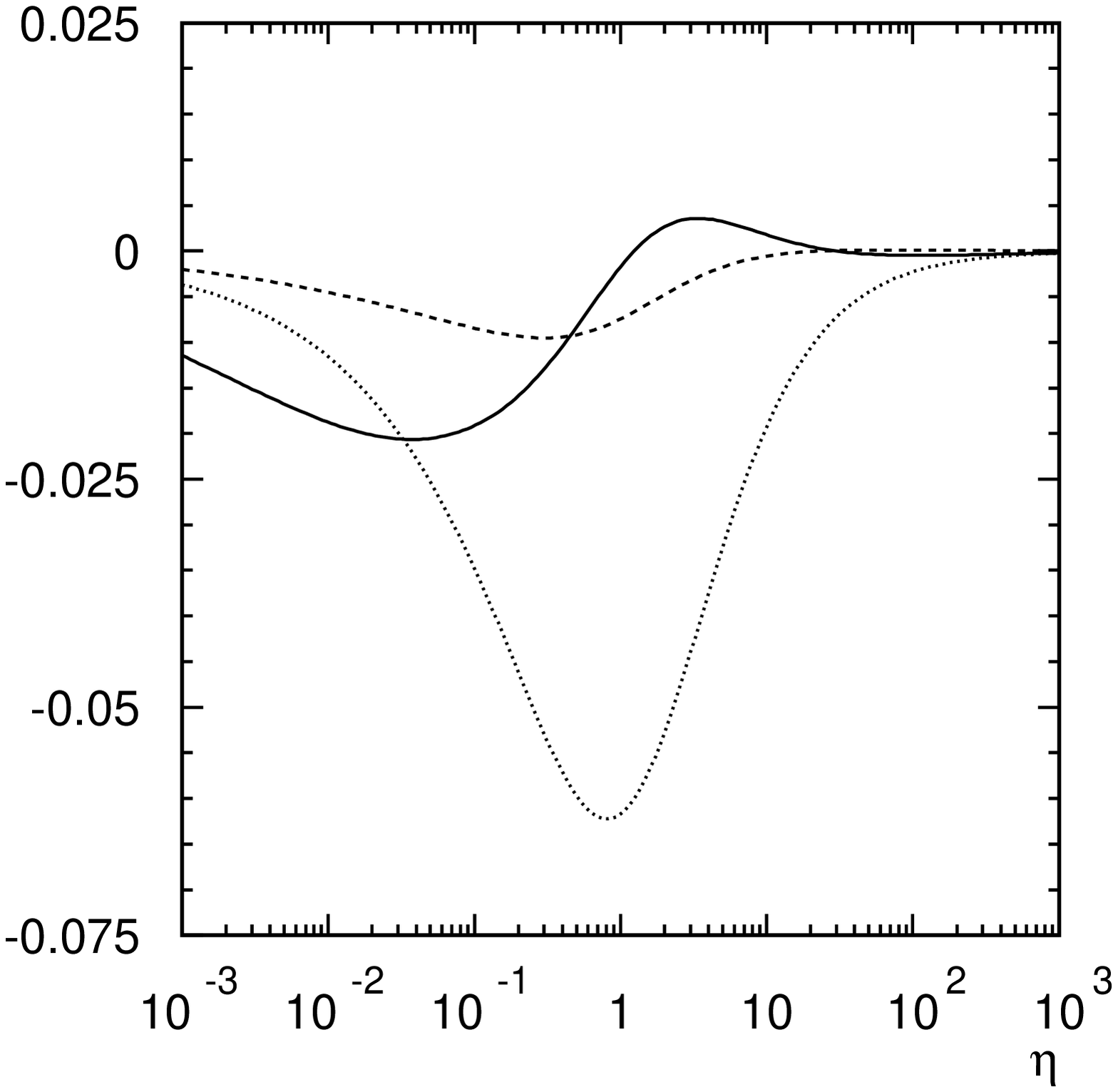,width=9cm}}
      \put(3.15,-3.2){\psfig{figure=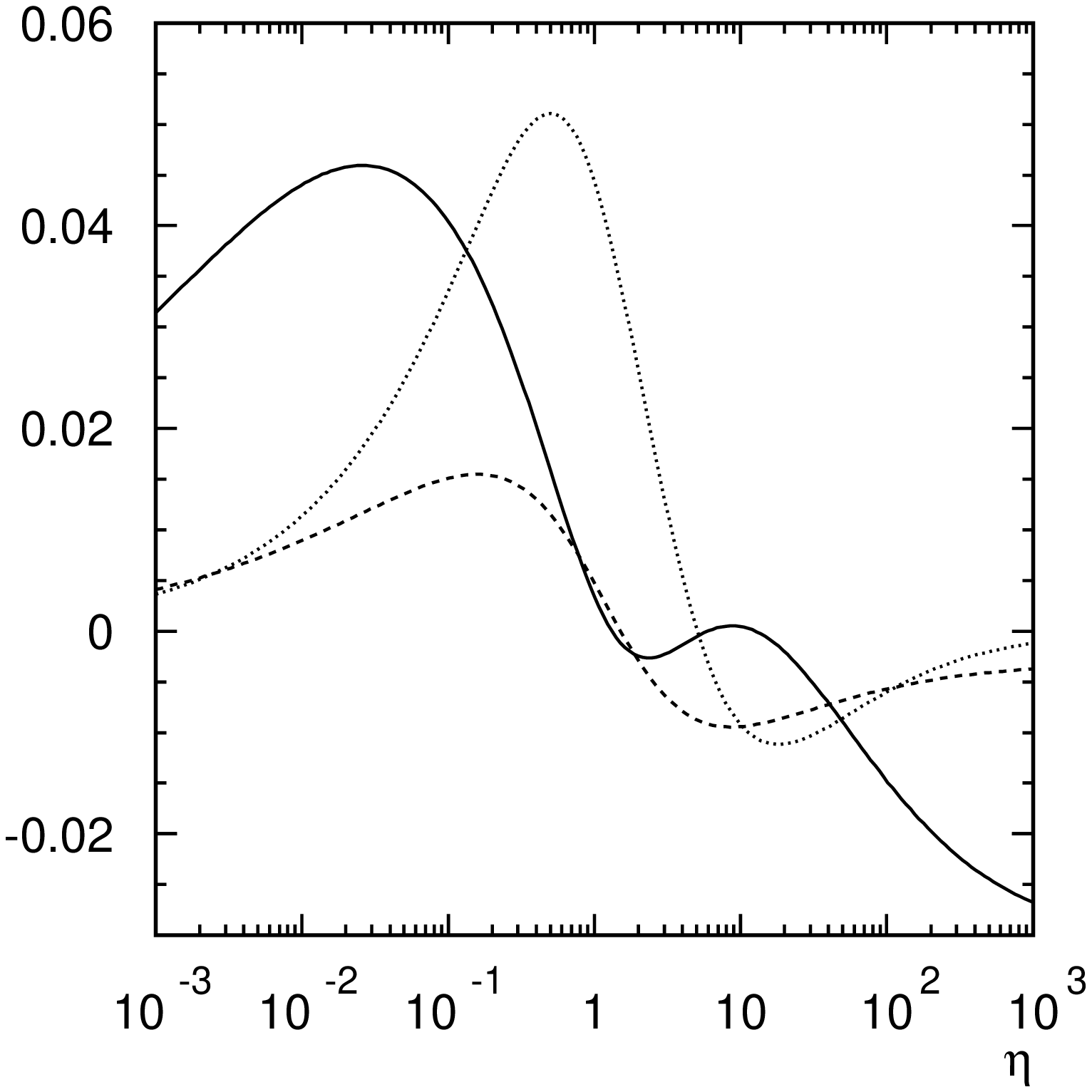,width=9cm}}
    \end{picture}
    \caption{\it Left: Scaling functions $g^{(0)}_{q\bar{q},2}
      (\eta)$ (dotted), 
      $g^{(1)}_{q\bar{q},2}(\eta)$ (full), 
      and $\tilde{g}^{(1)}_{q\bar{q},2}(\eta)$ (dashed).
      Right: The same 
      for the process $gg\to t\bar{t}(g)$.}
    \label{fig:o2}
  \end{center}
\end{figure}
\begin{figure}[ht!]
  \unitlength1.0cm
  \begin{center}
    \begin{picture}(7.5,7.5)
      \put(-5.85,-3.2){\psfig{figure=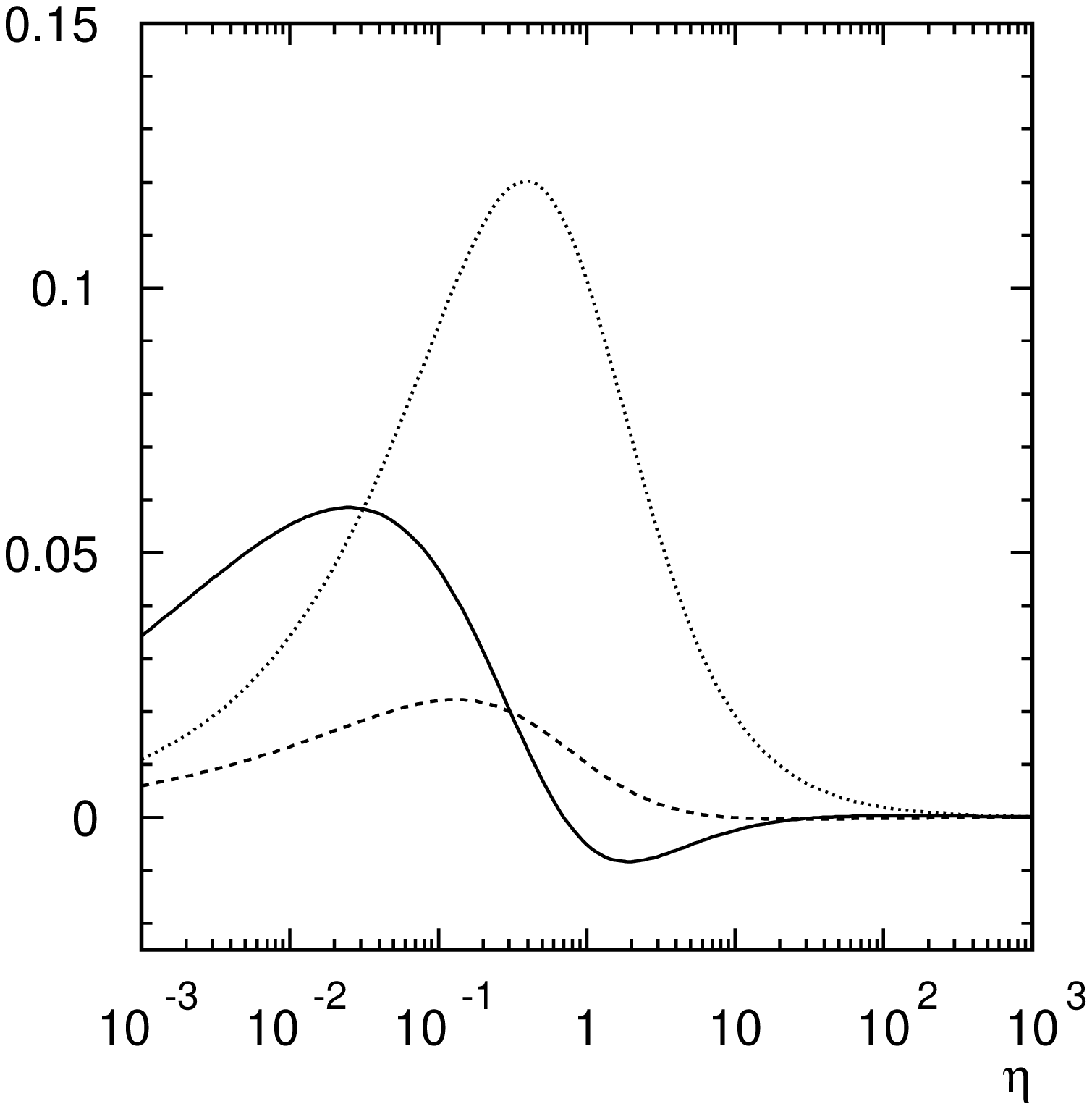,width=9cm}}
      \put(3.15,-3.2){\psfig{figure=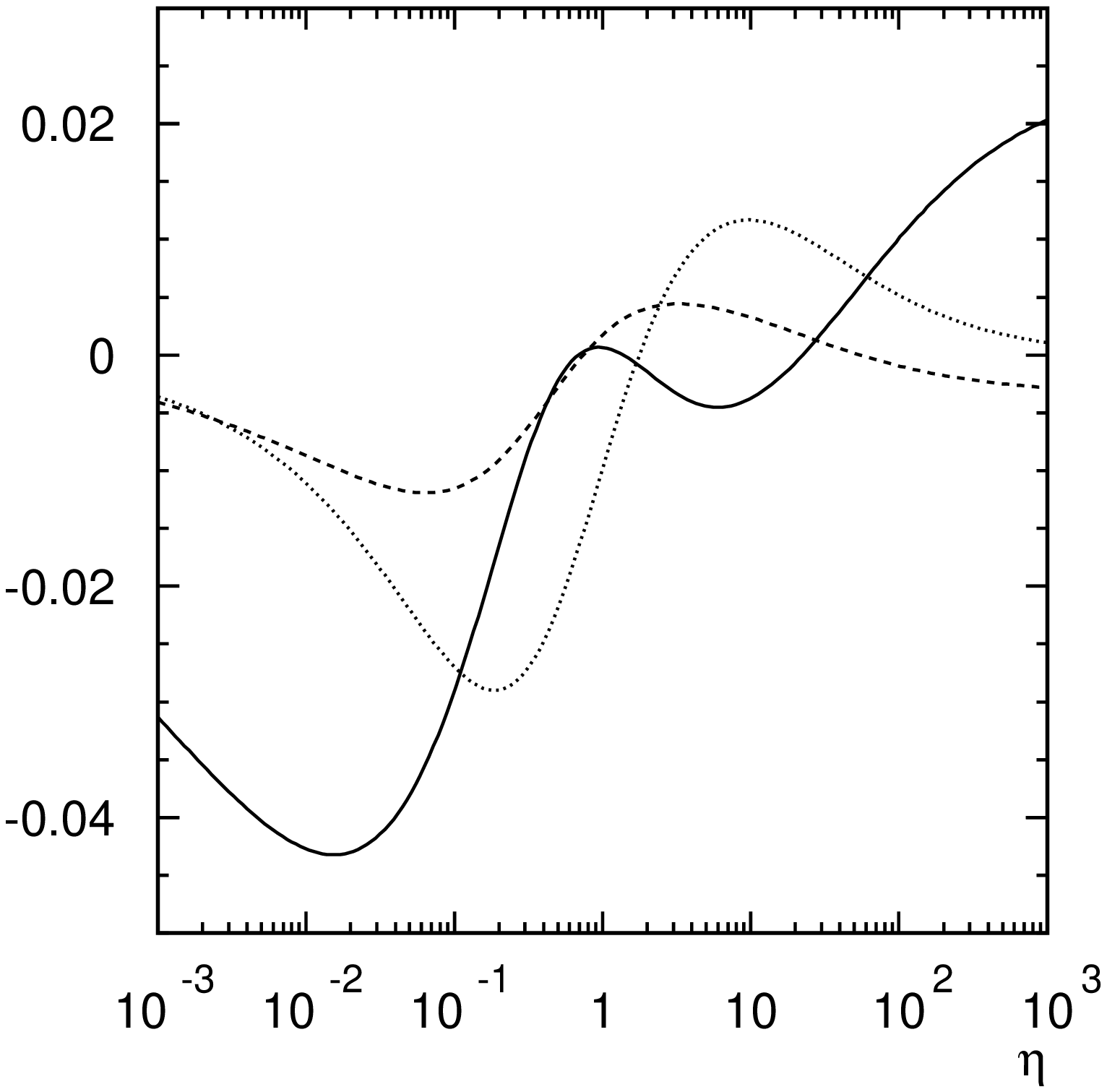,width=9cm}}
    \end{picture}
    \caption{\it Left: Scaling functions $g^{(0)}_{q\bar{q},3}
      (\eta)$ (dotted), 
      $g^{(1)}_{q\bar{q},3}(\eta)$ (full), 
      and $\tilde{g}^{(1)}_{q\bar{q},3}(\eta)$ (dashed).
      Right: The same 
      for the process $gg\to t\bar{t}(g)$.}
    \label{fig:o3}
  \end{center}
\end{figure}
\begin{figure}[ht!]
  \unitlength1.0cm
  \begin{center}
    \begin{picture}(7.5,7.5)
      \put(-5.85,-3.2){\psfig{figure=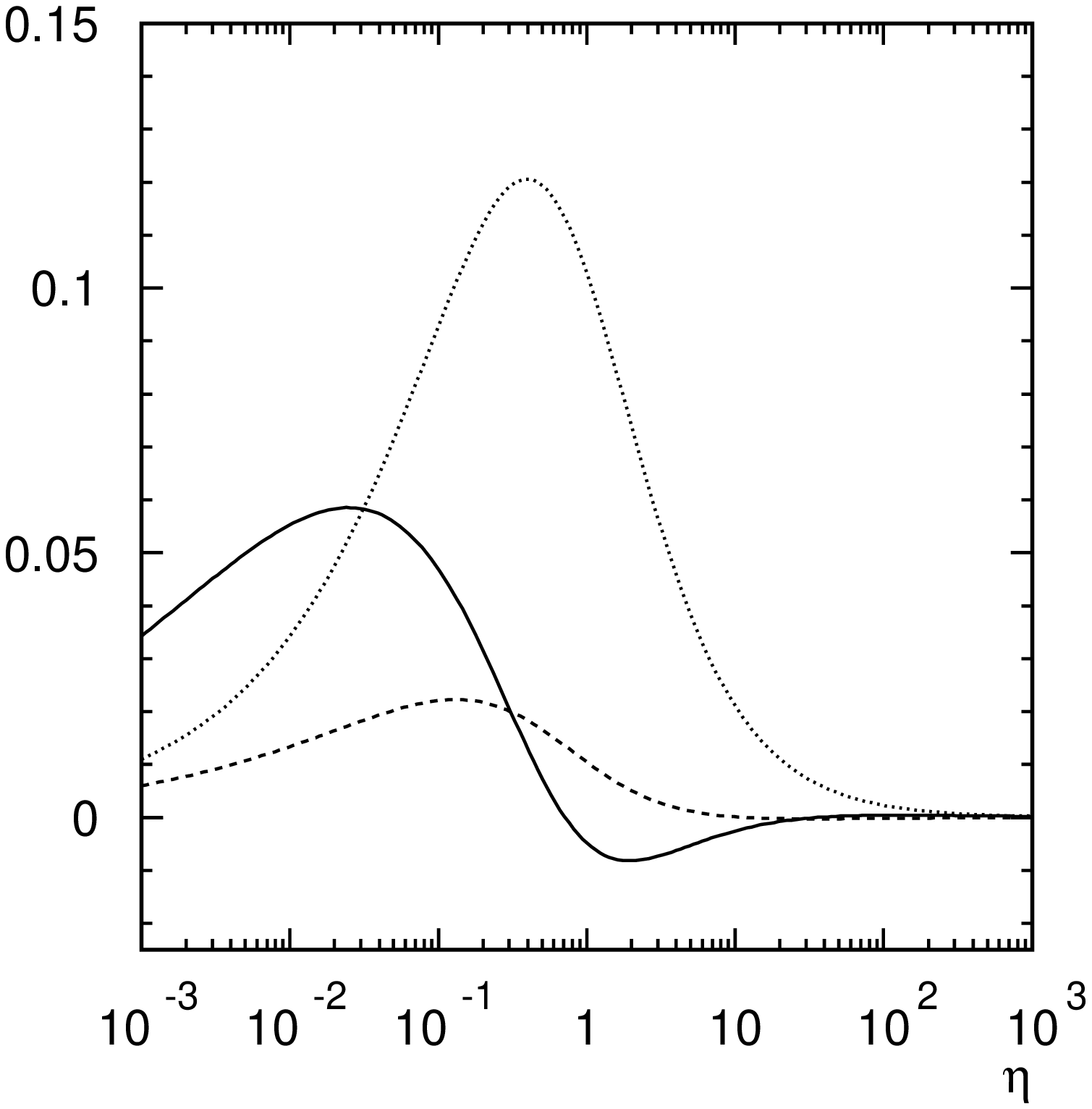,width=9cm}}
      \put(3.15,-3.2){\psfig{figure=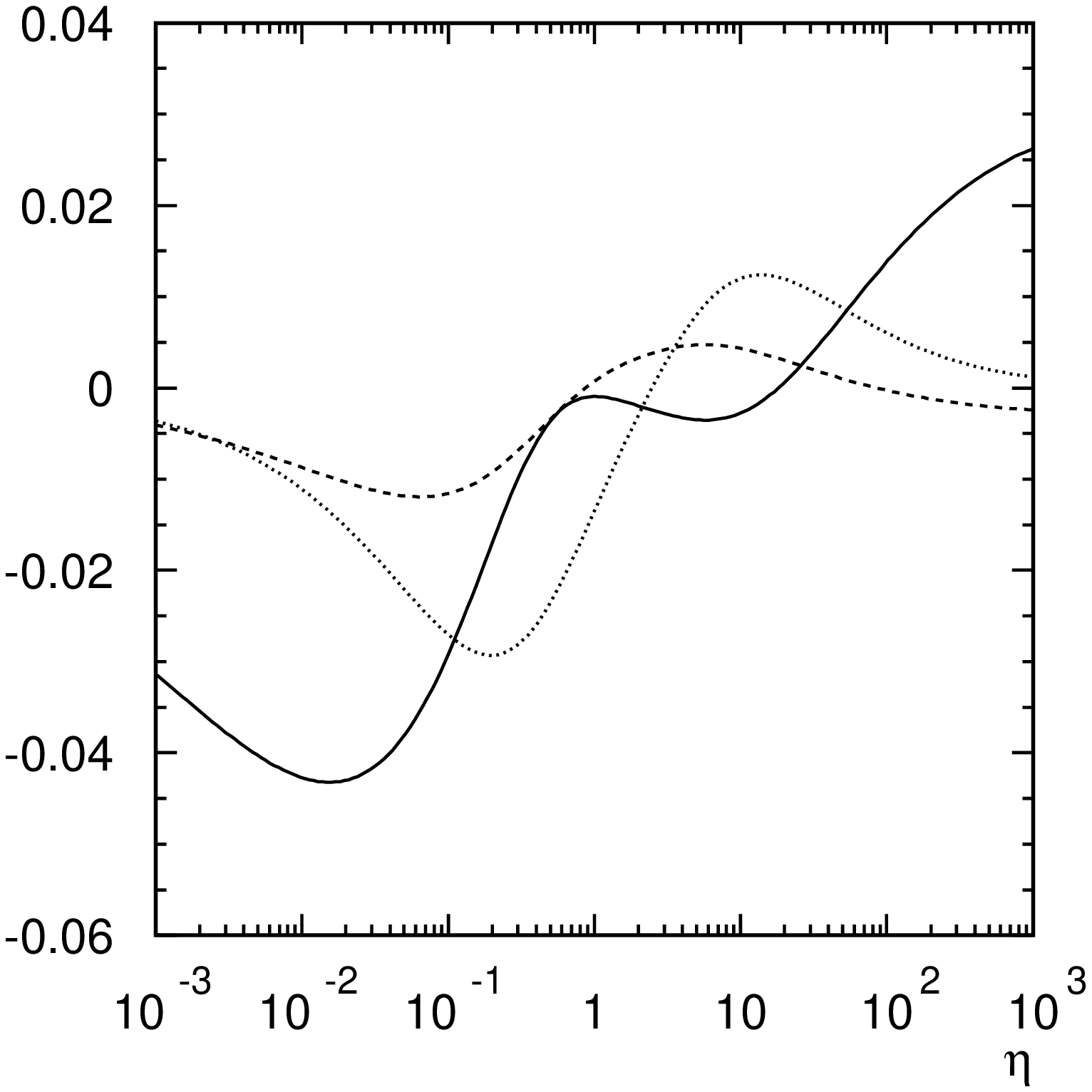,width=9cm}}
    \end{picture}
    \caption{\it Left: Scaling functions $g^{(0)}_{q\bar{q},4}
      (\eta)$ (dotted), 
      $g^{(1)}_{q\bar{q},4}(\eta)$ (full), 
      and $\tilde{g}^{(1)}_{q\bar{q},4}(\eta)$ (dashed).
      Right: The same 
      for the process $gg\to t\bar{t}(g)$.}
    \label{fig:o4}
  \end{center}
\end{figure}

The scaling functions 
$f^{(1)}_{i}$ and $g^{(1)}_{i,a}$  are
obtained by numerical integration.
As a check we computed the spin-averaged NLO $t \bar t$ production
cross sections 
$\sigma^{i}$ as functions of $\rho$. We
found  excellent  agreement with the results of 
Refs.~\cite{Nason:1987xz,Beenakker:1988bq,Beenakker:1990ma}. 
Our result for
${\tilde f}^{(1)}_{gq}$ also agrees
with the one of Ref.~\cite{Nason:1987xz}. In the comparison of 
${\tilde f}^{(1)}_{j}$ one has to take into account that 
Ref.~\cite{Nason:1987xz} uses a renormalization scheme where heavy and
light quark loops are subtracted differently according to the scheme
of Ref.~\cite{Collins:wz},  while we use the standard
${\overline{\rm MS}}$ definition of $\alpha_s$  of $N_f=6$
flavour QCD.  This implies that our scaling 
functions ${\tilde f}^{(1)}_{j}$ are related to the functions
$({\tilde f}^{(1)}_{j})_{\rm N}$
of Nason, Dawson and Ellis by
\begin{equation}
{\tilde f}^{(1)}_{j} = ({\tilde f}^{(1)}_{j})_{\rm N} -\frac{1}{12\pi^2}
{f}^{(0)}_{j} \, , \quad j=q {\bar q}, gg \, . 
\label{fconvers}
\end{equation}
Performing the conversion we again find  perfect agreement with  
Ref.~\cite{Nason:1987xz}. 

For the functions 
$f^{(1)}_{i}$ and $g^{(1)}_{i,a}$ 
we give fits 
to the results of our  numerical integrations in 
Appendix~\ref{sec:FitFunctions}.
To account for the richer
structure of the functions $g_{i,a}$ --- compared with the results for
the unpolarized cross sections --- we have used a more general ansatz
than in Ref.~\cite{Nason:1987xz}. 
All scaling functions for the initial states $q\bar{q}$ and $gg$
are plotted in Figs.~\ref{fig:o0}--\ref{fig:o4} as a function of
\begin{equation}
\eta = {1\over \rho}-1\, .
\end{equation}
For large values of $\eta$, the results for the functions
$g^{(1)}_{i,a}$ and  $\tilde{g}^{(1)}_{i,a}$ differ from the
corresponding results in 
Refs.~\cite{Bernreuther:2000yn,Bernreuther:2001bx}, since
in those papers the spin axes were defined using the
initial-state parton c.m.s. rather than the $t\bar{t}$-ZMF. 
\section{Decays of polarized top quarks}
\label{decay}
In the leading pole
approximation for the intermediate top and antitop quarks
the NLO  QCD analysis of 
the reactions Eq.~(\ref{eq:ttrec1})--Eq.~(\ref{eq:ttrec3}) involves
also the matrix elements
to order $\alpha_s$ of the main SM decay modes of the (anti)top 
quark in a given spin state, that is, the semileptonic modes
\begin{equation}
  t \to  b \ell^+ \nu_{\ell}, b \ell^+ \nu_{\ell} g
\end{equation}
$ (\ell=e,\mu,\tau)$,
and the non-leptonic decays 
\begin{equation}
  t \to  b q {\bar q}', b q {\bar q}' g,
\end{equation}
where  $q {\bar q}'= u {\bar d}, c {\bar s}$ for the dominant channels. 
Specifically, for computing the angular distributions
(\ref{eq:ddist1}) and (\ref{eq:ddist2}), we need the  
matrix elements for the parton reactions
\begin{equation}
  i \ {\buildrel
    t{\bar t}\over \longrightarrow} \ a_1 + a_2 + X \, ,
  \label{a12ink}
\end{equation}
where $a_1,a_2$ denotes a lepton or a jet.
Thus we require, in addition to the above production density matrices,
the one-particle inclusive angular distributions
$d\Gamma/d\cos\theta$ for the decays
\begin{eqnarray} 
  t(s_t) \to a_1(q_1) + X_1 \, ,\nonumber \\
  {\bar t}(s_{\bar t}) \to a_2(q_2) + X_2 \, ,
\end{eqnarray}
at NLO in $\alpha_s$. Here  $q_1$ and $q_2$
are the momenta  of $a_1$ and $a_2$, respectively, 
which we define in the rest frame
of the (anti)top quark, and $\theta$
is the angle between the polarization vector of the (anti)top quark
and the direction of flight of $a_1(a_2)$. 
For a fully polarized ensemble of top
quarks these distributions are of the form 
\begin{equation}
 {d\Gamma^{(1)}\over d\cos\theta}
={\Gamma^{(1)}\over 2}(1+\kappa^{(1)} \cos\theta) \, , 
\label{andist}
\end{equation}
where $\Gamma^{(1)}$ is the partial width of the
respective decay channel and $\kappa^{(1)}$ is the top-spin 
analysing power of $a_1$. 
For the case of the
standard $(V-A)$ charged current interactions these distributions
were computed to order $\alpha_s$ for the semileptonic and
non-leptonic channels in Refs.~\cite{Czarnecki:1990pe}
and \cite{Brandenburg:2002xr}, respectively. 
>From these results the corresponding unnormalized 
one-particle inclusive decay density
matrices $\rho$ and $\bar\rho$ integrated over the energies can be extracted. 
These  matrices have  the form
\begin{equation}
  \rho^{t\to a_1}_{\alpha'\alpha}
  = {\Gamma^{(1)}\over 2}(\one +{\kappa}^{(1)}\,\mathbf{\tau} \cdot
  {{\bf{\hat q}}_1})_{\alpha'\alpha},
  \label{rhoa1}
\end{equation}
\begin{equation}
  {\bar \rho}^{{\bar t}\to a_2}_{\alpha'\alpha}
  = {\Gamma^{(2)}\over 2}(\one - {\kappa}^{(2)}\,\mathbf{\tau} \cdot
  {{\bf{\hat q}}_2})_{\alpha'\alpha},
  \label{rhoa2}
\end{equation}
where ${\bf{\hat q}}_{1,2}$
are the directions of flight of $a_1$ and $a_2$ in the rest frame of the top
and antitop quarks, respectively.  
As we work to lowest order in the electroweak couplings,
${\Gamma^{(2)}} = \Gamma^{(1)}$ and 
${\kappa}_2 = {\kappa}_1$ to all orders in $\alpha_s$, if the channel
$a_2 + X_2$ is the
charge-conjugate of $a_1 + X_1$. 
\par
For  semileptonic top
decays, the charged lepton is the most efficient analyser of the spin
of the top quark.
In the case of   non-leptonic  decays $t \to b q {\bar q}'
(g)$ this role is taken by the weak isospin 
$I_W=-1/2$ quark from $W$ decay. Tagging of
the flavor of $q$ or ${\bar q}'$, besides those of the $b$ quark, is
feasible, as far as the dominant channels are concerned,   only
for $q=c$, albeit with low efficiency. The
non-$b$ jets in non-leptonic top decay may be used  
as analysers of the top spin
in the following, more efficient way: angular momentum conservation and  
the
$(V-A)$ structure 
enforce that the  
fermion with $I_W= -1/2$ from  $W$ decay is, on average, less
energetic than its partner with $I_W= +1/2$. This implies that the 
least-energetic light quark jet is a good top spin analyser.
As far as the $W$ boson is concerned notice 
that its analysing power is practically the same
as that of the $b$ quark: 
\begin{equation}
\kappa_W = - \kappa_b + {\cal O}(\alpha_s).
\end{equation}
\par
In Ref.~\cite{Brandenburg:2002xr}  
the coefficients $\kappa^{(f)}$ were given to  NLO 
accuracy. In Section \ref{angular} we  need, however, the
unnormalized density matrices (\ref{rhoa1}) and (\ref{rhoa2});
that is, we require, apart from the partial widths
\begin{equation}
\Gamma^{(f)} = a^{(f)}_0 + 4\*\pi\*\alpha_s \*a^{(f)}_1,
\label{expga}
\end{equation}
the dimensionful
coefficients
\begin{equation}
 \Gamma^{(f)} \kappa^{(f)} =
b^{(f)}_0 +4\*\pi\*\alpha_s\* b^{(f)}_1 \, .
\label{expka}
\end{equation}

For the determination of these coefficients we use the Fermi constant 
$G_F=1.16639\times 10^{-5}\mbox{ GeV}^{-2}$,
$m$ = 175 GeV, $m_W$ = 80.41 GeV,  $\Gamma_W$ = 2.06 GeV,
$m_b$ = 5 GeV, and all other quark and lepton masses are put to zero.
(We do not use the narrow width
approximation for the intermediate  $W$ boson.)
We obtain, for the hadronic (h) and semileptonic (sl) decays
$t\to bq\bar{q}'$, $b\ell\nu$,
putting the CKM matrix elements $|V_{tb}|=|V_{qq'}|=1$: 
\begin{eqnarray}
a_0^{\rm h}  &=& 0.50645 {\rm \ GeV}, \nn \\ 
a_0^{\rm sl} &=&  {a_0^{\rm h} \over N},\nn \\
a_1^{\rm h}  &=&  -0.01900(8) {\rm \ GeV}, \nn \\
a_1^{\rm sl} &=&  -0.01061(2) {\rm \ GeV}.
\end{eqnarray}
For the  LO  coefficients
$b_0$ we obtain:
\begin{eqnarray}
b_0^d&=& a_0^h=N\*b_0^{\ell}, \nn \\
b_0^{b,\rm h}&=& -0.20663 {\rm \ GeV}=N\; \*b_0^{b,\rm sl},\nn \\ 
b_0^{u}&=& -0.15819 {\rm \ GeV}=N\; \*b_0^{\nu},\nn \\
b_0^{j}&=&  \phantom{-}0.25840 {\rm \ GeV},\nn \\
b_0^{T}&=& -0.16040{\rm \ GeV}. 
\end{eqnarray}
Here $j$ stands for least energetic non-$b$ jet,  and $T$ for an oriented
thrust axis (for details, compare  Ref.~\cite{Brandenburg:2002xr}).
In Table~\ref{tab:decay} we give the corresponding NLO coefficients $b_1$. 
The numbers in the first columns are for `bare' quarks,
while in the second and third columns the E-algorithm and Durham
algorithm
were used, respectively, as jet clustering schemes
\cite{Brandenburg:2002xr}.
We demand that the 
four final-state partons in $t\to bq\bar{q}'g$ be always clustered
into three jets, i.e. no jet resolution parameter is involved.
This is possible because the leading order matrix elements
are free from soft and collinear singularities.

\begin{table}[htbp]
\caption{\it NLO coefficients $b_1$. Numbers are in units of GeV.
\label{tab:decay}}
\begin{center}\renewcommand{\arraystretch}{1.2}
\begin{tabular}{|c|c|c|c|c|}\hline
& Partons & Jets, E-algo. & Jets, D-algo. & Semileptonic decay\\
\hline \hline 
$b_1^d$  [GeV] & $-$0.03154(22)&$-$0.04218(22) &
$-$0.04410(22)&--\\
$b_1^{\ell}$ [GeV] &--&--&--& $-$0.01074(7)\\
$b_1^{b,\rm h}$ or $b_1^{b,\rm sl}$ [GeV] & 
\phantom{$-$}0.01354(23)& \phantom{$-$}0.01420(23) & \phantom{$-$}0.01412(23) & \phantom{$-$}0.00626(7)\\
$b_1^u$ or $b_1^{\nu}$ [GeV] & 
 \phantom{$-$}0.00431(21)& 
\phantom{$-$}0.00935(22) & 
\phantom{$-$}0.00852(22)& 
\phantom{$-$}0.00120(7)\\
$b_1^j$ [GeV] &--&$-$0.02336(23) & $-$0.02342(23)&-- \\
$b_1^T$ [GeV] &\phantom{$-$}0.00915(22)&-- &-- &-- \\
\hline
\end{tabular}
\vspace*{1em}
\end{center}
\end{table}
Higher-dimensional distributions for $t$ and/or $\bar{t}$ decays such as
$d\Gamma/(dx_id\cos\theta)$, where $x_i$ the scaled lepton or jet
energy are also known. 
Together with the production density matrices 
of Sections \ref{calc}--\ref{partonsection}, they can be used for predictions of 
higher-dimensional distributions
for the 2-particle inclusive reactions 
(\ref{eq:ttll})--(\ref{eq:ttjj}).

\section{Angular correlations in hadronic collisions}
\label{angular}
Let us now discuss  the hadronic  reactions 
Eqs.~(\ref{eq:ttll})--(\ref{eq:ttjj}). In particular we derive  the  
distributions of Eq.~(\ref{eq:ddist1}) and  Eq.~(\ref{eq:ddist2})
at NLO in $\alpha_s$, using the 
narrow-width approximation for the intermediate
top and antitop quarks and 
taking into account only the factorizable QCD 
corrections. Recall that in this approximation the squared matrix elements of
the parton reactions (\ref{eq:ttrec1})--(\ref{eq:ttrec3}) 
are of the form shown in Eq.~(\ref{eq:trace}). We postpone the discussion
of the non-factorizable QCD corrections until the end of
this section. There we show that, 
at NLO in $\alpha_s$,  they are irrelevant for the 
distributions defined in Eqs.~(\ref{eq:ddist1}) and (\ref{eq:ddist2}).
 
With the ingredients of 
Sections \ref{calc}--\ref{decay} we can first determine
the two-fold differential (\msbar-subtracted) cross sections 
\begin{equation}
  {d\sigma(i\to a_1 a_2 +
  X) \over d\cos\theta_1 d\cos\theta_2} 
\end{equation}
for the parton reactions 
$i\to a_1 a_2 +X$, where $\theta_1(\theta_2)$ refers to the 
angle between the direction of flight ${\bf \hat q}_1$ 
(${\bf \hat q}_2$) of particle $a_1$ $(a_2)$
in the $t$  $({\bar t})$ rest frame and an arbitrary direction ${
{\bf \hat a}}$ $({ {\bf \hat b}})$ that may be used as $z$ 
axis in the (anti)top rest frame.
Recall that we define the $t ({\bar t})$ rest frame by performing a
rotation-free Lorentz boost from the $t\bar{t}$-ZMF.
For  the reactions (\ref{eq:ttll})--(\ref{eq:ttjj}), i.e. 
\begin{equation}
h_1 (P_1) \; h_2(P_2) \to t {\bar t} + X \to a_1 \; a_2 + X \, ,
\label{hada12}
\end{equation}
where $h_1,h_2 = p$ or ${\bar p}$,  the differential cross sections are 
obtained in the usual way:
\begin{eqnarray}
  {d\sigma (h_1(P_1) h_2(P_2) \to a_1 a_2+X)
    \over {d\cos\theta_1 d\cos\theta_2}} &=&
  \sum\limits_{a,b}\int {dx_1} {dx_2} \;
  f_{a}^{h_1}(x_1,\mu_F) 
  f_{b}^{h_2}(x_2,\mu_F)\nn\\
  &&\times\quad{d\sigma (a(x_1 P_1) b(x_2 P_2)
    \to a_1 a_2 +X) \over {d\cos\theta_1 d\cos\theta_2}} \, ,
\label{sigt12hadr}
\end{eqnarray}
where $f_{a}^{h}(x,\mu_F)$ denotes the parton distribution function (PDF)
of parton $a$ in hadron $h$.
\par 
The following analysis is based on integrating over the full phase-space of
the particles in the final state.
What then is the structure of the  normalized
distributions 
\begin{equation}
  {1\over \sigma} {d\sigma \over  {d\cos\theta_1
      d\cos\theta_2}} 
\end{equation}
if we choose as reference axes 
${{\bf\hat a}}, {{\bf \hat b} }$ one of the sets 
Eq.~(\ref{helbasis})--Eq.~(\ref{offbasis})?
Each of the contributions to 
$d\sigma (i \to a_1 a_2 +X)$ is of the form ${\rm Tr}\;[\rho
R^i_s{\bar{\rho}}]$ $(s=B, V, {\rm soft, col, c, res},$ depending on $i$),
where the $R^i_s$ have the structure as given in Eq.~(\ref{eq:Rstruct})
and $\rho$ and ${\bar{\rho}}$ are  given by Eqs.~(\ref{rhoa1}) and
(\ref{rhoa2}), respectively. All contributions are
bilinear in ${\bf{\hat q}}_1$
and ${\bf{\hat q}}_2$. Thus  the differential cross
section Eq.~(\ref{sigt12hadr}) is bilinear in $\cos\theta_1$ and 
$\cos\theta_2$. Parity invariance of QCD dictates  that the top and
antitop quarks have no polarization along the above reference axes 
${{\bf \hat a}}$ and ${ {\bf\hat b}}$, see Eq. (\ref{sppol}).
This implies that in QCD
\begin{equation}
  \langle \cos\theta_1 \rangle = \langle \cos\theta_2 \rangle = 0 \, .
  \label{the12qcd}
\end{equation}
Here the expectation value refers to 
reaction (\ref{hada12}), i.e.
$\langle {\cal O}\rangle= \sigma^{-1}\int d\sigma{\cal O}.$
(Equation (\ref{the12qcd})
applies also to cuts that are parity-invariant.) Thus for these
reference axes the coefficients ${\rm B}_1$ and ${\rm B}_2$
in Eq.~(\ref{eq:Rstruct}) are absent in
the double distribution Eq.~(\ref{eq:ddist1}), and we obtain
\begin{equation}
  {1\over \sigma}{d\sigma (h_1 h_2 \to a_1 a_2+X)
    \over d\cos\theta_1 d\cos\theta_2} =
  {1\over 4} (1 - {\rm C}_i \cos\theta_1 \cos\theta_2)\,\, .
  \label{kap6dist1}
\end{equation} 
with $i=\mbox{hel}, \mbox{beam}, \mbox{off}$ for  the helicity, beam, and off-diagonal bases.
Non-zero ${\rm B}_{1,2}$ are generated by parity-violating contributions to 
$i\to t \bar t X.$ Computations indicate
\cite{Beenakker:1993yr,Kao:bs,Kao:1999kj}
that they are small within the standard model. 

Next we consider the distribution of the angle $\varphi$
between the
directions  of flight  ${\bf{\hat q}}_{1,2}$ of particle/jet  $a_1$ 
and  $a_2$, defined  in the $t$ and $\bar{t}$
rest frames, respectively. Using CP invariance of QCD and
arguments analogous to those above, we find that this distribution is
of the form 
\begin{equation}
{1\over \sigma}{d\sigma  (h_1 h_2 \to a_1 a_2+X)\over d\cos\varphi}=
{1\over 2} (1 - {\rm D} \cos\varphi).
\label{kap6dist2}
\end{equation}
Defining, for $\mu_R =\mu_F=m$,
\begin{eqnarray} 
  N_r &=& {\alpha_s^2\over m^2}{1\over \Gamma_t^2}
  \sum\limits_{a,b}\int {dx_1}{dx_2}\;
  f_{a}^{h_1}(x_1,\mu_F) 
  f_{b}^{h_2}(x_2,\mu_F) \nn \\ &\times&
  \left\{g_{ab,r}^{(0)} b_0^{(1)} b_0^{(2)} +
    4\pi\alpha_s [g_{ab,r}^{(1)}  b_0^{(1)} b_0^{(2)}
    + g_{ab,r}^{(0)}  b_1^{(1)} b_0^{(2)} + g_{ab,r}^{(0)} b_0^{(1)}
    b_1^{(2)} ]
  \right\} \, , 
  \label{Zrformel}
\end{eqnarray}
\begin{eqnarray} 
  \sigma &=& {\alpha_s^2\over m^2}
  {1\over \Gamma_t^2}\sum\limits_{a,b}\int {dx_1} {dx_2}\;
  f_{a}^{h_1}(x_1,\mu_F) 
  f_{b}^{h_2}(x_2,\mu_F) \nn \\ &\times& \left\{
    f_{ab}^{(0)}  a_0^{(1)} a_0^{(2)} + 4\pi\alpha_s [f_{ab}^{(1)} 
    a_0^{(1)} a_0^{(2)}
    +  f_{ab}^{(0)} a_1^{(1)} a_0^{(2)} + f_{ab}^{(0)}  
    a_0^{(1)} a_1^{(2)} ] \right\},
  \label{Nrformel}
\end{eqnarray}
we obtain to  NLO in the QCD coupling:
\begin{equation}
  D = {N_1\over \sigma},\quad
  C_{\rm hel} = {N_2\over \sigma},\quad
  C_{\rm beam} = {N_3\over \sigma},\quad
  C_{\rm off} = {N_4\over \sigma}.
  \label{crformel}  
\end{equation}

Here $f_i^{(0,1)}(\rho)$ and $g_{i,r}^{(0,1)}(\rho)$
are the scaling functions
defined in Eqs.~(\ref{eq:xsection}) and  (\ref{eq:expval}),
and collected in Appendix \ref{sec:FitFunctions}, 
with $\rho=4\*m^2/\hat{s}$,
where the partonic c.m. energy $\sqrt{\hat s}$ is given in terms of
the hadronic  c.m. energy  $\sqrt{s}$ through
\begin{equation}
  \hat s = x_1 x_2 s.
\end{equation}
The total top quark width is
denoted by $\Gamma_t$. 
\par 
Finally, we discuss the issue of non-factorizable corrections.
At NLO in $\alpha_s$, such corrections are
present in the one-loop  contributions to the $q \bar q$ and $gg$ fusion
processes Eq.~(\ref{eq:ttrec1}) and in the squared matrix elements for
the corresponding processes Eq.~(\ref{eq:ttrec2}) 
with real gluon radiation. 
These non-factorizable corrections correspond, at
this order in the QCD coupling,  to
gluon exchange that interconnects the different stages of the
off-shell $t \bar t$ production and decay process. These corrections
were studied in the semi-soft gluon approximation
in  Ref.~\cite{Beenakker:1999ya}. This approximation
consists in taking into account only virtual, respectively real
gluons with energies $E_g$ of the order of  $\Gamma_t$. 
This is adequate because the contribution
of a hard gluon to this correction is suppressed by $\Gamma_t/E_g.$
The contributions $d\sigma_{nf}^{q \bar q}$ and $d\sigma_{nf}^{gg}$
to the respective differential cross sections are
important for calculating a number of distributions, such as invariant
mass distributions of the $t$, $\bar t$ quarks  and of $t\bar t$
pairs. However, the non-factorizable corrections do not contribute to 
observables, which are inclusive in both the top and antitop invariant masses 
\cite{Fadin:1993dz,Fadin:1993kt,Melnikov:np,Beenakker:1997ir,
Beenakker:1999ya}. 
A well-known example for such a case is the total cross section.
  
It is straightforward to show
that this result applies also to the distributions shown in 
Eqs.~(\ref{kap6dist1}) and (\ref{kap6dist2}). 
The non-factorizable contributions
$d\sigma_{nf}^{q \bar q}$ and $d\sigma_{nf}^{gg}$ (which contain both
virtual and real gluon exchanges with momenta being integrated
over)  are proportional to the
respective differential Born cross sections, decomposed into colour
singlet and octet pieces. 
We represent the  Lorentz-invariant phase-space element
$d\Gamma_6$ for the 6-particle final states  in Eq.~(\ref{eq:ttrec1})
as 
\begin{equation}
  d\Gamma_6 \propto d\Gamma_{t\bar t} d\Gamma_t d\Gamma_{\bar t} dM^2_t
  dM^2_{\bar t}, 
\end{equation}
where $M_t$ $(M_{\bar t})$ is the invariant
mass of the top (anti)quark,
$d\Gamma_{t\bar t}$ is the phase-space
element of the intermediate $t \bar t$ state, 
and $d\Gamma_t$ $(d\Gamma_{\bar t})$ is the 
3-particle phase-space element of
$t$ $({\bar t})$ decay in its rest frame, where the $z$ axis is
identified with  the
reference axis  ${{\bf \hat a}}$ $({\bf \hat b})$. From this
decomposition, it is clear that  one has to integrate out the
top and antitop invariant masses
in calculating the contribution
of $d\sigma_{nf}^i$ to the distribution (\ref{kap6dist1}). 
But then the theorem of Refs.~\cite{Fadin:1993dz,Fadin:1993kt,Melnikov:np} 
applies.

In order to show that the opening angle distribution (\ref{kap6dist2})
receives no contribution from $d\sigma^i_{nf}$ either, we recall that
this distribution is due to 
$\langle\Sp \cdot \Sm\rangle$, and $\Sp \cdot \Sm$ can be
decomposed as given below Eq. (\ref{eq:sbasis}). 
Thus, to NLO in the QCD coupling the above angular distributions are
determined solely by the Born contributions and the factorizable
QCD corrections.
\section{Numerical results}
\label{numres}
In this section we present the numerical results for the angular distributions
(\ref{kap6dist1}) and  (\ref{kap6dist2}).
In the absence of cuts the spin correlation coefficients
can be computed in terms of expectation values:
\begin{eqnarray}\label{expval}
{\rm C}&=& -9\langle \cos\theta_1\cos\theta_2 \rangle,\nn \\ 
{\rm D}&=& -3  \langle \cos\varphi\rangle.
\end{eqnarray}

\par
\begin{table}[htbp!]
\caption{\it Results for the quantities $Z_i^{(0)}$ and 
$N_{i,r}^{(0)}$ defined in Eqs.~(\ref{n0z01}),\ (\ref{n0z02}) 
for different PDFs at the Tevatron and the LHC.
}\label{tab:nz0}
\begin{center}
\renewcommand{\arraystretch}{1.2}
\begin{tabular}{|c|c|c|c|}
\hline
\multicolumn{2}{|c|}{Tevatron} & $i=q\bar{q}$ & $i=gg$ \\ \hline
$Z^{(0)}_i$ &CTEQ6.1M      &  $\phantom{-}0.032162$
  & $\phantom{-}3.9358\cdot 10^{-3}$ \\ \cline{2-4}
        &MRST2003    & $\phantom{-}0.033193$   &  
$\phantom{-}3.9751\cdot 10^{-3}$ \\ \cline{2-4}
        &GRV98       & $\phantom{-}0.033528$   &  
$\phantom{-}5.0460\cdot 10^{-3}$ \\ \hline
$N_{i,1}^{(0)}$ & CTEQ6.1M  & $\phantom{-}0.010721$  
& $-1.6864\cdot 10^{-3}$ \\ \cline{2-4}
            &MRST2003 & $\phantom{-}0.011064$     
& $-1.7534\cdot 10^{-3}$ \\ \cline{2-4}
            &GRV98    & $\phantom{-}0.011176$    
& $-2.2477\cdot 10^{-3}$\\ \hline
$N_{i,2}^{(0)}$ & CTEQ6.1M     & $-0.016695$ 
& $\phantom{-}2.3820\cdot 10^{-3}$
 \\ \cline{2-4}
               &MRST2003    & $-0.017240$ & 
$\phantom{-}2.4585 \cdot 10^{-3}$ \\ \cline{2-4}
               &GRV98       & $-0.017436$ & 
$\phantom{-}3.1438\cdot 10^{-3}$\\ \hline
$N_{i,3}^{(0)}$ & CTEQ6.1M  & $\phantom{-}0.031849$ 
& $-1.2099\cdot 10^{-3}$\\ \cline{2-4}
               &MRST2003    & $\phantom{-}0.032869$ 
& $-1.2774\cdot 10^{-3}$ \\ \cline{2-4}
               &GRV98       & $\phantom{-}0.033199$ 
& $-1.6457 \cdot 10^{-3}$\\ \hline
 $N_{i,4}^{(0)}$ & CTEQ6.1M     & $\phantom{-}0.032162$ 
& $-1.2959\cdot 10^{-3}$\\ \cline{2-4}
                &MRST2003    & $\phantom{-}0.033193$ 
& $-1.3606\cdot 10^{-3}$ \\ \cline{2-4}
                &GRV98       & $\phantom{-}0.033528$ 
& $-1.7497\cdot 10^{-3}$ \\ \hline \hline
\multicolumn{2}{|c|}{LHC} & $i=q\bar{q}$ & $i=gg$  \\ \hline
$Z^{(0)}_i$ & CTEQ6.1M  & $\phantom{-}0.46482$  
&$\phantom{-}3.2048$ \\ \cline{2-4}
          &MRST2003 &$\phantom{-}0.49789$ 
&$\phantom{-}3.5616$ \\ \cline{2-4}
        &GRV98      &$\phantom{-}0.48075$ 
&$\phantom{-}4.1237$ \\ \hline
$N_{i,1}^{(0)}$ &CTEQ6.1M  &$\phantom{-}0.15494$ 
& $-0.92015$ \\ \cline{2-4}
&  MRST2003&$\phantom{-}0.16596$ & $-1.0163$ \\ \cline{2-4}
&GRV98     &$\phantom{-}0.16025$ & $-1.1620$  \\ \hline
$N_{i,2}^{(0)}$& CTEQ6.1M &$-0.26645$ & $\phantom{-}1.3931$\\ \cline{2-4}
               &MRST2003  &$-0.28604$ &$\phantom{-}1.5401$ \\ \cline{2-4}
             &GRV98       &$-0.27709$ &$\phantom{-}1.7621$ \\ \hline
\hline
\end{tabular}\end{center}
\end{table}
\begin{table}[htbp!]
\caption{\it Results for the quantities $Z_i^{(1)}$ and 
$N_{i,r}^{(1)}$ defined in Eqs.~(\ref{n0z01}),\ (\ref{n0z02}) 
for different PDFs at the Tevatron and the LHC.}\label{tab:nz1}
\begin{center}
\renewcommand{\arraystretch}{1.2}
\begin{tabular}{|c|c|c|c|c|} \hline
\multicolumn{2}{|c|}{Tevatron} & $i=q\bar{q}$ & $i=gg$ 
&$i=qg+g\bar{q}$\\ \hline
$Z^{(1)}_i$   &CTEQ6.1M      & $\phantom{-}4.699\cdot 10^{-3}$ 
& $\phantom{-}2.312\cdot 10^{-3}$ & $-3.627\cdot 10^{-4}$\\ \cline{2-5}
&MRST2003    &$\phantom{-}4.836\cdot 10^{-3}$ 
&$\phantom{-}2.423\cdot 10^{-3}$ 
& $-3.926\cdot 10^{-4}$\\ \cline{2-5}
&GRV98       &$\phantom{-}4.854\cdot 10^{-3}$ 
&$\phantom{-}3.113\cdot 10^{-3}$ & $-4.442\cdot 10^{-4}$\\ \hline
$N_{i,1}^{(1)}$ &CTEQ6.1M   &$\phantom{-}1.539\cdot 10^{-3}$  
&$-1.070\cdot 10^{-3}$ &$\phantom{-}1.216\cdot 10^{-4}$\\ \cline{2-5}
&MRST2003 & $\phantom{-}1.583\cdot 10^{-3}$ 
&$-1.143\cdot 10^{-3}$ &$\phantom{-}1.289\cdot 10^{-4}$\\ \cline{2-5}
&GRV98    &$\phantom{-}1.589\cdot 10^{-3}$ 
&$-1.478\cdot 10^{-3}$ 
&$\phantom{-}1.456\cdot 10^{-4}$\\ \hline
$N_{i,2}^{(1)}$ &CTEQ6.1M   & $-2.643\cdot 10^{-3}$ 
&$\phantom{-}1.458\cdot 10^{-3}$ & $-1.970\cdot 10^{-4}$\\ \cline{2-5}
&MRST2003 & $-2.721\cdot 10^{-3}$ & $\phantom{-}1.552\cdot 10^{-3}$ 
& $-2.065\cdot 10^{-4}$\\ \cline{2-5}
& GRV98   & $-2.734\cdot 10^{-3}$& $\phantom{-}2.004\cdot 10^{-3}$ 
& $-2.331\cdot 10^{-4}$\\ \hline
$N_{i,3}^{(1)}$  & CTEQ6.1M     & $\phantom{-}4.459\cdot 10^{-3}$  
&$-8.647\cdot 10^{-4}$ 
&$\phantom{-}2.066\cdot 10^{-5}$\\ \cline{2-5}
&MRST2003 & $\phantom{-}4.587\cdot 10^{-3}$  &$-9.246\cdot 10^{-4}$ 
 & $\phantom{-}1.995\cdot 10^{-5}$\\ \cline{2-5}
&GRV98       & $\phantom{-}4.602\cdot 10^{-3}$ 
&$-1.195\cdot 10^{-3}$ & $\phantom{-}2.258\cdot 10^{-5}$\\ \hline
$N_{i,4}^{(1)}$ & CTEQ6.1M &$\phantom{-}4.515\cdot 10^{-3}$ 
&$-9.117\cdot 10^{-4}$ 
&$\phantom{-}3.033\cdot 10^{-5}$\\ \cline{2-5}
&MRST2003 & $\phantom{-}4.645\cdot 10^{-3}$ 
&$-9.725\cdot 10^{-4}$  &$\phantom{-}2.975\cdot 10^{-5}$\\ \cline{2-5}
&GRV98  & $\phantom{-}4.660\cdot 10^{-3}$
&$-1.256\cdot 10^{-3}$ &$\phantom{-}3.363\cdot 10^{-5}$\\ \hline \hline
\multicolumn{2}{|c|}{LHC} & $i=q\bar{q}$ & $i=gg$ &$i=qg+g\bar{q}$\\ \hline
$Z^{(1)}_i$  & CTEQ6.1M & $\phantom{-}0.04197$ & 
$\phantom{-}1.277$ & $\phantom{-}0.04565$ \\ \cline{2-5}
&MRST2003  & $\phantom{-}0.04415$ &$\phantom{-}1.411$ 
&$\phantom{-}0.05120$ \\ \cline{2-5}
&GRV98    & $\phantom{-}0.04194$ &$\phantom{-}1.625$ 
&$\phantom{-}0.06018$\\ \hline
$N_{i,1}^{(1)}$ &  CTEQ6.1M & $\phantom{-}0.01329$
 &$-0.4330$ &  $\phantom{-}0.03077$ \\ \cline{2-5}
 &MRST2003 & $\phantom{-}0.01396$ &$-0.4760$ 
& $\phantom{-}0.03307$ \\ \cline{2-5}
&GRV98    &  $\phantom{-}0.01324$ &$-0.5423$ & 
$\phantom{-}0.03082$ \\ \hline
$N_{i,2}^{(1)}$ & CTEQ6M  &$-0.02417$ &$\phantom{-}0.5927$ 
&$-0.07905$\\ \cline{2-5}
&MRST2003 & $-0.02550$ &$\phantom{-}0.6513$ &$-0.08540$ \\ \cline{2-5}
&GRV98    & $-0.02419$ &$\phantom{-}0.7402$ &$-0.08181$\\ \hline
\end{tabular}
\end{center}
\end{table}
In the following we present the NLO results for the coefficients
${\rm C}$ and ${\rm D}$ in the absence of phase-space cuts, using 
Eq.~(\ref{Zrformel}) and Eq.~(\ref{Nrformel}). It is convenient
to rewrite these formulae:
\begin{eqnarray}\label{n0z01}
N_r &=& {\alpha_s^2\over m^2}{1\over \Gamma_t^2}
\sum\limits_{i}
 \left\{N_{i,r}^{(0)} b_0^{(1)} b_0^{(2)} +
4\pi\alpha_s [N_{i,r}^{(1)}  b_0^{(1)} b_0^{(2)}
+ N_{i,r}^{(0)}  b_1^{(1)} b_0^{(2)} + N_{i,r}^{(0)} b_0^{(1)}b_1^{(2)} ]
\right\},
\end{eqnarray}
\begin{eqnarray}\label{n0z02}
\sigma &=& {\alpha_s^2\over m^2}
{1\over \Gamma_t^2}\sum\limits_{i}\left\{
Z_i^{(0)}  a_0^{(1)} a_0^{(2)} + 4\pi\alpha_s [Z_i^{(1)} a_0^{(1)} a_0^{(2)}
+  Z_i^{(0)} a_1^{(1)} a_0^{(2)} + Z_i^{(0)}  a_0^{(1)} a_1^{(2)} ] \right\}, 
\end{eqnarray}
where $i=q\bar{q},gg,gq,g\bar{q}$. For the evaluation of Eqs.~(\ref{n0z01}),
(\ref{n0z02})
we use the NLO parton distribution functions of Ref.~\cite{Pumplin:2002vw}
(CTEQ6.1M) and of Ref.~\cite{Martin:2003sk} (MRST2003). 
In addition
we also use the PDFs of Ref.~\cite{Gluck:1998xa} (GRV98), to illustrate
the dependence on the gluon distribution function. We put
$\mu_R=\mu_F=m_t=175$ GeV, $N_f=5$ and use the following
values for $\alpha_s^{N_f=5}$:
$\alpha_s(m_t)=0.1074$ for CTEQ6.1M, 
$\alpha_s(m_t)=0.1062$ 
for MRST2003, and
$\alpha_s(m_t)=0.1041$ for GRV98. (The values of $\alpha_s$ are those
used by the different groups when fitting the PDFs.)
In Table \ref{tab:nz0} we present the values of $Z_i^{(0)}$ and 
$N_{i,r}^{(0)}$ for $p\bar{p}$ collisions at $\sqrt{s}=1.96$ GeV (Tevatron)
and $pp$ collisions at $\sqrt{s}=14$ TeV (LHC). In the latter case,
we only show results for the ${\rm D}$ coefficient and the ${\rm
  C_{\rm hel}}$
coefficient, since the spin correlations in the
beam and the
off-diagonal basis are very small at the LHC 
\cite{Bernreuther:2001rq}. Table \ref{tab:nz1} lists the NLO coefficients
$Z_i^{(1)}$ and $N_{i,r}^{(1)}$. We observe the following generic features:
For the coefficients $N_{i,a}$  the two initial states
$q\bar{q}$ and $gg$ always contribute with opposite signs.
For the $gg$
initial state the GRV98 PDF systematically gives a larger contribution,
while the other two PDFs agree quite well in most of the
cases.
\par
Adding all contributions as in Eqs.~(\ref{n0z01}),~(\ref{n0z02}) 
and taking the 
ratios $N_r/\sigma$, we obtain LO and NLO predictions
for the spin correlation coefficients. The results are listed in 
Table~\ref{tab:cdtev} for the Tevatron and in Table~\ref{tab:cdlhc}
for the LHC, for the different decay modes
of the top and antitop quarks. 
In these two tables we use the CTEQ6L (LO) and CTEQ6.1M
(NLO) PDFs. As spin analysers we use the charged lepton and/or the
least energetic non-$b$-quark jet defined in the Durham jet clustering
scheme.
\par
At the Tevatron, the helicity basis is not the best choice,
since the $t$ and $\bar{t}$ are only moderately relativistic
in this case. The spin correlations are large both in the beam and
off-diagonal basis. In fact, since the results are almost 
the same in these two bases, the beam basis is probably the best choice
since it is easier to determine experimentally. The QCD corrections
decrease the LO results at the Tevatron. The relative size of the corrections 
varies between  $\sim$ 16\% (${\rm C}_{\rm beam,off}$)
and $\sim$ 28\% (${\rm D}$) for the dilepton channel 
and between $\sim$ 27\% (${\rm C}_{\rm beam,off}$) and 
$\sim$ 38\% (${\rm D}$) in the all-jets channel.
\par 
At the LHC, where the $gg$ initial state gives the dominant contribution,
the helicity correlation is the best choice from the four 
distributions. Here the spin correlations
in the dilepton channel are of the order of 30\%. The QCD corrections
are much smaller than at the Tevatron. In the dilepton channel they 
enhance the spin correlation by $\sim$ 2\% in the helicity basis 
and by $\sim$ 10\% for the opening angle correlation coefficient $\rm D$.
In the all-jet channel, the QCD corrections decrease the LO results
for ${\rm C}_{\rm hel}$  by $\sim$ 8\% and for $\rm D$ by $\sim$ 1\%. 
\begin{table}[htbp!]
\caption{\it LO and NLO results for the spin correlation coefficients
${\rm C}$ and ${\rm D}$ of the distributions 
(\ref{kap6dist1}) and  (\ref{kap6dist2}) in the case of $p\bar{p}$ collisions
at $\sqrt{s}=1.96$ TeV for different $t\bar{t}$ decay modes.
The PDF  CTEQ6L (LO) and CTEQ6.1M
(NLO) of Ref.~\cite{Pumplin:2002vw} were used, and 
$\mu_F=\mu_R=m_t$.}\label{tab:cdtev}
\begin{center}
\renewcommand{\arraystretch}{1.2}
\begin{tabular}{|ccccc|} \hline  
          &    &  Dilepton  &Lepton--jet  & Jet--jet  \\ \hline 
${\rm C}_{\rm hel}$ &LO  & $-0.471$    & $-0.240$ & $-0.123$ \\ 
          &NLO & $-0.352$    & $-0.168$ &  $-0.080$ \\ \hline
${\rm C}_{\rm beam}$ &LO &  $\phantom{-}0.928$    
&  $\phantom{-}0.474$ &  
$\phantom{-}0.242$ \\
           &NLO&  $\phantom{-}0.777$    &  $\phantom{-}0.370$ 
&  $\phantom{-}0.176$ \\ \hline
${\rm C}_{\rm off}$   &LO&  $\phantom{-}0.937$    
&  $\phantom{-}0.478$ 
&  $\phantom{-}0.244$ \\
           &NLO&  $\phantom{-}0.782$    &   $\phantom{-}0.372$
&  $\phantom{-}0.177$ \\ \hline
${\rm D}$        &LO &  $\phantom{-}0.297$    
&  $\phantom{-}0.151$
 &  $\phantom{-}0.0773$ \\
           &NLO&  $\phantom{-}0.213$    &  $\phantom{-}0.101$
 &  $\phantom{-}0.0480$ \\ \hline
\end{tabular}
\end{center}
\end{table}
\begin{table}[htbp!]
\caption{\it Results for ${\rm C}_{\rm hel}$ and ${\rm D}$ for
$pp$ collisions at $\sqrt{s}=14$ TeV using the same PDF and parameters
as in Table \ref{tab:cdtev}.}\label{tab:cdlhc}
\begin{center}
\renewcommand{\arraystretch}{1.2}
\begin{tabular}{|ccccc|} \hline  
          &    &  Dilepton  &Lepton--jet  & Jet--jet  \\ \hline 
${\rm C}_{\rm hel}$ &LO  &   $\phantom{-}0.319$   
&  $\phantom{-}0.163$ 
&  $\phantom{-}0.083$ \\
          &NLO &   $\phantom{-}0.326$   &  
$\phantom{-}0.158$ 
&   $\phantom{-}0.076$ \\ \hline
${\rm D}$        &LO &  $-0.217$   & $-0.111$ & $-0.0567$ \\
           &NLO&  $-0.237$   & $-0.115$ & $-0.0560$ \\ \hline 
\end{tabular}
\end{center}
\end{table}
\par
In Table \ref{tab:cdpdf} we compare the NLO results for the spin correlation
coefficients evaluated for the 3 different PDFs that we used.
The CTEQ6.1M and MRST2003 distributions give almost identical results, which
is remarkable since the individual contributions from the $q\bar{q}$ and $gg$
initial states differ in some cases significantly (see Tables \ref{tab:nz0},
\ref{tab:nz1}). (For previous versions of these distribution functions,
the agreement for the spin correlations was not so striking 
\cite{Bernreuther:2001rq}.)
Using the GRV98 PDF, however, leads to
spin correlations that are up to $\sim$ 13\% smaller at the Tevatron
and  up to $\sim$ 4\% larger at the LHC as compared to the other
two PDFs. 
This shows that the spin correlations are very sensitive to the
relative quark and gluon content of the proton. Future measurements
of these correlations may help to pin down the parton distributions further.
\begin{table}
\caption{\it Spin correlation coefficients at NLO for different PDFs for
the Tevatron (upper part) and the LHC (lower part) for dilepton final
states.}\label{tab:cdpdf}
\begin{center}
\renewcommand{\arraystretch}{1.2}
\begin{tabular}{|cccc|} \hline  
\multicolumn{4}{|c|}{Tevatron} \\ \hline
              &  CTEQ6.1M  &MRST2003       &GRV98\\ \hline
${\rm C}_{\rm hel}$ & $-0.352$    &  $-0.352$   & $-0.313$  \\  \hline
${\rm C}_{\rm beam}$  &  $\phantom{-}0.777$    
&   $\phantom{-}0.777$   
& $\phantom{-}0.732$   \\ \hline
${\rm C}_{\rm off}$  &  $\phantom{-}0.782$    
& $\phantom{-}0.782$   & $\phantom{-}0.736$ \\ \hline
${\rm D}$    &  $\phantom{-}0.213$    &    $\phantom{-}0.212$   
&  $\phantom{-}0.185$  \\ \hline \hline
\multicolumn{4}{|c|}{LHC} \\ \hline
${\rm C}_{\rm hel}$ &    $\phantom{-}0.326$   &    
$\phantom{-}0.327$   &  $\phantom{-}0.339$ \\ \hline
${\rm D}$       &  $-0.237$   &  $-0.237$   &$-0.243$ \\ \hline
\end{tabular}
\end{center}
\end{table}
\begin{figure}[htbp!]
  \unitlength1.0cm
  \begin{center}
    \begin{picture}(7.5,7.5)
      \put(-5.85,-3.2){\psfig{figure=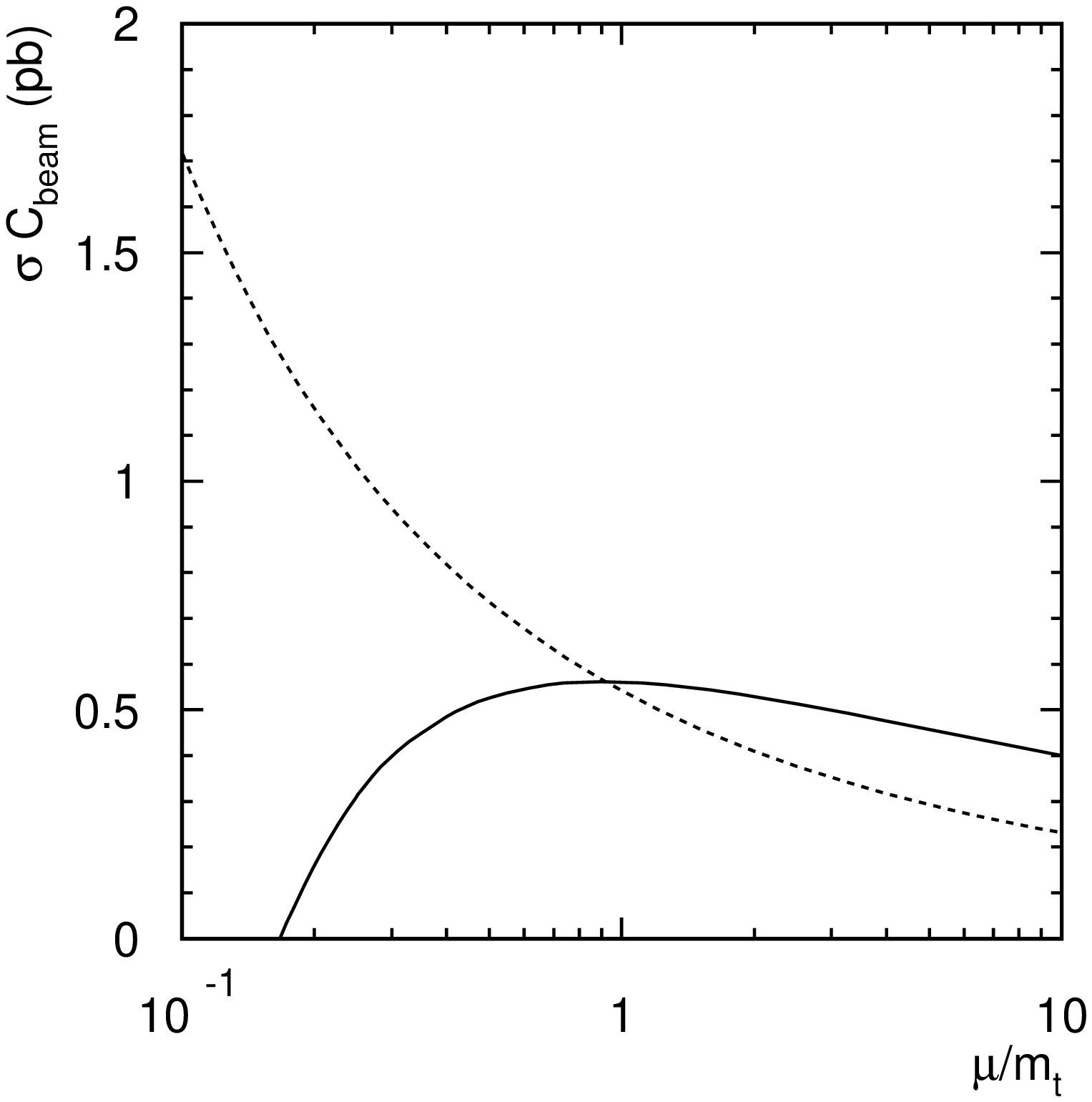,width=9cm}}
      \put(3.15,-3.2){\psfig{figure=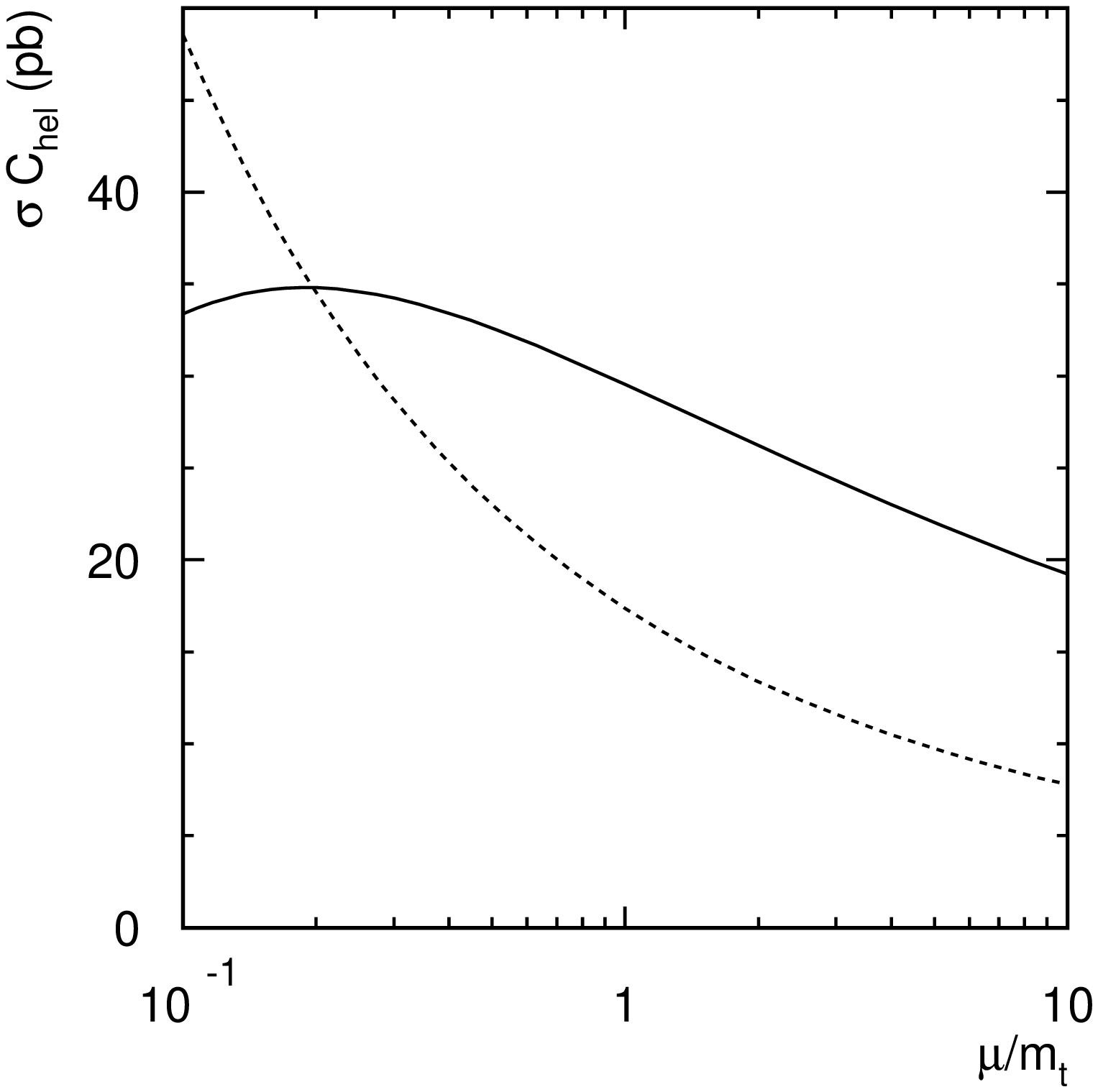,width=9cm}}
    \end{picture}
    \caption{\it Left: Dependence of $\sigma\*{\rm C}_{\rm beam}$ 
    on $\mu=\mu_F=\mu_R$ for $p\bar{p}$ collisions
at $\sqrt{s}=1.96$ TeV in the dilepton decay 
channel. The dashed line shows the LO result, the full
line shows the result at NLO. Right: The same for 
$\sigma\*{\rm C}_{\rm hel}$ for $pp$ collisions
at $\sqrt{s}=14$ TeV.}
    \label{fig:mu}
  \end{center}
\end{figure}
\begin{table}[ht]
\caption{\it  
Dependence of the correlation coefficients, computed with the PDF CTEQ6.1M, 
on $\mu=\mu_R=\mu_F$ at NLO for the dilepton decay channel.}\label{tab:mu}
\begin{center}
\renewcommand{\arraystretch}{1.2}
\begin{tabular}{|c|cccc|cc|} \hline 
&\multicolumn{4}{|c}{Tevatron}
&\multicolumn{2}{|c|}{LHC} \\
$\mu_R=\mu_F$ &${\rm C}_{\rm hel}$ &${\rm C}_{\rm beam}$ & 
${\rm C}_{\rm off}$ &${\rm D}$   &${\rm C}_{\rm hel}$ & ${\rm D}$ \\ \hline
$m_t/2$       &$-0.326$  &0.735 &  0.740 &0.191 &0.329 &$-0.244$ \\
$m_t$         &$-0.352$  &0.777 &  0.782 &0.213 &0.326 &$-0.237$ \\
$2m_t$        &$-0.370$  &0.804 &  0.810 &0.227 &0.324 &$-0.232$ \\ \hline
\end{tabular}
\end{center}
\end{table}
\par
To estimate higher order corrections to our predictions, we now
study the dependence on the renormalization and factorization
scales. For simplicity, we keep $\mu=\mu_F=\mu_R$ and vary
$\mu$ between $m_t/10$ and $10m_t$. The results for the
quantities $\sigma {\rm C}_{\rm beam}$ (Tevatron) and
$\sigma {\rm C}_{\rm hel}$ (LHC) are shown in Fig.~\ref{fig:mu}
for the dilepton decay channel, where we sum over $\ell=e,\mu,\tau$.
The inclusion of the QCD corrections reduces the scale dependence 
significantly. The correlation coefficients ${\rm C}$ and ${\rm D}$
are ratios in which a factor $\alpha_s^2$ drops out. Therefore, at
LO, the coefficients only depend on $\mu_F$ but not on $\mu_R$.
A comparison between the scale dependence at LO and NLO is thus
meaningless for these coefficients. Table \ref{tab:mu} therefore only
lists the NLO scale dependence. At the Tevatron, a variation
of $\mu$ between $m_t/2$ and $2m_t$ changes the results at 
$\mu=m_t$  by $\sim \pm$ (5--10)\%, while at the LHC
the change of ${\rm C}_{\rm hel}$ is less than a per cent and 
${\rm D}$ varies by $\sim \pm$ 3\%.  
\par
Before closing this section, we summarize how 
an experimental measurement of the distributions 
(\ref{kap6dist1}) and (\ref{kap6dist2}) that matches our predictions
should proceed:
\begin{enumerate}
\item Reconstruct the top and antitop 4-momenta in the laboratory frame
  ($=$ c.m. frame of the colliding hadrons).
\item Perform a rotation-free boost from the laboratory frame to the 
  $t\bar{t}$-ZMF. Compute ${\bf \hat a}$ and ${\bf \hat b}$ in that frame.
\item Perform rotation-free boosts from the $t\bar{t}$-ZMF
  to the top quark rest frame and 
the antitop quark rest frame. Compute the direction
  ${\bf \hat q}_1$  of the top quark decay product $a_1$ 
in the $t$ rest frame 
and the direction ${\bf\hat q}_2$ of the antitop quark decay
product $a_2$ in the $\bar{t}$ rest frame.  Finally, compute
  $\cos\theta_1={\bf \hat a}\cdot{\bf \hat q}_1,\  
  \cos\theta_2={\bf \hat b}\cdot{\bf \hat q}_2$ and $\cos\phi
  ={\bf\hat q}_1\cdot{\bf\hat q}_2$.
\end{enumerate}
Note that in this prescription 
the $t$ and $\bar{t}$ rest frames are obtained by first boosting
into the $t\bar{t}$-ZMF. If this step is left out, and the 
$t$ and $\bar{t}$ rest frames are constructed 
by directly boosting from the lab frame, a Wigner rotation has to
be taken into account.
\par
The results in this paper were obtained without imposing kinematic cuts.
Such cuts will in general distort the distributions, i.e. ${\rm C}$
and ${\rm D}$ will in general depend on the angles $\theta_1,\theta_2$
and $\varphi$, respectively.
The expectation values on the r.h.s. of Eqs.~(\ref{expval}) 
may still be used as measures of the $t\bar{t}$ spin correlations in the
presence of cuts.  They  are then  no longer directly
related to these differential distributions. 
One strategy, 
followed for instance in Ref.~\cite{Slabospitsky:2002ag}, is to correct for 
these distortions by Monte Carlo methods before extracting the
spin correlation coefficient and comparing with theoretical predictions.
A future aim will be to directly include the cuts
in an NLO event generator to be constructed with the above results
for all relevant  $2\to 6$ and $2\to 7$ processes.
\newpage
\section{Conclusions}
\label{conclusions}

We have determined  the differential cross sections for  
top quark pair production in a general spin configuration
by $q\bar q$ annihilation, $g g$, $q g$, and ${\bar q} g$ fusion
to order $\alpha_s^3$. These cross sections provide, together
with the differential rates of polarized
top and antitop decays at order $\alpha_s$, the NLO 
factorizable  contributions  to the reactions given in
Eqs.~(\ref{eq:ttrec1}) to Eq.~(\ref{eq:ttrec3}).
We have evaluated the corresponding spin density matrices for on-shell
intermediate top and antitop quarks. However, our results can also be 
employed as building blocks for taking the finite top quark width at NLO
into account.
(For studies at the Born level, see for instance
 Refs.~\cite{Kauer:2001sp,Kauer:2002sn}.)

As an application of the above results we have studied a number of
distributions due to top--antitop spin correlations. 
The QCD-induced spin correlations 
are large effects, which are easily visible in 
distributions of the final-state
particles. Given the size of the spin correlations, 
they are expected to become 
a good tool for  analysing in detail top quark
pair production and decay dynamics. 
They  can
be studied at the Tevatron and --- in view of the expected large $t\bar
t$ data samples --- especially at the LHC, in the dilepton, 
single lepton and all-hadronic 
decay channels, by measuring suitably defined 
double angular distributions. In fact, a first measurement 
of spin correlations in the off-diagonal basis was already
performed by the D0 collaboration for the
dilepton decay channel \cite{Abbott:2000dt}. While this
analysis was limited by the very small event sample, it clearly
demonstrated the possibility of the experimental study of 
these effects. For the LHC a simulation was performed
\cite{Slabospitsky:2002ag}, showing that the experimental
accuracy of a measurement of ${\rm C}_{\rm hel}$ at the 
LHC can be expected
to be better than 10\%. 
\par
On the theoretical side, we have shown in this paper that the 
NLO QCD corrections to the double angular distributions are
of the order of 15 to 40\% for the Tevatron, and below 10\%
for the LHC. A study of the scale dependence of our NLO 
predictions
indicates that the residual theoretical uncertainty due
to higher order corrections should be $\lesssim$ 10\%. 
Work on soft gluon and threshold resummations will further
reduce the theoretical uncertainties. 
\par
In particular 
for $pp$ collisions at $\sqrt{s}$ = 14 TeV, the theoretical uncertainties
are quite small. 
By the time the 
LHC will start operating it may therefore be expected that 
further theoretical progress 
will have turned
top quark spin correlations  
into a precision tool for the analysis of $t\bar t$ events.
\subsubsection*{Acknowledgements}
We wish to thank A. Chapovsky for discussions about his work on
non-factorizable corrections.
\newpage
\appendix
\section{One-loop integrals}
\label{sec:LoopIntegrals}
In this appendix, we collect a number of functions that appear
in the virtual corrections given in  Section
\ref{sec:VirtualCorrections} 
and in Appendices \ref{sec:VirtResultsqq}, \ref{sec:VirtResultsgg}.
Only the real
parts of the integrals are given.
It proves useful to express the integrals in terms of the following variables:
\begin{eqnarray}
y_{\pm}&=&{1 \over 2}(1\pm\beta), \nn \\
\yt &=& {1 \over 2}(1-\beta \* y),\nn \\
\xi&=&{m^2 \over \hat{s}},\nn \\
x&=&{1-\beta \over 1+\beta}.
\end{eqnarray}

1)  The six-dimensional box integrals are defined by
\begin{eqnarray}
&&D^6_0(q_1,q_2,q_3,m_1,m_2,m_3,m_4)\nonumber \\&=&{1\over i\*\pi^2}
\*\int {d^{6}l\over
(l^2-m_1^2)\*((l+q_1)^2-m_2^2)\*((l+q_1+q_2)^2-m_3^2)\*
((l+q_1+q_2+q_3)^2-m_4^2)}.
\end{eqnarray}

The following integrals appear in our results:
\begin{eqnarray} 
\Dsixa  &=& {\pi \over  \hats\* (\yt -1)}
\* \Bigg\{{1 \over 1-{\xi \over \yt \*(1-\yt )}}\*\left[-\Li2\left(1-{\xi \over \yt }\right) +
  {1 \over 2}\*\ln^2(\yt )\right]\nn \\ 
&+& {\ym  \over (\yp -\ym )}\*{1 \over (1-{\yp  \over 1-\yt })}\*
\left[-\Li2(\ym )+{1 \over 2}\*\ln^2(\ym )\right] \nn \\
&+&{\yp  \over (\ym  - \yp )} \* {1 \over (1-{\ym  \over 1-\yt })}
\*\left[-\Li2(\yp )+{1 \over 2}\*\ln^2(\yp )\right]\Bigg\}.
\\
\Dsixone &=&{2\*\pi \over \hats\* \beta\* (1-y^2)}\*\Bigg\{
{(1+\beta)\*(1-y)\over 4}\*\left[{\pi^2\over 3}-\ln^2(x)\right]\nn \\
&+&2\*\yt\*\left[
\Li2(-x)-\ln(x)\*\ln\left({\yt\over \ym}\right)
\right]+ (\beta-y)\*
\Li2\left(1-{\yt \over \xi}\right)
\Bigg\}.
\\
\Dsixthree &=&{\pi \over \hats }\*\Bigg\{
{\pi^2 \over 2}+{4\*\xi\over \beta^2 \* (1-y^2)}\*
\Bigg[{\pi^2 \over 6}+
{\yt\*(1-\yt) \over 2\*\xi}
\*\ln^2\left({\yt \over 1-\yt}\right)  \nn \\
&+&\Li2\left(1-{\yt\over \xi}\right)
+\Li2\left(1-{1-\yt \over \xi}\right)
\Bigg]\Bigg\}.\\
\Dsixb  &=&  \Dsixa |_{y \rightarrow -y}.\\
\Dsixcone &=& \Dsixone |_{y\rightarrow -y}.
\end{eqnarray}
2) The 3-point integrals are defined by
\begin{equation}
C_0(q_1,q_2,m_1,m_2,m_3)={1\over i\*\pi^2}
\*\int {(2\*\pi\*\mu)^{2\*\epsilon}\*d^{4-2\*\epsilon}l\over
(l^2-m_1^2)\*((l+q_1)^2-m_2^2)\*((l+q_1+q_2)^2-m_3^2)}.
\end{equation}
The following integrals appear in our results 
(they agree with those listed in \cite{Beenakker:1988bq}):

\begin{eqnarray}
C_0(p_1,p_2,0,0,0) &=& {C_{\epsilon} \over \hats } \* \left[
{1\over \epsilon^2}+{1\over \epsilon}\*\ln(\xi)\right]+\C(p_1,p_2,0,0,0)
,\nonumber \\
\C(p_1,p_2,0,0,0)&=&
{1\over \hat{s}}\*\left[
{1 \over 2}\* \ln^2\left(\xi\right)-{7\*\pi^2 \over 12}\right].\\
C_0(-p_1,k_1,0,0,m) &=& -{C_{\epsilon} \over \yt\*\hats}
\* \left[
{1\over 2\*\epsilon^2}-{1\over \epsilon}
\*\ln\left({\yt\over \xi}\right)\right]
+\C(-p_1,k_1,0,0,m),\nonumber \\
\C(-p_1,k_1,0,0,m)&=&-{1\over \yt\*\hats }\*\left[
\ln^2\left({\yt \over \xi}\right)
+\Li2\left(1-{\yt \over \xi}\right)
+{\pi^2 \over 24}\right]. \\
C_0(-p_1,k_2,0,0,m) &=&
C_0(-p_1,k_1,0,0,m)|_{y\rightarrow -y}.\\
C_0(k_1,k_2,0,m,0)&=&{1 \over \hats \* \beta} \* \left[
2\* \Li2(-x)+{1 \over 2}\* \ln^2(x)+{\pi^2 \over 6}\right].\\
C_0(k_1,k_2,m,0,m) &=&
{C_{\epsilon} \over \hats \* \beta}\* \left[
{1\over \epsilon}\*\ln(x)\right]+\C(k_1,k_2,m,0,m),\nn \\
\C(k_1,k_2,m,0,m) &=&
{1\over \hats \* \beta}\* \left[
-2\* \ln(x)\*\ln(1-x)-
2\* \Li2(x)+{1 \over 2}\* \ln^2(x)-{2\*\pi^2 \over 3}\right]. \\
C_0(k_1,-p_1,0,m,m)&=&{1 \over \yt \*\hats }\*
\left[{\pi^2 \over 6}-\Li2\left(1-{\yt\over \xi}\right)\right]. \\
C_0(k_2,-p_1,0,m,m)&=&C_0(k_1,-p_1,0,m,m)|_{y\rightarrow -y}.\\
C_0(p_1,p_2,m,m,m))&=&{1 \over 2\* \hats }\*\left[\ln^2(x)-\pi^2\right].
\end{eqnarray}
3) The two-point integrals are defined by
\begin{equation}
B_0(q,m_1,m_2)={1\over i\*\pi^2}
\*\int {(2\*\pi\*\mu)^{2\*\epsilon}\*d^{4-2\*\epsilon}l\over
(l^2-m_1^2)\*((l+q)^2-m_2^2)}={C_{\epsilon}\over \epsilon}+\B(q,m_1,m_2).
\end{equation}
The following finite parts of these integrals are needed:
\begin{eqnarray}
\B(k_1+k_2,0,0)&=& 2+\ln\left(\xi\right).\\
\B(k_1,0,m)&=& 2.\\
\B(k_1-p_1,0,m)&=& 2+{\yt\over \yt-\xi}\*
\ln\left({\xi\over \yt}\right).\\
\B(k_2-p_1,0,m)&=&
\B(k_1-p_1,0,m)|_{y\rightarrow -y}.\\
\B(k_1+k_2,m,m)&=&2+\beta\* \ln(x).
\end{eqnarray}
4) Finally, the one-point integral is given by
\begin{equation}
A_0(m)={1\over i\*\pi^2}
\*\int {(2\*\pi\*\mu)^{2\*\epsilon}\*d^{4-2\*\epsilon}l\over
(l^2-m^2)}={C_{\epsilon}\over \epsilon}\*m^2+m^2.
\end{equation}

\section{Virtual corrections to $\mathbf{q\bar{q}\to t\bar{t}}$}
\label{sec:VirtResultsqq}
Here we list the coefficients $A_V^{q\bar{q}}$ and $e_a^{V}$, 
defined in (\ref{calf2}), which appear in the one-loop contributions
to $q{\bar q} \to t{\bar t}$ (note that the contributions proportional 
to the finite part of the 
one-point integral ($=m^2$) and those of $\B(k_1,0,m)=2$ 
are added to those terms that are 
not multiplied by an $n$-point integral):

\begin{eqnarray}
A_V^{q\bar{q}}&=& e_0^{V}={\Dsixa\over \pi} \*
\shat \* \beta^2\* (1-y^2)\* 
{N^2-2 \over 4\* N}\* \bigg({\beta^2-3\over 1-\beta\*y}+2\*\beta\*y
\bigg)\nn \\
&+& {\Dsixb\over \pi} \*
\shat\* \beta^2\* (1-y^2)\* 
{1 \over 2\*N}\* \bigg({\beta^2-3\over 1+\beta\*y}-2\*\beta\*y\bigg) \nn \\
&-&{m^2 \over \beta^2} \* C_0(k_1,k_2,0,m,0)
\* \Bigg\{{N\over 2}\*
(2\*y^2\*\beta^2-3\*y^2+2\*\beta^3\*y-2\*\beta\*y+1)
+{4\*\beta\*y\*(1-\beta^2)\over N} \nn \\ 
&+&\beta^2\*(\beta^2-3)\*\left[{N^2-2\over N\*(1-\beta\*y)}
+{2\over N\*(1+\beta\*y)}\right]\Bigg\}\nn \\
&-&{\B(k_1+k_2,0,0) \over 4\* {\beta}^2}\* 
      \Big\{
N\*(1-3\*y^2-\beta^4+y^2\*\beta^4-2\*\beta\*y+4\*y^2\*\beta^2)
\nn \\ &-&{\beta\over N}\*(-8\*y+6\*\beta-3\*\beta^3+3\*y^2\*\beta^3)
  \Big\}
+ {N^2-2 \over 2\* N}\* 
\B(k_1-p_1,0,m)\* 
\Bigg\{
{2\*(1-\beta^2)\over 1-\beta\*y}-(1+\beta\*y)
\Bigg\}\nn \\
&+& {1 \over N} \* \B(k_2-p_1,0,m)\*
\Bigg\{{2\*(1-\beta^2)\over 1+\beta\*y}-(1-\beta\*y)\Bigg\}
 \nn \\
&+& {1 \over 4\* N} \* \B(k_1+k_2,m,m)\* 
(y^2+2 \* y^2\* \beta^2+5-2\* \beta^2)\nn \\
&+& {1\over 2\*\beta^2}\*\Bigg\{
(2\*\beta^3\*y+5\*y^2\*\beta^2+\beta^2-2\*\beta\*y+1-3\*y^2)\*N
+{\beta\*(1-\beta^2)\over N}\*(\beta-\beta\*y^2+8\*y)\nn \\
&-&4\*\beta^2\*(1-\beta^2)\*\left[{N^2-2\over N\*(1-\beta\*y)}
+{2\over N\*(1+\beta\*y)}\right]
\Bigg\}.
\end{eqnarray}

\begin{eqnarray}
e_1^{V}&=&-{\shat\*\Dsixa\over\pi}
     \*{N^2 -2 \over 2\*N}\*
    \left(2-{1-\beta^2\over 1-\beta\*y}\right) \nn \\
&-&  {\shat \*\Dsixb\over \pi\*N}\* 
      \left(2-{1-\beta^2\over 1+\beta\*y} \right) \nn \\
&+& {m^2\over \beta^2}\*C_0(k_1,k_2,0,m,0)
 \*\left\{N+2\*\beta^2\*\left[{N^2-2\over N\*(1-\beta\*y)}
+{2\over N\*(1+\beta\* y)}
\right]\right\}  \nn \\
&+&{\B(k_1+k_2,0,0)\over 2\*\beta^2}\* 
      \left\{(1-\beta^2)\*N+{3\*\beta^2\over N}\right\}
-{\B(k_1-p_1,0,m)   \* \beta \* 
    \left(N^2 -2\right) \* 
    \left( \beta - y \right)  \over N\*
    \left( 1 - \beta \* y \right) }
\nn \\
&-&{2 \* \B(k_2-p_1,0,m) \*
\beta \*  \left( \beta + y \right)  \over N
    \* \left( 1 + \beta \* y \right) } 
+{3 \* \B(k_1+k_2,m,m) \over 2\* N}
\nn \\
&+& {1 \over \beta^2}\* \Bigg\{
N\*(3\*\beta^2-1)
-2\*\beta^2\*(1-\beta^2)\*\left[{N^2-2\over N\*(1-\beta\*y)}
+{2\over N\*(1+\beta\* y)}\right]
\Bigg\}.
\end{eqnarray}

\begin{eqnarray}
e_2^{V}&=&-{\shat\* \Dsixa\over \pi}\*
    {N^2-2  \over 2 \* N}\*   
  \left\{1+\beta^2-3\*\beta\*y+3\*\beta^3\*y-2\*\beta^3\*y^3
+{(1-\beta^2)^2\over 1-\beta\*y}
\right\} \nn \\
&-& {\shat\* \Dsixb\over \pi\*N}\*
\left\{1+\beta^2+3\*\beta\*y-3\*\beta^3\*y+2\*\beta^3\*y^3+
{(1-\beta^2)^2\over 1+\beta\*y}
\right\}  \nn \\
&+& {m^2\over \beta^2}\*C_0(k_1,k_2,0,m,0)\* 
 \Bigg\{N\*(2\*y^2\*\beta^2-3\*y^2+2\*\beta^3\*y-2\*\beta\*y+3)
+{8\*y\*\beta\*(1-\beta^2)\over N}
\nonumber \\ &-& 2\*\beta^2\*(1-\beta^2)\*\left[{N^2-2\over N\*(1-\beta\*y)}
+{2\over N\*(1+\beta\* y)}\right]
             \Bigg\} \nn \\
&+&{\B(k_1+k_2,0,0) \over 2\* {\beta}^2}\* 
\left\{N\*
(3-2\*\beta\*y-3\*y^2-\beta^4+y^2\*\beta^4-2\*\beta^2+4\*y^2\*\beta^2)
+{\beta\over N}\*(8\*y+3\*\beta^3-3\*y^2\*\beta^3)
   \right\} \nn \\
&-&{\B(k_1-p_1,0,m)\* 
     \left(N^2-2 \right) \* 
    \left( 1 - \beta \* y \right)  \over N} 
-{2 \* \B(k_2-p_1,0,m)\* 
\left( 1 + \beta \* y \right)  \over N } \nn \\
&+&{\B(k_1+k_2,m,m) \* \left( 1 + 2 \* {\beta}^2 \right)\*
  \left( 1 - y^2 \right)  \over 2\* N} \nn \\
&-&{1\over \beta^2}\*\Bigg\{
(3-2\*\beta\*y+2\*\beta^3\*y-3\*y^2-5\*\beta^2+5\*y^2\*\beta^2)\*N+
{\beta\over N}\*(1-\beta^2)\*(\beta-\beta\*y^2+8\*y)
\Bigg\}.
\end{eqnarray}
\begin{eqnarray}
e_3^{V}&=& \sqrt{(1-y^2)\*(1-\beta^2)}\*\Bigg\{
{m^2\over 2\*\beta^2}\*C_0(k_1,k_2,0,m,0)\* 
\left[
-(2\*\beta+3\*y)\*N+{8\*\beta\over N}
\right]
\nn \\
&+&
{\B(k_1+k_2,0,0)\over 4\* \beta^2 }\* 
\left[
(-2\*\beta-3\*y+2\*\beta^2\*y)\*N+{8\*\beta\over N}
\right]
\nn \\
&+&{\B(k_1-p_1,0,m)\* 
      \left(N^2-2\right)\* \beta \over 2 \* {N}
    \* \left( 1 - \beta \* y \right) }
- {\B(k_2-p_1,0,m)\* 
\beta \over N\* (1+ \beta \* y)} 
-{\B(k_1+k_2,m,m)\* y  \over 4 \* N}
\nn \\
&-&{1 \over 2 \* {\beta}^2} \* \Bigg[
N\*(3\*\beta^2\*y-3\*y-2\*\beta)+{\beta\*(8-\beta\*y)\over N}
+2\*\beta^3\*\left(
{N^2-2\over N\*(1-\beta\*y)}
-{2\over N\*(1+\beta\* y)}
\right) \Bigg]\Bigg\}.
\end{eqnarray}

\newpage
\section{Virtual corrections to $\mathbf{gg\to t\bar{t}}$}
\label{sec:VirtResultsgg}
We list here the 
coefficients $A_V^{gg},C_V^{gg},D_V^{gg}$ and $E_V^{gg}$
of Eq.  (\ref{fgg}), which
are part of the IR-finite terms of the one-loop corrections to
$gg \to t {\bar t}$.

\begin{eqnarray}
A_V^{gg}&=&{{\beta}^2 \* \Dsixa \*  {N}^2\*  
     \shat \* \left( 1 - y^2 \right)  \over 2\*\pi} \* \Bigg\{
-1-\beta\*y+{4-\beta^2\over 1-\beta\*y}-2\*
{2-\beta^2\over (1-\beta\*y)^2}
+{2\*(1-\beta^2)^2\over (1-\beta\*y)^3}
\Bigg\} \nn \\
&+&{\Dsixone \* \shat \* 
    \left(2 + N^2\*  
       \left(1 - \beta \* y \right)  \right)  \over 4\*\pi\*  
    N^2 \*\beta^2\*  
    \left( 1 + \beta \* y \right) }\* \Bigg\{
\beta^2\*y^2+\beta^5\*y-3\*\beta^3\*y+2\*\beta\*y-\beta^4-2\*\beta^2+6
\nn \\
&-&{\beta^6+12-2\*\beta^4-3\*\beta^2\over 1-\beta\*y}
+{2\*(4-3\*\beta^2)\*(1-\beta^4)\over (1-\beta\*y)^2}
-{2\*(1+\beta^2)\*(1-\beta^2)^3\over (1-\beta\*y)^3}
\Bigg\} \nn \\
&+&{2 \* {\beta}^4\*  {\Dsixthree } \* \shat\*  
    \left(1 - y^2 \right)  \over \pi\*{\left(1 - 
       {\beta}^2 \* y ^2 \right) }^3} \* \Big\{
-4 + 3 \* y ^2 + {\beta}^4 \* y ^2 + 
  {\beta}^2 \* (1-y^2)\*(3-y^2)
\Big\} \nn \\
&+&{ m^2\* C_0(k_1,k_2,0,m,0)\* 
    N^2 \over 
    2\*\beta^4} \* \Bigg\{
2\*(2+y^2)\*\beta^4+3\*(1-y^2)\*\beta^2-3
+{6\*\beta^6-22\*\beta^4-3\*\beta^2+3\over 1-\beta\*y}
\nn \\
&+&{8\*\beta^4\*(3-2\*\beta^2)\over (1-\beta\*y)^2}
-{8\*\beta^4\*(1-\beta^2)^2\over (1-\beta\*y)^3}
\Bigg\} \nn \\
&-&{m^2\*  
C_0(k_1,-p_1,0,m,m) \over
 \beta^2 \* N^2} \* \Bigg\{
(\beta^3\*y-\beta\*y+\beta^2\*y^2-\beta^4+4)\*N^2-2\*\beta^2-2\*\beta\*y \nn \\
&+& {(-1+\beta^4+3\*\beta^6-7\*\beta^2)
\*N^2-4\*\beta^4+4\*\beta^2+5-\beta^6\over 1-\beta\*y}
+{2\*(1-\beta^2)\*
(\beta^2\*(1-2\*\beta^2)\*N^2+\beta^4+\beta^2-1)\over (1-\beta\*y)^2}\nn \\
&+& {(-3-5\*\beta^4+3\*\beta^6+17\*\beta^2)\*N^2-2\*\beta^4+2\*\beta^2
-3-\beta^6
\over 1+\beta\*y}
+{4\*N^2\*(2\*\beta^2+\beta^4-5)\*\beta^2\over 
(1+\beta\*y)^2}
 \nn \\
&+&{8\*N^2\*(1-\beta^2)^2\*\beta^2\over (1+\beta\*y)^3}
\Bigg\}
+{m^2
\* C_0(p_1,p_2,m,m,m)\over
   N^2 \* \beta^2}
\* \Bigg\{-\beta^2\*(1-\beta^2)\*N^3-N^2\*\beta^2+2
\nn \\
&+&{\beta^2\*(1-\beta^2)\*N^3+(1-\beta^2)\*(1+4\*\beta^2+\beta^4)\*N^2
-\beta^6+\beta^2-3\*\beta^4-5\over 1-\beta\*y}
\nn \\
&-&{(1+\beta^2)\*((1-\beta^2)^2\*N^2+\beta^4+2\*\beta^2-5)\over (1-\beta\*y)^2}
-{2\*(1+\beta^2)\*(1-\beta^2)^2\over (1-\beta\*y)^3}
\Bigg\} \nn \\
&+& {\B(k_1-p_1,0,m) \* \left[2 - 
      N^2\*( 1 + \beta \* y) 
      \right]  \over 8 \* N^2 \*     
    \left( 1 + \beta \* y \right) } \* \Bigg\{
-(10\*\beta\*y+31-9\*\beta^2)\*N^2+2\*\beta\*y+23-9\*\beta^2 \nn \\
&+&{(1-\beta^2)\*(\beta^2-5)\*(N^2-1)\over 1-2\*\beta\*y+\beta^2}
+4\*
{(\beta^4-13\*\beta^2+18)\*N^2-14-\beta^4+9\*\beta^2\over 1-\beta\*y}
\nn \\
&+& {8\*(1-\beta^2)\*((3\*\beta^2-5)\*N^2-3\*\beta^2+3)\over (1-\beta\*y)^2}
+{8\*(1-\beta^2)^2\*((1-\beta^2)\*N^2+\beta^2)\over (1-\beta\*y)^3}
\Bigg\} \nn \\
&+&{\B(k_1+k_2,0,0) \* N^2\* (1 - y^2)\*  
    \left( -2 + \left( -3 + 4 \* {\beta}^2 + 2\*  {\beta}^4
         \right) \* y^2 \right)  \over 4\* 
    \left(1 - \beta^2 \* y^2 \right) } \nn \\
&+&{\B(k_1+k_2,m,m)\over 
4 \* N^2 \* \left(1 -\beta^2\* y^2 \right)} \* \Big\{
2\*\beta^2\*y^2\*(1-\beta^2)^2\*N^3+(1-y^2)\*
(2-\beta^2\*y^2-2\*\beta^4\*y^2)\*N^2+4\*(1-y^2)
 \Big\} \nn \\
&+&{1 \over 24 \* N^2\*\beta^4} 
\* \Bigg\{
(-60\*\beta^4\*y^2+23\*\beta^6+36\*\beta^2\*y^2-41\*\beta^4-72\*\beta^2+36)
\*N^4\nn \\ 
&-&3\*\beta^2\*(9\*\beta^4-43\*\beta^2+4-4\*\beta^2\*y^2)\*N^2
+12\*\beta^4-48\*\beta^2 
+{4\*N^4\*(6\*\beta^8-56\*\beta^6+59\*\beta^4+18\*\beta^2-9)\over 
1-\beta\*y}\nn \\
&-&{24\*\beta^4\*(1-\beta^2)\*
(4\*(2-\beta^2)\*N^4+(\beta^4+6\*\beta^2-13)\*N^2-\beta^4-3\*\beta^2+6)
\over (1-\beta\*y)^2}\nn \\
&+&{48\*\beta^4\*(N^2-1)\*(1-\beta^2)^2\*
((1-\beta^2)\*N^2+\beta^2)\over (1-\beta\*y)^3}
\nn \\
&+&{12\*\beta^2\*(
(2\*\beta^{10}+17\*\beta^6+3+2\*\beta^8-115\*\beta^2+43\*\beta^4)\*N^2
+24\*\beta^2-4\*\beta^6+12-2\*\beta^{10}-6\*\beta^8)\over (3+\beta^2)
\*(1-\beta\*y)} \nn \\
&+&{3\*\beta^4\*(N^2-1)\*(1-\beta^2)\*(\beta^2-5)\*
(N^2\*(3+\beta^2)-4)\over (3+\beta^2)\*(1+\beta^2+2\*\beta\*y)}
+{4\*{\beta}^6 \* \left(1 - {\beta}^2 \right) \* N^3 \* N_f \* 
y^2 \over   
    \left(1 - {\beta}^2 \* y ^2 \right) }
\Bigg\} +(y\to -y).
\end{eqnarray}

\begin{eqnarray}
C_V^{gg}&=&{2\*\Dsixa  \* N^2\* m^2 \over \pi}\* \Bigg\{
{10-3\*\beta^2\over 1-\beta\*y} 
-{6\*(1-\beta^2)\over (1-\beta\*y)^2}+
{2\*(1-\beta^2)^2\over (1-\beta\*y)^3} \nn \\
&+& {-\beta^3\*y^3+2\*\beta^3\*y
-\beta^2\*y^2+4\*\beta^2-4\*\beta\*y-4
\over 1-\beta^2} 
\Bigg\} \nn \\
&+&{\Dsixone \*  m^2\*  
    \left(2+ N^2\*(1-\beta \* y) \right)\*(1+\beta^2)  
\over \pi\*\beta^2\*  
    N^2 \*  ( 1 + \beta \* y) }\*  \Bigg\{
{8-\beta^2\over 1-\beta\*y} - 
{6\*(1-\beta^2)\over (1-\beta\*y)^2}
\nn \\
&+&{2\*(1-\beta^2)^2\over (1-\beta\*y)^3}
+ {-\beta^5\*y+\beta^4+\beta^3\*y-\beta^2\*y^2
-2\*\beta^2-2\*\beta\*y-4 \over 1-\beta^4}
\Bigg\} \nn \\
&+&{4 \* \Dsixthree  \* m^2\over \pi} \* \Bigg\{
-{2\over 1-\beta^2}+{2\*\beta^4+5\over 1-\beta\*y}
-{2\*(1-\beta^2)\*(2+\beta^2)\over (1-\beta\*y)^2}
+{2\*(1-\beta^2)^2\over (1-\beta\*y)^3}
\Bigg\} \nn \\
&+&{ C_0(k_1,k_2,0,m,0)\* 
    N^2 \* m^2 \over 2\*\beta^4} \* \Bigg\{
-2\*(2+y^2)\*\beta^4+(3\*y^2-7)\*\beta^2+3
-{3+6\*\beta^6-14\*\beta^4-7\*\beta^2\over 1-\beta\*y}
\nn \\
&-&{16\*\beta^4\*(1-\beta^2)\over (1-\beta\*y)^2}
+{8\*\beta^4\*(1-\beta^2)^2\over (1-\beta\*y)^3} 
\Bigg\} \nn \\
&+&{  C_0(k_1,-p_1,0,m,m) \* m^2 \over  \beta^2 \* N^2} 
\* \Bigg\{
(\beta^3\*y-\beta\*y+\beta^2\*y^2+2-\beta^4)\*N^2-2\*\beta^2-2\*\beta\*y
\nn \\
&-&{(1-\beta^4)\*((1+3\*\beta^2)\*N^2-3-\beta^2)\over 1-\beta\*y}
+{2\*(1-\beta^2)^2\*(2\*N^2\*\beta^2-\beta^2-1)\over (1-\beta\*y)^2}
+{8\*N^2\*(1-\beta^2)^2\*\beta^2\over (1+\beta\*y)^3} \nn \\
&+& {(3\*\beta^6+11\*\beta^2-5\*\beta^4-1)\*N^2-\beta^2-\beta^6-\beta^4-1 
\over 1+\beta\*y}
-{4\*N^2\*(1-\beta^2)\*(3+\beta^2)\*\beta^2\over (1+\beta\*y)^2}
\Bigg\} \nn \\
&+& 
{C_0(p_1,p_2,m,m,m)\* m^2\over
  \beta^2\* N^2 } \* \Bigg\{
\beta^2\*(1-\beta^2)\*N^3+N^2\*\beta^2-2 \nn \\
&+&{-\beta^2\*(1-\beta^2)\*N^3+(3\*\beta^4-\beta^2+\beta^6-1)\*N^2
+3+\beta^2+3\*\beta^4+\beta^6\over 1-\beta\*y} \nn \\
&+&{(1-\beta^4)\*(N^2\*(1-\beta^2)-3-\beta^2)\over (1-\beta\*y)^2}
+{2\*(1+\beta^2)\*(1-\beta^2)^2\over (1-\beta\*y)^3} 
\Bigg\} \nn \\
&+& {\B(k_1-p_1,0,m) \* (2 - 
      N^2 \* ( 1 + \beta \* y))\*(1-\beta^2) \over
    8 \* N^2\*  
    \left( 1 + \beta \* y \right) } \* \Bigg\{
{(-9\*\beta^2+11+10\*\beta\*y)\*N^2-2\*\beta\*y-19+9\*\beta^2 \over 1-\beta^2}
\nn \\
&+& {(1-\beta^2)\*(N^2-1)\over 1-2\*\beta\*y+\beta^2}
+{4\*(\beta^2-8)\*(N^2-1)\over 1-\beta\*y}
+{24\*(1-\beta^2)\*(N^2-1)\over (1-\beta\*y)^2}
-{8\*(1-\beta^2)^2\*(N^2-1)\over (1-\beta\*y)^3}
\Bigg\}  \nn \\
&-&{ \B(k_1+k_2,0,0)\*  N^2\*  
      \left( 1 - y^2 \right) \*  
      \left( 2 + \left( -3 + 4 \* {\beta}^2 + 2 \* {\beta}^4 \right)\* 
            y^2 \right)  \over 4\*  
    \left(1 - {\beta}^2 \* y ^2 \right) } \nn \\
&-&{\B(k_1+k_2,m,m)\over
  4 \* N^2 \* \left(1 - {\beta}^2 \* y ^2 \right)} \* \Bigg\{
2 \* {\beta}^2 \*y^2\* (1 - {\beta}^2)^2\*  
   N^3
 +  (-2 - ( 2 + 5 \* {\beta}^2 + 2 \* {\beta}^4)\* y^2 
 + \beta^2\*(1 + 2 \* {\beta}^2)\* y^4)\*N^2 \nn \\
 &-& 4 \* \left( 1 + y^2 \right) 
\Bigg\} 
+{1-\beta^2 \over 24\*\beta^4\*N^2} \*  \Bigg\{
{(60\*\beta^4\*y^2-43\*\beta^4+120\*\beta^2-23\*\beta^6
-36\*\beta^2\*y^2-36)\*N^4\over 1-\beta^2}
\nn \\ 
&-&{3\*\beta^2\*(-9\*\beta^4+31\*\beta^2+4\*\beta^2
\*y^2-4)\*N^2+12\*\beta^4-48\*\beta^2\over 1-\beta^2}
+{4\*N^4\*(6\*\beta^6-26\*\beta^4-21\*\beta^2+9)\over 1-\beta\*y} \nn \\
&+&12\*\beta^2\*{(4\*\beta^6+13\*\beta^4+54\*\beta^2-3+2\*\beta^8)\*N^2
-2\*\beta^8-12-38\*\beta^2-6\*\beta^6-12\*\beta^4\over 
(3+\beta^2)\*(1-\beta\*y)}
\nn \\
&+&{3\*\beta^4\*(1-\beta^2)\*(N^2-1)\*
(3\*N^2+N^2\*\beta^2-4)\over (3+\beta^2)\*(1+\beta^2+2\*\beta\*y)}
+{24\*\beta^4\*(1-\beta^2)\*(N^2-1)\*(4\*N^2-\beta^2-3)\over (1-\beta\*y)^2}
 \nn \\
&-&{48\*\beta^4\*(1-\beta^2)^2\*(N^2-1)^2\over (1-\beta\*y)^3}
-{4\*\beta^6\*N^3\*N_f\*y^2\over 1-\beta^2\*y^2}
\Bigg\} +(y\to -y).
\end{eqnarray}

\begin{eqnarray} 
D_V^{gg}&=&(1-\beta^2)\*\Bigg\{
{\Dsixa\*N^2 \over \pi\*\beta^2}\*\Bigg[{\beta^2\over 1-\beta\*y}
-{2\over 1-y^2}\Bigg] \nn \\
&+&
 { \Dsixone 
  \*( 2 + N^2 \* ( 1 - \beta \* y))  \over \pi\*\beta^2 
  \* N^2 \*( 1 + \beta \* y)} 
  \* \Bigg[
{2\*\beta^2-1\over 1-\beta\*y}+{1-\beta^2\over (1-\beta\*y)^2}
-{2\*\beta\*y\over 1-y^2}
  \Bigg] \nn \\
&-& {2\*(1 - \beta^2)^2 \*\Dsixthree \over
 \pi\* \beta^2\*( 1 - y^2)  
  \*( 1 - \beta^2\* y^2)^2} \nn \\
&+& {C_0(k_1,k_2,0,m,0)\* 
    N^2 \over
    4\* {\beta}^4}\*\Bigg[
3\*(1-\beta^2)+{(3+\beta^2)\*(2\*\beta^2-1)\over 1-\beta\*y}
-{4\*\beta^2\over 1-y}
\Bigg] \nn \\
&+&
{C_0(k_1,-p_1,0,m,m)\over  4\*\beta^2\* N^2}\* \Bigg[
{(1-\beta^2)\*(2\*N^2\*\beta^2+\beta^2-4)\over 1-\beta\*y}
-{2\*(1-\beta^2)\*(N^2-1)\*\beta^2 \over (1-\beta\*y)^2} \nn \\
&-&{2\*(2-3\*\beta^2+\beta^4)\*N^2+\beta^4+3\*\beta^2+4\over 1+\beta\*y}
+{4\*(1-\beta^2)\*N^2\*\beta^2\over (1+\beta\*y)^2}
-4\*{(1-\beta^2)\*N^2+2\*\beta\*y-2\over 1-y^2}
\Bigg] \nn \\
&+&{ 
   C_0(p_1,p_2,m,m,m) \over \beta^2\* N^2}\*\Bigg[
{N^2+\beta^2+1\over 1-\beta\*y}+{1-\beta^2\over (1-\beta\*y)^2}
-{N^2+2\over 1-y}
\Bigg]  \nn \\
&+& {\B(k_1-p_1,0,m)   
    \*(2 - N^2\*( 1 + \beta \* y))  \over \beta^2 
    \*N^2\* \hats\*(1 + \beta \* y)}\* \Bigg[
{N^2-\beta\over 1-y}+{N^2+\beta\over 1+y}
+{2\*\beta^2\over 1-\beta\*y}-{(1+N^2)\*\beta^2\over (1-\beta\*y)^2} \nn \\
&+&{(1-\beta^2)\*(2\*N^2-1)\*\beta^2\over (1-\beta\*y)^3}
 \Bigg] 
+{\B(k_1+k_2,0,0)\*N^2\over 
 2\*\beta^4\* \hats}\*\Bigg[
3-2\*\beta^2+{4\*\beta^2-3\over 1-\beta\*y}
\Bigg] \nn \\
&+&{\B(k_1+k_2,m,m) \over
 2\*\beta^2\* N^2\* \hats}\* \Bigg[
N^2+{(1+7\*\beta^2)\*N^2+4\*\beta^2+4\over (1-\beta^2)\*(1-\beta\*y)}
-4\*{N^2\*\beta^2+N^2+2\over (1-\beta^2)\*(1-y)}
\Bigg] \nn \\
&+&{1 \over
 N^2\* \beta^4\*\hats }\* \Bigg[
-N^2\*(3\*N^2\*(1-\beta^2)+\beta^2) 
+ {4\*(N^4-N^2+2)\*\beta^2 \over 1+y}
\nn \\
&-& 
{(5\*\beta^2-3)\*N^4-\beta^2\*(2\*\beta^4+6\*\beta^2-1)\*N^2
+5\*\beta^4+\beta^6+4\*\beta^2
\over 1-\beta\*y}\nn \\
&-&{\beta^4\*(4\*N^4-2\*N^2\*\beta^2-3+\beta^2)\over (1-\beta\*y)^2}
+{2\*\beta^4\*(1-\beta^2)\*(N^2-1)\*(2\*N^2-1)\over (1-\beta\*y)^3}
    \Bigg]
+(y\to -y) \Bigg\}.
\end{eqnarray}

\begin{eqnarray}
E_V^{gg}&=&{N^2\*\Dsixa \over \pi} \* \Bigg\{
2\*\beta^2-2\*\beta^2\*y^2-4\*\beta\*y-10
-2\*{1+2\*\beta\*y+\beta^2\over \beta^2\*(1-y^2)}
-{7\*\beta^2-19\over 1-\beta\*y} \nn \\
&-&{(1-\beta^2)\*(9-\beta^2)\over (1-\beta\*y)^2}
+4\*{(1-\beta^2)^2\over (1-\beta\*y)^3}
\Bigg\}
+ {N^2\*\Dsixb \over \pi} \*\Bigg\{ 
-2\*\beta^2+2\*\beta^2\*y^2+6
\nn \\
&-&5\*{1-\beta^2\over 1+\beta\*y}
+{(1-\beta^2)^2\over (1+\beta\*y)^2}
-2\*{1+2\*\beta\*y+\beta^2\over \beta^2\*(1-y^2)}
\Bigg\} \nn \\
&-&  {\Dsixone
\*(2 + N^2\* ( 1 - \beta \* y))\over 
  4\*\pi\* \beta^2\*N^2\*( 1 - \beta \* y )}\*
\Bigg\{
2\*\beta^2-4\*\beta\*y-6
+{(1-\beta^2)\*(3\*\beta^4+4\*\beta^2+9)
  \over 1-\beta\*y} \nn \\
&-&4\*{(1-\beta^4)\*(1-\beta^2)\over (1-\beta\*y)^2}
-{(1+\beta^2)\*(3\*\beta^4+6\*\beta^2-1)\over 1+\beta\*y}
+8\*{(1-\beta^2)\*\beta\*y\over 1-y^2}
\Bigg\} \nn \\
&+&
{\Dsixcone  
\*( 2 + N^2\*( 1 + \beta \* y))
\over 4\*\pi\* \beta^2\*N^2\*( 1+\beta \* y )}
\*
\Bigg\{
2\*\beta^2+4\*\beta\*y+2
+{(3-\beta^2)\*(1-\beta^2)^2\over (1+\beta\*y)}
\nn \\
&+&
{(1+\beta^2)\*(\beta^6-15\*\beta^4-13\*\beta^2-5)
\over (1-\beta^2)\*(1-\beta\*y)} 
+4\*{\beta\*(1+\beta)^3\over (1-\beta)\*(1-y)}
-4\*{(1-\beta)^3\*\beta\over (1+\beta)\*(1+y)}\Bigg\}
 \nn \\
&+&
{2\*\Dsixthree\over \pi}\*
\Bigg\{
{2\*\beta^4+\beta^2+9\over 1-\beta\*y}
-{(1-\beta^2)\*(5+3\*\beta^2)\over (1-\beta\*y)^2}
+4\*{(1-\beta^2)^2\over (1-\beta\*y)^3}
 \nn \\
&-& {(1-\beta^2)\*(3+2\*\beta^2)\over 1+\beta\*y}
+ {(1-\beta^2)^2\over (1+\beta\*y)^2}
-2\*{1+2\*\beta\*y+\beta^2\over \beta^2\*(1-y^2)}
\Bigg\}
 \nn \\
&-& {(1 - \beta^2)\*C_0(k_1,k_2,0,m,0) 
    \*N^2 \over 4\*\beta^4}\*
\Bigg\{
-2\*(3-\beta^2)\*(1+\beta\*y)
+ {15\*\beta^6-13\*\beta^2-24\*\beta^4+6\over  
(1-\beta^2)\*(1-\beta\*y)}
\nn \\ 
&+&4\*\beta^4\*{5-\beta^2\over (1-\beta\*y)^2}
-16\*\beta^4\*{1-\beta^2\over (1-\beta\*y)^3}
+\beta^2\*{1+11\*\beta^2\over 1+\beta\*y}
-4\*\beta^4\*{1-\beta^2\over (1+\beta\*y)^2}
+8\*\beta^2\*{1+2\*\beta\*y+\beta^2\over (1-\beta^2)\*(1-y^2)}
\Bigg\}
\nn \\
&+& 
{( 1 - {\beta}^2)\*C_0(k_1,-p_1,0,m,m)\over 2 \* {\beta}^2\* N^2 } \*
\Bigg\{
N^2\*(1+2\*\beta\*y+\beta^2)
-4
-{(2+3\*\beta^2+3\*\beta^4)\*N^2
-5-2\*\beta^2-\beta^4\over (1-\beta\*y)}
\nn\\
&-&{2\*(1-\beta^2)\*(2+\beta^2-3\*\beta^2\*N^2)\over (1-\beta\*y)^2}
-{(1+\beta^2)\*(1+3\*\beta^2)\*N^2+1-4\*\beta^2-\beta^4
\over 1+\beta\*y} \nn \\
&+&{4\*\beta^2\*N^2\*(1-\beta^2)\over (1+\beta\*y)^2}
-2\*{N^2\*(1+2\*\beta\*y+\beta^2)-2\*(1+\beta\*y)\over 
1-y^2}
\Bigg\}
 \nn \\
&+&
{C_0(k_2,-p_1,0,m,m)\over 2\* {\beta}^2\*N^2 
   }\*
\Bigg\{
(1-\beta^2)\*(3+2\*\beta\*y-\beta^2)\*N^2+4\*(1-\beta^2)
-3\*{(1-\beta^2)^2\*(1-N^2\*\beta^2)\over (1+\beta\*y)}
\nn\\
&-&{(5-3\*\beta^2)\*(1-\beta^2)^2\*N^2
+5+2\*\beta^2+9\*\beta^4\over 1-\beta\*y}
-{4\*\beta^2\*N^2\*(1-\beta^2)\*(3+\beta^2)\over (1-\beta\*y)^2}\nn\\
&+&{16\*\beta^2\*N^2\*(1-\beta^2)^2\over (1-\beta\*y)^3}
-2\*{(1-\beta^2)\*(1+2\*\beta\*y+\beta^2)\*N^2 
-2-2\*\beta \* (3\*y+3\*\beta+\beta^2\*y)\over 1-y^2}
\Bigg\}
\nn \\
&+&
{C_0(p_1,p_2,m,m,m)\*( 1 - {\beta}^2 )\over 2\*N^2\*\beta^2} \*
\Bigg\{-{4\*(2+N^2)\*(1+2\*\beta\*y+\beta^2)\over
  (1-\beta^2)\*(1-y^2)} +{(1-\beta^2)\*(3-\beta^2)\over (1+\beta\*y)^2}\nn \\
&-&
{\beta^2\*(1-\beta^2)^2\*N^3
-\beta^2\*(1+\beta^2)\*(3+\beta^2)\*N^2-7-3\*\beta^2-5\*\beta^4-\beta^6
\over (1-\beta^2)\*(1-\beta\*y)}+4\*{(1-\beta^4)\over (1-\beta\*y)^3}
\nn\\
&-&
{5+3\*\beta^4-2\*N^2\*(1-\beta^4)\over (1-\beta\*y)^2}
+{(N^3\*\beta^2\*(1-\beta^2)
+(2-3\*\beta^2-\beta^4)\*N^2-1-6\*\beta^2-\beta^4)\over (1+\beta\*y)}
\Bigg\}
\nn \\
&+& 
{\B(k_1-p_1,0,m)\*( 2 - N^2 
     \*( 1 + \beta \* y))\over 
    4\*N^2\* \hats }\*
\Bigg\{
-{(25-2\*\beta^2+\beta^4)\*N^2-17-\beta^4-6\*\beta^2\over
  1-\beta\*y}
\nn\\
&+&
{2\*(1-\beta^2)\*((9+\beta^2)\*N^2-\beta^2-5)\over (1-\beta\*y)^2}
+{4\*(1-2\*N^2)\*(1-\beta^2)^2\over (1-\beta\*y)^3}
+{32\*(N^2-1)\over (3+\beta^2)\*(1-2\*\beta\*y+\beta^2)}
\nn \\
&-& {(-25+\beta^4-9\*\beta^2+\beta^6)\*N^2
+1-23\*\beta^2-9\*\beta^4-\beta^6\over (3+\beta^2)\*(1+\beta\*y)}
+8\*{N^2\*(1+\beta\*y)-\beta\*(\beta+y)\over \beta^2\*(1-y^2)}
\Bigg\}
\nn \\
&+&
{\B(k_2-p_1,0,m)\*(2 -N^2\* ( 1 - \beta \* y ) )  \over N^2\* \hats  
    \*( 1 - \beta \* y)}\*
\Bigg\{
1-5\*N^2+{(5\*N^2-3)\*(1-\beta^2)\over 1+\beta\*y}
-{N^2\*(1-\beta^2)^2\over (1+\beta\*y)^2}\nn \\
&+&2\*{N^2\*(1+2\*\beta\*y+\beta^2)+\beta\*(\beta^2\*y+2\*\beta+y)
\over \beta^2\*(1-y^2)}
\Bigg\} \nn \\
&+&{\B(k_1+k_2,0,0)\*N^2\over 2\*\beta^4 \*\hats }\*
\Bigg\{
6\*(1+\beta^2)\*(1-2\*\beta^2)\*(1+\beta\*y)
-{(6-11\*\beta^2-19\*\beta^4+4\*\beta^6)\over 1-\beta\*y} \nn \\
&-&{\beta^2\*(1-\beta^2)\*(1+4\*\beta^2)\over 1+\beta\*y}
\Bigg\}
+{\B(k_1+k_2,m,m) \over 2\*\beta^2\* N^2 \*\hats}\*
\Bigg\{
2\*N^2\*(1+5\*\beta^2)\*(1+\beta\*y) \nn \\
&+&\beta^2\*{
2\*(1-\beta^2)^2\*N^3-(3+5\*\beta^2)\*N^2-8\over (1+\beta\*y)}
-{8\*(N^2\*\beta^2+2+N^2)\*(1+2\*\beta\*y+\beta^2)\over (1-\beta^2)\*(1-y^2)}
\nn\\
&+&{(-2\*\beta^2\*(1-\beta^2)^3\*N^3
+(2+11\*\beta^2+24\*\beta^4-5\*\beta^6)\*N^2+8\*(1+3\*\beta^2))
\over (1-\beta^2)\*(1-\beta\*y)}
\Bigg\}
\nn \\
&+&
{1\over 3\*\hats\*N^2\*\beta^4}\*
\Bigg\{
6\*N^2\*(1+\beta\*y)\*((3\*\beta^4+4\*\beta^2-3)\*N^2-\beta^2-\beta^4)
+{(32\*\beta^6-107\*\beta^4-39\*\beta^2+18)\*N^4\over 1-\beta\*y}
\nn \\ &+&
{3\*\beta^2
\*(39\*\beta^2+3\*\beta^4+2\*\beta^6-2)\*N^2
-3\*\beta^2\*(8+27\*\beta^2+10\*\beta^4+\beta^6)
\over 1-\beta\*y} \nn \\
&-& {6\*\beta^4\*(1-\beta^2)\*((\beta^2-7)\*N^4+(\beta^2+9)\*N^2-\beta^2-3)
\over (1-\beta\*y)^2}-{12\*\beta^4\*(1-\beta^2)^2\*(N^2-1)\*(2\*N^2-1)
\over (1-\beta\*y)^3} \nn \\
&+&{(1-\beta^2)\*\beta^2\*N^4\*(32\*\beta^2+3)\over 1+\beta\*y}
-{3\*(1-\beta^2)\*\beta^4\*
((\beta^2+5)\*(2\*\beta^2+7)\*N^2-\beta^4-14\*\beta^2-35)\over 
(3+\beta^2)\*(1+\beta\*y)} \nn \\
&-&{6\*N^2\*\beta^4\*(1-\beta^2)^2\*(N^2-1)\over (1+\beta\*y)^2}
+{24\*(N^4-N^2+2)\*\beta^2\*(1+\beta^2+2\*\beta\*y)\over 1-y^2} \nn \\
&+&{12\*\beta^4\*(N^2-1)\*(N^2\*\beta^2+3\*N^2-4)\over (3+\beta^2)\*
(1-2\*\beta\*y+\beta^2)} 
-{2\*\beta^5\*(1-\beta^2)\*N^3\*N_f\*y \over 1-\beta^2\*y^2}
\Bigg\}
\, .
\end{eqnarray}

\section{Helicity amplitudes for the real emission processes}
\label{sec:HelAmps}
In this section we give explicit results in terms of helicity amplitudes
for all real emission processes needed to compute the 
$2\to 3$ spin density matrices $R^i_{\rm res}$,
$i=q\bar{q},gg,qg,\bar{q}g$, discussed in sections
\ref{real} and \ref{partonsection}. Generically,
the matrices $R^i_{\rm res}$ can be obtained from 
\begin{equation}
\label{density}
\sum|{\cal{T}}|^2 \propto {\rm Tr} \left [ R^i_{\rm res} 
(\one + {\bf\hat s}_t\cdot\mathbf{\tau})
\otimes (\one + {\bf\hat s}_{\bar t}\cdot\mathbf{\tau})  \right ] \, ,
\end{equation}
where the amplitudes ${\cal T}$ are given below and the sum stands for
averaging or summing over colours and unobserved spins of the
initial- or final-state particles. For the evaluation of the
spin correlation observables introduced in Eq.~(\ref{genob})
one can also use the right-hand side of Eq.~(\ref{double}).
\newcommand{\bra}{\langle}
\newcommand{\ket}{\rangle}
\def\tb{{\bar t}}
\def\Qb{{\bar Q}}
\def\Q{Q}
\def\qb{{\bar q}}
\def\vb{\bar v}
\def\ub{\bar u}
\def\rem#1#2{}
\def\isandpp#1,#2,#3{\bra{#1}^+|#2|{#3}^+\ket}
\def\sandpp#1{\expandafter\isandpp#1}
\def\qb{{\bar q}}
\def\kt{{k_1}}
\def\ktb{{k_2}}
\def\kq{{k_1}}
\def\kqb{{k_2}}
\def\k#1{{p_#1}}
\def\r#1{{r_#1}}
\def\q#1{{q_#1}}
\def\cT{{\cal T}}
\def\maple#1{}
\def\idotp#1,#2{(#1\cdot#2)}
\def\dotp#1{\expandafter\idotp#1}
\def\ispa#1,#2{\bra #1#2\ket }
\def\spa#1{\expandafter\ispa#1}
\def\ispb#1,#2{[#1#2] }
\def\spb#1{\expandafter\ispb#1}

\def\st{{s_t}}
\def\stb{{s_\tb}}
\def\sandm(#1,#2,#3){\bra #1^-|#2|#3^-\ket}
\def\isandpp#1,#2,#3{\bra{#1}^+|#2|{#3}^+\ket}
\def\sandpp#1{\expandafter\isandpp#1}
\newcommand{\psl}{p\!\!\!/}
\subsection{The process $\mathbf{gg\to t \bar t g}$}
In this subsection we present the  Born amplitudes for the process
\begin{equation}
   g(\k1)g(\k2)g(\k3) \to t(\kt)\tb(\ktb)
\label{d2gggtt}
\end{equation}
for all possible gluon helicity, and for
$t$ and $\bar t$ spin configurations in 4 dimensions. 
The gluon 4-momenta in (\ref{d2gggtt}) are incoming, and the $t, \bar
t$ momenta are outgoing. 
The amplitudes for the process $gg \to t\tb g$ can be easily
obtained by crossing from the amplitudes given below.
It is convenient to decompose the amplitude for (\ref{d2gggtt})
with respect to its  colour structure. Performing this decomposition
the  ${\cal T}$
matrix element is then of the form:
\begin{equation}
  \cT(ggg\to t\tb ) = \sum_{\{i,j,k\}\in P(1,2,3)}
  (T_{a_i}T_{a_j}T_{a_k})_{c_t c_\tb} \, A(p_i,p_j,p_k,\kt,\st,\ktb,\stb),
\end{equation}
where the  sum runs over all permutations of the indices $\{1, 2,
3\}$. The $T_a$ are the generators of
colour SU$(N)$ in the fundamental
representation.   
The indices $a_i, $$c_t,$ and $ c_\tb$ label the colour
of the gluons,  quarks, and antiquarks, respectively. 
The amplitudes $A(\k1,\k2,\k3,\kt,\st,\ktb,\stb)$ are the so-called
colour-ordered  subamplitudes. They can be computed in a very compact
way by using spinor helicity methods
 \cite{Xu:1986xb,Kleiss:1985yh,Gunion:vc,Mangano:1990by}.
We use the following
notation  \cite{Xu:1986xb}
for  massless spinors $u,v$ of helicity $\pm {1\over 2}$:
\begin{eqnarray}
  u(p,\pm) &=& v(p,\mp) = |p\pm\ket,\nn\\
  \bar u(p,\pm)&=&\bar v(p,\mp)=\bra p\pm|.
\end{eqnarray}
For the spinor products we use the short-hand notation:
\begin{eqnarray}
  \spa{{p_i},{p_j}} &=& \bra p_i-|p_j+\ket,\nn\\
  \spb{{p_i},{p_j}} &=& \bra p_i+|p_j-\ket,\nn\\
  \bra p_i\pm|p_j|p_k\pm\ket &=&\bra p_i\pm|\psl_j|p_k\pm\ket
  = \bra p_i\pm|p_j\mp\ket \bra p_j\mp|p_k\pm\ket,
\end{eqnarray}
for $p_i^2=p_j^2=p_k^2=0$. The polarization vector of a gluon with
momentum $k$ can be written in the 4-dimensional helicity scheme
as follows:
\begin{eqnarray}
  \varepsilon_\mu^\pm(k,q) = \pm {\bra q^\mp|\gamma_\mu|k^\mp\ket\over
    \sqrt{2}\bra q^\mp|k^\pm\ket}.
\end{eqnarray}
The momentum $q$ denotes a `massless' ($q^2=0$) reference momentum,
which is otherwise arbitrary.
For a top quark with mass $m$ and spin vector $s_t$ 
we use the formalism of Ref.~\cite{Kleiss:1985yh}:
\begin{equation}
  u(\kt,\st) = {\spb{q_1,{q_2}}\over m} |q^+_1\ket +|q_2^-\ket,
\end{equation}

with
\begin{equation}
  \kt = q_1+q_2,\quad\mbox{and}\quad \st = (q_1-q_2)/m
\end{equation}
and
\begin{equation}
  v(\ktb,\stb)  = {\spb{r_1,{r_2}}\over m} |r^+_1\ket -|r_2^-\ket,
\end{equation}
with
\begin{equation}
  \ktb = r_1+r_2,\quad\mbox{and}\quad
  \stb = (r_2-r_1)/m .
\end{equation}
Note that $\q1,\q2,\r1,$ and $\r2$ are massless momenta with
\begin{equation}
  2\dotp{q_1,{q_2}}=2\dotp{r_1,{r_2}}=m^2.
\end{equation}
We use the short-hand notation
\begin{equation}
\bra p_i\pm|\sum_j p_j|p_k\pm\ket \equiv 
\sum_j\bra p_i\pm|p_j|p_k\pm\ket . 
\end{equation}
(If the momentum of the top or antitop quark is sandwiched between two
spinors one must first
decompose it into the sum of  massless momenta $q_1+q_2$ or $r_1+r_2$.)
For specific gluon helicity states we obtain: 
\def\None{\dotp{{\kt},{\ktb}}+ m^2}
\def\Ntwo{\dotp{{\k1},{\kt}}}
\def\Nthree{\dotp{{\k3},{\ktb}}}
\begin{eqnarray}
  &&A(1^-,2^-,3^-)
  = 
  {\spb{{\q1},{\r2}}\*
    \sandpp{{\q1},{\k1+\k2+\k3},{\q2}}
    \over \spb{{\k1},{\q1}}\*\spb{{\k1},{\k2}}
    \*\spb{{\k2},{\k3}}\*\spb{{\k3},{\q1}}}
  +
  {\spb{{\q1},{\r2}}
    \over \spb{{\k1},{\q1}}
    \*\spb{{\k2},{\q1}}\*\spb{{\k3},{\q1}}}\*
  \bigg(\nn\\
  &&{1\over   4\*\dotp{{\k1},{\kt}}\*\dotp{{\k3},{\ktb}}}\*
  \sandpp{{\q1},{\ktb},{\k3}}
  \*\bigg[\spa{{\k2},{\q2}}\*\sandpp{{\q1},{\kt},{\k1}}
  -\spa{{\k1},{\k2}}\*\sandpp{{\q1},{\k1},{\q2}}
  \bigg]\nn\\
  &&
  - {1\over 2\*\dotp{{\k3},{\ktb}}}
  \*{1\over\spb{{\k1},{\k2}}}
  \*\sandpp{{\q1},{\k1+\k2},{\q2}}\*\sandpp{{\q1},{\ktb},{\k3}}
  \nn\\ &&
  +{1\over 2\*\dotp{{\k1},{\kt}} }
  \*{1\over \spb{{\k2},{\k3}}}\*
  \bigg[
  \sandpp{{\q1},{\k3+\k2},{\q2}}\*\sandpp{{\q1},{\kt},{\k1}}
  +\sandpp{{\q1},{\k1},{\q2}} \*
  \sandpp{{\q1},{\k2+\k3},{\k1}}
  \bigg]\bigg),\\
  && A(1^+,2^-,3^-) =
  -{1 \over 2}
  \*{\spa{{\k2},{\k3}}\*
    (\spa{{\r1},{\r2}}\*\spb{{\k1},{\r2}}\*\sandpp{{\k1},{\k2+\k3},{\q2}}
    -\spa{{\q1},{\q2}}
    \*\spb{{\k1},{\q1}}\*\sandpp{{\k1},{\k2+\k3},{\r1}})
    \over (m^2+\dotp{\kt,\ktb})\*
    \spa{{\k1},{\k2}}\*\spb{{\k1},{\k2}}\*\spb{{\k2},{\k3}}\*\spa{{\r1},{\r2}}}
  \nn\\
  &&
  +{1 \over 4} \*{\sandpp{{\k1},{\kt},{\k2}}\over
    \dotp{{\k1},{\kt}}\*\spa{{\k1},{\k2}}\*\spb{{\k1},{\k2}}
    \*\spb{{\k1},{\k3}}}
  \*\bigg(
  {2\over
    \spb{{\k2},{\k3}}
   }
  \*{\spa{{\q1},{\q2}}\over \spa{{\r1},{\r2}}}\*\spb{{\k1},{\q1}}
  \*
  \sandpp{{\k1},{\k2+\k3},{\r1}}
  \nn\\
  &&
  +
  {1\over
    \dotp{{\k3},{\ktb}}}\*{\spa{{\q1},{\q2}}\over 
\spa{{\r1},{\r2}}}\*\spb{{\k1},{\q1}}
  \*(\spa{{\k2},{\k3}}\*\spa{{\k3},{\r1}}\*\spb{{\k1},{\k3}}
  +\spa{{\k2},{\r1}}\*\sandpp{{\k1},{\ktb},{\k3}})\nn\\
  &&
  -
  {1
    \over \dotp{{\k3},{\ktb}}
   }
  \*\spa{{\k2},{\q2}}\*\spb{{\k1},{\r2}}
  \*\sandpp{{\k1},{\ktb},{\k3}}
 -{2
    \over
    \spb{{\k2},{\k3}}}
  \*\spb{{\k1},{\r2}}
  \*\sandpp{{\k1},{\k2+\k3},{\q2}}
  \bigg),\\
  &&A(1^-,2^+,3^-)
  =
  -{1 \over 2} \*{\spa{{\k1},{\k3}}^2\*(\spb{{\k2},{\r2}}
    \*\sandm(\q2,\k1+\k3,\k2)\*\spa{{\r1},{\r2}}-\spa{{\q1},{\q2}}
    \*\spb{{\k2},{\q1}}\*\sandm(\r1,\k1+\k3,\k2))\over
    \spa{{\k1},{\k2}}\*\spa{{\k2},{\k3}}\*\spb{{\k2},{\k3}}\*\spa{{\r1},{\r2}}
    \*\spb{{\k1},{\k2}}\*\None}\nn\\&&
  +{1 \over 2} \*{\spa{{\k1},{\k3}}\over
    \spa{{\k1},{\k2}}\*\spa{{\k2},{\k3}}\*\spb{{\k2},{\k3}}
    \*\spa{{\r1},{\r2}}\*\spb{{\k1},{\k2}}\*\Ntwo}
  \*\bigg(
  \spa{{\k1},{\q1}}\*\spa{{\k3},{\q2}}\*\spb{{\k2},{\q1}}
  \*\spb{{\k2},{\r2}}\*\spa{{\r1},{\r2}}\nn\\
  &&+\spa{{\k1},{\q2}}\*\spb{{\k2},{\r2}}\*
  \sandm(\k3,\k1-\q2,\k2)\*\spa{{\r1},{\r2}}
  +\spa{{\k3},{\r1}}\*\spa{{\q1},{\q2}}
  \*\spb{{\k2},{\q1}}\*\sandm(\k1,\q1+\q2,\k2)\bigg)
  \nn\\
  &&
  +{1 \over 4} \*{\sandm(\k1,\kt,\k2)\over \spb{{\k1},{\k2}}
    \*\spa{{\k1},{\k2}}\*\spb{{\k2},{\k3}}\*\spa{{\r1},{\r2}}\*\Ntwo\*\Nthree}
  \*\bigg(\spa{{\k1},{\k3}}\*\spa{{\k3},{\r1}}\*\spa{{\q1},{\q2}}
  \*\spb{{\k2},{\q1}}\*\spb{{\k2},{\k3}}\nn\\
  &&
  -\sandm(\k3,\ktb,\k2)\*(\spa{{\k1},{\q2}}
  \*\spb{{\k2},{\r2}}\*\spa{{\r1},{\r2}}
  -\spb{{\k2},{\q1}}\*
  \spa{{\q1},{\q2}}\*\spa{{\k1},{\r1}})\bigg),\\
  &&A(1^-,2^-,3^+) =
  -{1 \over 2} \*{\spa{{\k1},{\k2}}\*(
    \spb{{\k3},{\r2}}\*\sandm(\q2,\k1+\k2,\k3)\*\spa{{\r1},{\r2}}
    -\spa{{\q1},{\q2}}\*\spb{{\k3},{\q1}}\*\sandm(\r1,\k1+\k2,\k3))\over
    \spb{{\k2},{\k3}}\*\spa{{\k2},{\k3}}\*\spb{{\k1},{\k2}}
    \*\spa{{\r1},{\r2}}\*\None}
  \nn\\
  &&
  -{1 \over 2} \*{\sandm(\k2,\ktb,\k3)\*(
    \spb{{\k3},{\r2}}\*\sandm(\q2,\k1+\k2,\k3)\*\spa{{\r1},{\r2}}
    -\spa{{\q1},{\q2}}\*\spb{{\k3},{\q1}}\*\sandm(\r1,\k1+\k2,\k3))\over
    \spb{{\k1},{\k3}}
    \*\spa{{\k2},{\k3}}\*\spb{{\k1},{\k2}}\*\spa{{\r1},{\r2}}
    \*\spb{{\k2},{\k3}}
    \*\Nthree
    }\nn\\
  &&
  +{1 \over 4} \*{\sandm(\k2,\ktb,\k3)\over
    \spb{{\k1},{\k3}}\*\spb{{\k2},{\k3}}\*\spa{{\k2},{\k3}}\*
    \spa{{\r1},{\r2}}\*\Ntwo\*\Nthree}
  \*(
  \spb{{\k3},{\r2}}\*\spa{{\k1},{\k2}}\*\spa{{\k1},{\q2}}
  \*\spb{{\k1},{\k3}}\*\spa{{\r1},{\r2}}\nn\\
  &&
  +\sandm(\k1,\kt,\k3)\*(\spb{{\k3},{\r2}}\*\spa{{\k2},{\q2}}
  \*\spa{{\r1},{\r2}}-\spa{{\k2},{\r1}}\*\spa{{\q1},{\q2}}\*\spb{{\k3},{\q1}}
  )).  
\end{eqnarray}
The subamplitudes
for the remaining gluon helicity configurations can be obtained by 
exploiting, for example, CPT invariance.
Calculating the squared amplitude in terms of colour-ordered
subamplitudes, we have:
\begin{eqnarray}
  |\cT(ggg\to t\tb )|^2 &=& N^3 {N^2-1\over N}\Bigg\{
   \sum_{\{i,j,k\}\in P(1,2,3)}  \left|A(i,j,k)\right|^2\nn\\
 && \hspace{-3.5cm} -{1\over N^2} \sum_{\{i,j,k\}\in P(1,2,3)}
  \left|A(i,j,k)+A(i,k,j)+A(k,i,j)\right|^2
  +{N^2+1\over N^4} \left|\sum_{\{i,j,k\}\in P(1,2,3)} A(i,j,k)\right|^2
  \Bigg\}.
  \label{eq:ReconstructSquaredAmplitude}
\end{eqnarray}
The structure of (\ref{eq:ReconstructSquaredAmplitude})
must be the same as that of the 
${\cal T}$ matrix element squared of the reaction $Z,\gamma \to t\tb
ggg $,  which was computed in Ref.~\cite{Campbell:1997hg}.  
We find agreement with this result.

\subsection{The process $\mathbf{q \qb \to t\tb g }$}
Here  we give the Born amplitudes  for the reaction
\begin{equation}
  q(\k1) \qb(\k2) \to t(\kq)\tb(\kqb) g(\k3). 
\end{equation}
The colour decomposition of the corresponding ${\cal T}$ matrix element
is given by
\begin{equation}
  \cT( q\qb \to t\tb g) 
  = 
  {1\over N}\* \delta_{\qb\, q}\* (T_a)_{t\,\tb }\* A_1
  + {1\over N}\* \delta_{t\,\tb }\* (T_a)_{\qb\, q}\* A_2
  + \delta_{\qb \tb }\*(T_a)_{t\,q}\* A_3
  + \delta_{tq} \*(T_a)_{\qb\, \tb }\* A_4,
\label{qqttghel}
\end{equation}
with $q,\qb$ denoting the colour indices of the massless  quarks, while
$t,\tb$ denote the colour indices of the top quarks.
The amplitudes for the processes 
\begin{equation}
  qg \to t\tb q, \quad\qb g \to t\tb \qb
\end{equation}
can be obtained  by crossing from (\ref{qqttghel}).
Using the above colour decomposition, the squared matrix element is given by
\begin{eqnarray}
  |\cT( q\qb \to t\tb g)|^2 = 
\rem{3}{
    N\* (N^2-1)\* (|A_3|^2 + |A_4|^2 )\nn\\
    &+&{1\over N}\*(N^2-1)\*(A_1\* A_1^* + A_2\* A_2^* + A_1^*\* A_3 
    + A_2^*\* A_3 + A_1\* A_3^* + A_2\* A_3^* + A_1^*\* A_4 
    +A_2^* \*A_4 + A_1\* A_4^* + A_2\* A_4^*)\nn\\
    &=&}
  (N^2-1)\*\bigg(N\*  \left(|A_3|^2 + |A_4|^2 \right)
  -{1\over N}\*\left( A_1\*A_1^*
  +A_2\*A_2^*+2\*A_1^*\*A_2+2\*A_2^*\*A_1\right)\bigg),
\label{tsqqtg}
\end{eqnarray}
where we have used 
\begin{equation}
  A_1 + A_2 + A_3 + A_4=0
\end{equation}
in order to simplify the term that is subleading in the number of colours.
Using again the 4-dimensional spinor helicity methods 
 as in 
the previous section, we obtain the following results for specific
helicity configurations of the massless partons:
\def\den(#1){{1\over #1}}%
\begin{eqnarray}
&&  A_1(\k1^-,\k2^-,\k3^+) =
  + {1\over 4}\*\den(\dotp{{\k1},{\k2}})\*\den(\dotp{{\k3},{\kq}}) \* 
  \bigg(
  -
  \den(\spa{{\k1},{\k3}})\*\sandpp{{\k3},{\kq},{\k1}}\*\spa{{\k1},{\q2}}
  \*\spb{{\k2},{\r2}}\nn\\&&
  +
  \den(\spa{{\k1},{\k3}})\*\den(\spa{{\r1},{\r2}})\*\sandpp{{\k3},{\kq},{\k1}}
  \*\spa{{\k1},{\r1}}\*\spa{{\q1},{\q2}}\*\spb{{\k2},{\q1}}
  - \den(\spa{{\r1},{\r2}})\*\spa{{\k1},{\r1}}\*\spa{{\q1},{\q2}}
  \*\spb{{\k2},{\k3}}\*\spb{{\k3},{\q1}}
  \bigg)\nn\\&&
   + {1\over 4}\*\den(\dotp{{\k1},{\k2}})
  \*\den(\dotp{{\k3},{\kqb}}) 
  \* \bigg(
  - \spa{{\k1},{\q2}}\*\spb{{\k2},{\k3}}\*\spb{{\k3},{\r2}}
  +\den(\spa{{\k1},{\k3}})\*\sandpp{{\k3},{\kqb},{\k1}}\*\spa{{\k1},{\q2}}
  \*\spb{{\k2},{\r2}}\nn\\&&
  -
  \den(\spa{{\k1},{\k3}})\*\den(\spa{{\r1},{\r2}})\*
  \sandpp{{\k3},{\kqb},{\k1}}
  \*\spa{{\k1},{\r1}}\*\spa{{\q1},{\q2}}\*\spb{{\k2},{\q1}}
  \bigg),\\
  && A_2(\k1^-,\k2^-,\k3^+) =
  + {1\over 4}\*\den(\dotp{{\k2},{\k3}})\*\den(\dotp{{\kq},{\kqb}}+m^2 ) 
  \* \bigg(
  \den(\spa{{\k1},{\k3}})\*\den(\spa{{\r1},{\r2}})\*\spa{{\k1},{\k2}}
  \*\spa{{\k1},{\r1}}\*\spa{{\q1},{\q2}}\*\spb{{\k2},{\k3}}\*\spb{{\k2},{\q1}}
  \nn\\&&
  + \spa{{\k1},{\q2}}\*\spb{{\k2},{\k3}}\*\spb{{\k3},{\r2}}
  - \den(\spa{{\k1},{\k3}})\*\spa{{\k1},{\k2}}\*\spa{{\k1},{\q2}}
  \*\spb{{\k2},{\k3}}\*\spb{{\k2},{\r2}} \nn \\ &&
  - \den(\spa{{\r1},{\r2}})\*\spa{{\k1},{\r1}}\*\spa{{\q1},{\q2}}
  \*\spb{{\k2},{\k3}}\*\spb{{\k3},{\q1}}
  \bigg), \\
  && A_3(\k1^-,\k2^-,\k3^+) =
  + {1\over 4}\*
  \den(\dotp{{\k1},{\k2}})\*\den( \dotp{{\kq},{\kqb}}+m^2) 
  \* \bigg(
  -
  {\spa{{\k1},{\k2}}
  \*\spa{{\k1},{\r1}}\over \spa{{\k1},{\k3}}\*\spa{{\r1},{\r2}}}
  \*\spa{{\q1},{\q2}}\*\spb{{\k2},{\k3}}\*\spb{{\k2},{\q1}}
  \nn\\&&
  - \spa{{\k1},{\q2}}\*\spb{{\k2},{\k3}}\*\spb{{\k3},{\r2}}
  + \den(\spa{{\k1},{\k3}})\*\spa{{\k1},{\k2}}\*\spa{{\k1},{\q2}}\*
  \spb{{\k2},{\k3}}\*\spb{{\k2},{\r2}}
  + \den(\spa{{\r1},{\r2}})\*\spa{{\k1},{\r1}}\*\spa{{\q1},{\q2}}\*
  \spb{{\k2},{\k3}}\*\spb{{\k3},{\q1}}
  \bigg)\nn\\&&
  + {1\over4}\*
  \den(\dotp{{\k1},{\k2}})\*\den(\dotp{{\k3},{\kq}}) \* \bigg(
  +
  \den(\spa{{\k1},{\k3}})\*\sandpp{{\k3},{\kq},{\k1}}\*\spa{{\k1},{\q2}}
  \*\spb{{\k2},{\r2}}\nn\\&&
  - \den(\spa{{\k1},{\k3}})\*\den(\spa{{\r1},{\r2}})\*
  \sandpp{{\k3},{\kq},{\k1}}\*\spa{{\k1},{\r1}}\*\spa{{\q1},{\q2}}
  \*\spb{{\k2},{\q1}}
  + \den(\spa{{\r1},{\r2}})\*\spa{{\k1},{\r1}}
  \*\spa{{\q1},{\q2}}\*\spb{{\k2},{\k3}}\*\spb{{\k3},{\q1}}
  \bigg),\\
  && A_1(\k1^-,\k2^-,\k3^-) =
  + {1\over 4}\*\den(\dotp{{\k1},{\k2}})\*\den(\dotp{{\k3},{\kq}}) \* \bigg(
  - \spa{{\k1},{\k3}}\*\spa{{\k3},{\q2}}\*\spb{{\k2},{\r2}}\nn\\&&
  +
  \den(\spb{{\k2},{\k3}})\*\sandpp{{\k2},{\kq},{\k3}}\*\spa{{\k1},{\q2}}
  \*\spb{{\k2},{\r2}}
  -
  \den(\spb{{\k2},{\k3}})\*\den(\spa{{\r1},{\r2}})\*\sandpp{{\k2},{\kq},{\k3}}
  \*\spa{{\k1},{\r1}}\*\spa{{\q1},{\q2}}\*\spb{{\k2},{\q1}}
  \bigg)\nn\\ &&
  + {1\over 4}\*\den(\dotp{{\k1},{\k2}})\*\den(\dotp{{\k3},{\kqb}}) \* \bigg(
  - \den(\spa{{\r1},{\r2}})\*\spa{{\k1},{\k3}}\*\spa{{\k3},{\r1}}
  \*\spa{{\q1},{\q2}}\*\spb{{\k2},{\q1}}\nn\\&&
  - \den(\spb{{\k2},{\k3}})\*\sandpp{{\k2},{\kqb},{\k3}}\*
  \spa{{\k1},{\q2}}\*\spb{{\k2},{\r2}}
  + \den(\spb{{\k2},{\k3}})\*\den(\spa{{\r1},{\r2}})\*
  \sandpp{{\k2},{\kqb},{\k3}}\*\spa{{\k1},{\r1}}\*\spa{{\q1},{\q2}}
  \*\spb{{\k2},{\q1}}
  \bigg),\\
  && A_2(\k1^-,\k2^-,\k3^-) =
   {1\over 4}\*
  \den(\dotp{{\k1},{\k3}})\*\den(\dotp{{\kq},{\kqb}} + m^2 ) 
  \* \bigg(  
  - \den(\spb{{\k2},{\k3}})\*\den(\spa{{\r1},{\r2}})\*
  \spa{{\k1},{\k3}}\*\spa{{\k1},{\r1}}\*\spa{{\q1},{\q2}}
  \*\spb{{\k1},{\k2}}\*\spb{{\k2},{\q1}}\nn\\&&
  + \spa{{\k1},{\k3}}\*\spa{{\k3},{\q2}}\*\spb{{\k2},{\r2}}
  - \den(\spa{{\r1},{\r2}})\*\spa{{\k1},{\k3}}\*
  \spa{{\k3},{\r1}}\*\spa{{\q1},{\q2}}\*\spb{{\k2},{\q1}}
  \nn \\ &&
  + \den(\spb{{\k2},{\k3}})\*\spa{{\k1},{\k3}}\*
  \spa{{\k1},{\q2}}\*\spb{{\k1},{\k2}}\*\spb{{\k2},{\r2}}
  \bigg), \\
  && A_3(\k1^-,\k2^-,\k3^-) =
   {1\over 4}\*
  \den(\dotp{{\k1},{\k2}})\*\den( \dotp{{\kq},{\kqb}} + m^2)
  \* 
  \bigg(  - \den(\spb{{\k2},{\k3}})\*\den(\spa{{\r1},{\r2}})\*\spa{{\k1},{\k3}}
  \*\spa{{\k1},{\r1}}\*\spa{{\q1},{\q2}}
  \*\spb{{\k1},{\k2}}\*\spb{{\k2},{\q1}}
   \nn\\&&
  - \den(\spa{{\r1},{\r2}})\*\spa{{\k1},{\k3}}\*\spa{{\k3},{\r1}}
  \*\spa{{\q1},{\q2}}\*\spb{{\k2},{\q1}}
  + \den(\spb{{\k2},{\k3}})\*\spa{{\k1},{\k3}}\*\spa{{\k1},{\q2}}
  \*\spb{{\k1},{\k2}}\*\spb{{\k2},{\r2}}
  +\spa{{\k1},{\k3}}\*\spa{{\k3},{\q2}}\*\spb{{\k2},{\r2}}
  \bigg)\nn\\&&
  + \den(2\*\dotp{{\k1},{\k2}})\*\den(2\*\dotp{{\k3},{\kq}}) \* \bigg(
  + \spa{{\k1},{\k3}}\*\spa{{\k3},{\q2}}\*\spb{{\k2},{\r2}}
  - \den(\spb{{\k2},{\k3}})\*\sandpp{{\k2},{\kq},{\k3}}
  \*\spa{{\k1},{\q2}}\*\spb{{\k2},{\r2}}\nn\\&&
  + \den(\spb{{\k2},{\k3}})\*\den(\spa{{\r1},{\r2}})\*
  \sandpp{{\k2},{\kq},{\k3}}\*\spa{{\k1},{\r1}}\*\spa{{\q1},{\q2}}
  \*\spb{{\k2},{\q1}}\bigg)\nn\\ &&
  + {1\over 4}\*\den(\dotp{{\k1},{\k3}})\*\den(\dotp{{\kq},{\kqb}}+m^2  )
  \* 
  \bigg(
  - \spa{{\k1},{\k3}}\*\spa{{\k3},{\q2}}\*\spb{{\k2},{\r2}}
  + \den(\spa{{\r1},{\r2}})\*\spa{{\k1},{\k3}}\*\spa{{\k3},{\r1}}
  \*\spa{{\q1},{\q2}}\*\spb{{\k2},{\q1}}\nn\\&&
  - \den(\spb{{\k2},{\k3}})\*\spa{{\k1},{\k3}}\*\spa{{\k1},{\q2}}
  \*\spb{{\k1},{\k2}}\*\spb{{\k2},{\r2}}
  + \den(\spb{{\k2},{\k3}})\*\den(\spa{{\r1},{\r2}})\*\spa{{\k1},{\k3}}
  \*\spa{{\k1},{\r1}}\*\spa{{\q1},{\q2}}
  \*\spb{{\k1},{\k2}}\*\spb{{\k2},{\q1}}
  \bigg).
 \end{eqnarray}
The remaining two helicity configurations can be obtained 
by exploiting the parity invariance of QCD.
The amplitudes $A$ determined in this
way may get an additional phase,
which cancels, however,  when calculating (\ref{tsqqtg}).
%
\section{Fits to NLO results}
\label{sec:FitFunctions}
\def\nlf{n_{lf}}%
In the following we give results in the form of fits for those functions
discussed in Section \ref{partonsection} that are not given 
in analytic form. These fits can be used to obtain 
predictions at the hadron level by convoluting with the corresponding
parton distribution functions. At small $\beta$ values, the
accuracy of the fits is at the per cent level. For larger $\beta$ values,
they are less precise --- but still precise enough for all
phenomenological applications.
Note that because of cancellations between different contributions, the 8
digits of precision of the parameters $a_i$ are sometimes important.
 
For the scale-independent part in the reaction $gg\to t \bar t X$ we 
choose an ansatz similar to that used in Ref.~\cite{Nason:1987xz}:
\def\fggconsti{c_{gg,i}}
\begin{eqnarray}\label{eq:fitgg}
{f}^{(1)}_{gg}, {g}^{(1)}_{gg,a}  &=& \fggconsti\*
  {7\over 1536\*\pi}\* \left[ 12\* \beta \* \ln(8\*\beta^2)^2 
  -{366\over 7}\*\beta\*\ln(8\*\beta^2) + {11\over 42}\*\pi^2 \right]  
  \nn\\
  &+& \beta \*\big[ a_1 + \beta^2\* ( a_2\*\ln(8\*\beta^2) + a_3 )
  + a_4 \* \beta^4 \* \ln(8\*\beta^2) + \rho^2 \*(a_5\*\ln(\rho)
  + a_6\*\ln(\rho)^2 ) \nn\\
  &+& \rho\*(a_7\*\ln(\rho) + a_8 \* \ln(\rho)^2 )\big]\nn\\
  &&+  a_9\*\beta^5  + a_{10} \*\beta^7 \*  \ln(\beta)
          + \rho^3 \* ( a_{11}\*\ln(\rho)+a_{12}\*\ln(\rho)^2).
\end{eqnarray}
The coefficients $\fggconsti$ are determined from the behaviour 
for small $\beta$.
Fitting this function to our theoretical prediction, we obtain the
values given in Table \ref{tab:ggttX}.
\def\fqqconsti{c_{q\bar q,i}}
\begin{table}[htbp]
\caption{\it Fit parameters  as defined 
in Eq.~(\ref{eq:fitgg}) for the functions 
${f}^{(1)}_{gg},{g}^{(1)}_{gg,a}$.}\label{tab:ggttX}
        \begin{center}
\renewcommand{\arraystretch}{1.2}
                \begin{tabular}[h]{|c|*{5}{c|}}
\hline                   &${f}^{(1)}_{gg}$ & ${g}^{(1)}_{gg,1}$ & ${g}^{(1)}_{gg,2}$ & ${g}^{(1)}_{gg,3}$ & ${g}^{(1)}_{gg,4}$ \\ \hline
$\fggconsti$ & 1 &$-3$ &1 &$-1$ &$-1$ \\ \hline
                         $a_{1}$ & $0.10981761$ & $-0.31527984$ & $0.10618585$ & $-0.10654108$ & $-0.10708321$\\ \hline
                         $a_{2}$ & $-1.0630975$ & $-0.27483778$ & $-0.1722542$ & $-0.091362874$ & $0.062373685$\\ \hline
                         $a_{3}$ & $-1.2553496$ & $-0.98582485$ & $0.053571485$ & $-1.4480832$ & $-0.31938862$\\ \hline
                         $a_{4}$ & $-9.4872673$ & $-4.7958679$ & $0.46063565$ & $-1.7773284$ & $0.34555702$\\ \hline
                         $a_{5}$ & $-3.6880817$ & $-1.9966529$ & $0.40949176$ & $-1.5264395$ & $0.038824512$\\ \hline
                         $a_{6}$ & $-5.4736051$ & $-2.8306013$ & $1.3268284$ & $-1.3615869$ & $0.052584228$\\ \hline
                         $a_{7}$ & $2.0430787$ & $1.0567063$ & $-0.66333002$ & $0.12455129$ & $-0.35180567$\\ \hline
                         $a_{8}$ & $6.8070238\cdot 10^{-3}$ & $-9.6117859\cdot 10^{-3}$ & $0.034668583$ & $-0.082780239$ & $-0.13230861$\\ \hline
                         $a_{9}$ & $23.269583$ & $11.630659$ & $-0.71127314$ & $5.3857605$ & $-0.46984803$\\ \hline
                         $a_{10}$ & $-0.91120803$ & $-1.5749174$ & $4.1212869$ & $-2.14052$ & $-0.4187379$\\ \hline
                         $a_{11}$ & $0.2999526$ & $-0.32863816$ & $0.12551445$ & $-0.17836516$ & $-0.21912773$\\ \hline
                         $a_{12}$ & $-5.3509101$ & $-2.2770031$ & $-0.66992889$ & $-0.64180163$ & $0.22150197$\\ \hline
                \end{tabular}                
        \end{center}
\end{table}
\begin{table}[!htbp]
\caption{\it Fit parameters  as defined 
in Eq.~(\ref{eq:fitqq}) for the functions 
${f}^{(1)}_{q\bar{q}},{g}^{(1)}_{q\bar{q},a}$.}\label{tab:qqttX}
        \begin{center}
\renewcommand{\arraystretch}{1.2}
                \begin{tabular}[h]{|c|*{5}{c|}}
\hline                   &${f}^{(1)}_{q\bar{q}}$ & ${g}^{(1)}_{g\bar{q},1}$ & ${g}^{(1)}_{q\bar{q},2}$ & ${g}^{(1)}_{q\bar{q},3}$ & ${g}^{(1)}_{q\bar{q},4}$ \\ \hline
$\fqqconsti$&1 &1 &$-1/3$ &1 &1 \\ \hline
                         $a_{1}$ & $0.18272533$ & $0.18271262$ & $-0.061287765$ & $0.18253142$ & $0.18262171$\\ \hline
                         $a_{2}$ & $0.18635401$ & $0.18719675$ & $0.015656292$ & $0.1756192$ & $0.18665363$\\ \hline
                         $a_{3}$ & $-0.16337335$ & $-0.15894226$ & $-0.07245742$ & $-0.17432802$ & $-0.14086689$\\ \hline
                         $a_{4}$ & $0.26381224$ & $0.27417865$ & $-0.089475628$ & $0.22222852$ & $0.27777792$\\ \hline
                         $a_{5}$ & $-0.76180191$ & $-0.78525268$ & $0.21303639$ & $-0.6620887$ & $-0.78622279$\\ \hline
                         $a_{6}$ & $-0.030248112$ & $-0.034023724$ & $0.020746502$ & $-0.021337505$ & $-0.033873638$\\ \hline
                         $a_{7}$ & $0.012349472$ & $0.013843671$ & $-0.022717427$ & $0.014090792$ & $0.030078002$\\ \hline
                         $a_{8}$ & $7.7073036\cdot 10^{-3}$ & $7.85052\cdot 10^{-3}$ & $-8.1364483\cdot 10^{-3}$ & $5.9573824\cdot 10^{-3}$ & $8.8545826\cdot 10^{-3}$\\ \hline
                \end{tabular}
        \end{center}
\end{table}
\begin{table}[!htbp]
\caption{\it Fit parameters  as defined 
in Eq.~(\ref{eq:fitqg}) for the functions 
${f}^{(1)}_{qg},{g}^{(1)}_{qg,1,2}$.}\label{tab:qgttX}
        \begin{center}
\renewcommand{\arraystretch}{1.2}
                \begin{tabular}[h]{|c|*{3}{c|}}
\hline                   &${f}^{(1)}_{qg}$ & ${g}^{(1)}_{qg,1}$ & ${g}^{(1)}_{qg,2}$ \\ \hline
                         $a_{1}$ & $0.01532878$ & $-7.5038708\cdot 10^{-3}$ & $2.1881981\cdot 10^{-3}$ \\ \hline
                         $a_{2}$ & $-0.45170812$ & $-0.29554484$ & $0.23254702$\\ \hline
                         $a_{3}$ & $-0.1287014$ & $-0.048548285$ & $0.022275599$ \\ \hline
                         $a_{4}$ & $0.52370186$ & $0.30790838$ & $-0.23459259$ \\ \hline
                         $a_{5}$ & $-0.26811216$ & $-0.15956101$ & $0.12225428$ \\ \hline
                         $a_{6}$ & $0.022340876$ & $0.018846548$ & $-0.017627$ \\ \hline
                         $a_{7}$ & $-0.19533856$ & $-0.14177144$ & $0.11350417$\\ \hline
                         $a_{8}$ & $-0.054689213$ & $-0.02584021$ & $0.021535027$ \\ \hline
                         $a_{9}$ & $-0.045358797$ & $-0.013097499$ & $1.5210124\cdot 10^{-3}$ \\ \hline
                         $a_{10}$ & $-7.7692908\cdot 10^{-4}$ & $1.3753108\cdot 10^{-3}$ & $-1.5369012\cdot 10^{-3}$ \\ \hline
                         $a_{11}$ & $5.6252092\cdot 10^{-5}$ & $-8.811281\cdot 10^{-5}$ & $9.6359945\cdot 10^{-5}$ \\ \hline
                         $a_{12}$ & $-8.8107179\cdot 10^{-5}$ & $-5.0153667\cdot 10^{-5}$ & $2.1289443\cdot 10^{-5}$ \\ \hline
                         $a_{13}$ & $-0.03501338$ & $-0.013967759$ & $7.2211643\cdot 10^{-3}$ \\ \hline
                \end{tabular}
        \end{center}
\end{table}
\begin{table}[htbp!]  
\caption{\it  Fit parameters  as defined 
in Eq.~(\ref{eq:fitqg}) for the functions ${g}^{(1)}_{qg,3,4}$ 
${g}^{(1)}_{gq,3,4}$ 
discussed in the text.}\label{tab:gqttX}
        \begin{center}\renewcommand{\arraystretch}{1.2}
                \begin{tabular}[h]{|c|*{4}{c|}}
\hline            & ${g}^{(1)}_{qg,3}$ & ${g}^{(1)}_{qg,4}$       
& ${g}^{(1)}_{gq,3}$ & ${g}^{(1)}_{gq,4}$ \\ \hline
                         $a_{1}$ 
& $1.8239099\cdot 10^{-3}$ & $2.0186389\cdot 10^{-3}$
& $1.6574875\cdot 10^{-3}$ & $1.8413674\cdot 10^{-3}$\\ \hline
                         $a_{2}$ 
& $0.011452419$ & $-0.025864066$
& $-0.25682763$ & $-0.26685522$\\ \hline
                         $a_{3}$ 
& $-0.015540536$ & $-0.022761156$
& $-0.019640631$ & $-0.025869178$\\ \hline
                         $a_{4}$ 
& $-6.5191297\cdot 10^{-3}$ & $0.031656197$
& $0.25770922$ & $0.2732$\\ \hline
                         $a_{5}$ 
& $0.012121019$ & $-7.5126124\cdot 10^{-3}$
& $-0.13514839$ & $-0.13361839$\\ \hline
                         $a_{6}$ 
& $-4.8820211\cdot 10^{-3}$ & $-3.8773834\cdot 10^{-3}$
& $0.024057235$ & $0.018832676$\\ \hline
                         $a_{7}$ 
 & $-2.9631254\cdot 10^{-3}$ & $-0.021304475$
& $-0.12327878$ & $-0.13543713$\\ \hline
                         $a_{8}$ 
& $-6.6094881\cdot 10^{-3}$ & $-0.010139352$
& $-0.018603959$ & $-0.025245484$\\ \hline
                         $a_{9}$ 
& $-3.8199452\cdot 10^{-3}$ & $-4.3853099\cdot 10^{-3}$
& $-4.0261913\cdot 10^{-3}$ & $-6.6003001\cdot 10^{-3}$\\ \hline
                         $a_{10}$ 
& $1.8172768\cdot 10^{-3}$ & $1.8737765\cdot 10^{-3}$
& $1.065926\cdot 10^{-3}$ & $9.7391848\cdot 10^{-4}$\\ \hline
                         $a_{11}$ 
& $-1.0969205\cdot 10^{-4}$ & $-1.1280442\cdot 10^{-4}$
& $-5.9556692\cdot 10^{-5}$ & $-5.5377984\cdot 10^{-5}$\\ \hline
                         $a_{12}$ 
& $-2.2922513\cdot 10^{-5}$ & $-2.6895291\cdot 10^{-5}$
& $-2.100388\cdot 10^{-5}$ & $-2.4470893\cdot 10^{-5}$\\ \hline
                         $a_{13}$ 
& $-2.8989325\cdot 10^{-3}$ & $-5.9168202\cdot 10^{-3}$
& $-4.9177402\cdot 10^{-3}$ & $-6.3797473\cdot 10^{-3}$\\ \hline
                \end{tabular}
               \end{center}
\end{table}
\par
For the reaction $q\bar q \to t\bar t X$ we chose an ansatz similar to 
that of Ref.~\cite{Nason:1987xz}:
\begin{eqnarray}\label{eq:fitqq}
 {f}^{(1)}_{q\bar{q}},{g}^{(1)}_{q\bar{q},a}  &=& \fqqconsti\*
  {\rho \over 72\*\pi} \*
  \bigg[{16\over 3}\*\beta\*\ln(8\*\beta^2)^2 
  - {82\over 3}\*\beta\*\ln(8\*\beta^2)-{\pi^2\over 6} \bigg]\nn\\
  &+&\beta\*\rho\*[a_1 + \beta^2\*(a_2\*\ln(8\*\beta^2)+a_3) 
  + \beta^4\*(a_4\*\ln(8\*\beta^2)+a_5)\nn\\
  &+& a_6\*\beta^6\*\ln(8\*\beta^2)+ a_7\*\ln(\rho) + a_8\*
  \ln(\rho)^2].
\end{eqnarray}
The fitted parameters are given in Table~\ref{tab:qqttX}.
For the reaction $qg\to t\bar t X$ one has to consider two possibilities:
either the quark or the gluon direction of flight in the
$t\bar{t}$-ZMS corresponds to the  direction
used to define the axes ${\bf\hat a},{\bf\hat b}$. In the case of the beam and 
off-diagonal bases, these two choices give different results at the parton
level for the contribution from hard gluon emission.
We denote the scaling functions for the first case by   
${g}^{(1)}_{qg,3,4}$ and for the second case by 
${g}^{(1)}_{gq,3,4}$. As fit ansatz   
we choose 
\begin{eqnarray}\label{eq:fitqg}
 {f}^{(1)}_{qg},{g}^{(1)}_{qg(gq),a} &=&
  \beta \* \big[ \beta^2\*(a_1\*\ln(\beta) + a_2)
  + \beta^4\*(a_3\*\ln(\beta)+a_4) 
  + \rho^2\*(a_5\*\ln(\rho) + a_6\*\ln(\rho)^2) \nn\\
  &+& \rho\*(a_7\*\ln(\rho)+a_8\*\ln(\rho)^2) \big]
   + \beta^4 \* \big[a_9+a_{10}\*\ln(\eta)+a_{11}\*\ln(\eta)^2\big] 
  \nn \\ &+& \rho^3\*\big[a_{12}\*\ln(\rho)+a_{13}\*\ln(\rho)^2\big].
\end{eqnarray}
The fitted values are shown in Tables~\ref{tab:qgttX} and 
\ref{tab:gqttX}.
\par
>From the analytic expressions for the leading-order results, the 
scaling functions determining the $\mu$
dependence can be obtained by a simple 
convolution with the corresponding Altarelli--Parisi evolution kernels,
see Eq. (\ref{RGE}). In most cases the convolution integrals can be done
analytically. For the remaining functions we use 
as ansatz for the process $gg\to t\bar t X$:
\begin{eqnarray}\label{eq:fitggmu}
 \tilde{g}^{(1)}_{gg,3,4}  &=& a_{1} 
   + a_{2} \* \beta         
   + a_{3} \* \beta^2
   + a_{4} \* \beta^3
   + a_{5} \* \beta^4 
   + a_{6} \* \beta \* \ln(\beta)
   + a_{7} \* \beta^2 \* \ln(\beta)
   + a_{8} \* \beta \* \ln(\beta)^2
   + a_{9} \* \beta^2 \* \ln(\beta)^2  \nn\\       
  &+& a_{10} \*  \beta^3 \* \ln(\beta)
   + a_{11} \* \beta^3 \* \ln(\beta)^2
   + a_{12} \* \rho\*\ln(\rho)
   + a_{13} \* \rho^2\*\ln(\rho)
   + a_{14} \* \rho\*\ln(\rho)^2\nn\\&&
   + a_{15} \* \rho^2\*\ln(\rho)^2
   + a_{16} \* \beta\*\rho\*\ln(\rho).
\end{eqnarray}
The fitted values for the two functions $\tilde{g}^{(1)}_{gg,3,4}$
are given in Table \ref{tab:ggttXmu}.
\par
For the scale-dependent part in the reaction $qg\to t\bar t X$ we use
\begin{eqnarray}\label{eq:fitqgmu}
\tilde{g}^{(1)}_{qg,3,4}=\tilde{g}^{(1)}_{gq,3,4} 
&=&  \beta\*(\beta^2\*(a_{1}\*\ln(\beta)+a_{2})
 +\beta^4\*(a_{3}\*\ln(\beta)+a_{4})
 +\rho^2\*(a_{5}\*\ln(\rho)+a_{6}\*\ln(\rho)^2)
\nn \\ 
&+&\rho\*(a_{7}\*\ln(\rho)+a_{8}\*\ln(\rho)^2))
 + \beta^4 \* a_{9} 
 + \beta^4 \* a_{10} \* \ln(\eta)
 + \beta^4 \* a_{11} \* \ln(\eta)^2
\nn \\
 &+& \rho^3\*(a_{12}\*\ln(\rho)+a_{13}\*\ln(\rho)^2) 
 + a_{14} \* \beta \* \ln(\eta)
 + a_{15} \* \beta^3 \* \ln(\eta)
 + a_{16} \* \beta \* \ln(\eta)^2 \nn \\
 &+& a_{17} \* \beta^3 \* \ln(\eta)^2 
 + a_{18} \* \beta \* \rho \*  \ln(\eta)
 + a_{19} \* \beta^3 \* \rho \* \ln(\eta)
 + a_{20} \* \beta \* \rho \*  \ln(\eta)^2. 
\end{eqnarray}
The fitted values for the two functions $\tilde{g}^{(1)}_{gq,3,4}$
are shown in Table~\ref{tab:qgttXmu}.
\begin{table}[htbp]
\caption{\it 
Fit parameters  as defined 
in Eq.~(\ref{eq:fitggmu}) for the two functions 
$\tilde{g}^{(1)}_{gg,3,4}$.}\label{tab:ggttXmu}
        \begin{center}\renewcommand{\arraystretch}{1.2}
                \begin{tabular}[h]{|c|*{2}{c|}}
\hline                   & $\tilde{g}^{(1)}_{gg,3}$ & $\tilde{g}^{(1)}_{gg,4}$ \\ \hline
                         $a_{1}$ & $3.600512\cdot 10^{-4}$ & $9.8339551\cdot 10^{-4}$\\ \hline
                         $a_{2}$ & $-0.89103071$ & $-2.1615978$\\ \hline
                         $a_{3}$ & $4.1432564$ & $-2.4383759$\\ \hline
                         $a_{4}$ & $-5.652246$ & $-1.7863628$\\ \hline
                         $a_{5}$ & $2.396262$ & $6.3822522$\\ \hline
                         $a_{6}$ & $-0.26139117$ & $-0.7174869$\\ \hline
                         $a_{7}$ & $0.55643278$ & $-4.634966$\\ \hline
                         $a_{8}$ & $-0.029004255$ & $-0.076074349$\\ \hline
                         $a_{9}$ & $0.44754783$ & $0.26832936$\\ \hline
                         $a_{10}$ & $-0.39236517$ & $-8.4049829$\\ \hline
                         $a_{11}$ & $-0.52419439$ & $1.3153541$\\ \hline
                         $a_{12}$ & $1.9909471$ & $7.6847524$\\ \hline
                         $a_{13}$ & $-0.8099637$ & $-2.9175181$\\ \hline
                         $a_{14}$ & $0.013395378$ & $0.023018999$\\ \hline
                         $a_{15}$ & $0.031263952$ & $0.035390606$\\ \hline
                         $a_{16}$ & $-1.9746581$ & $-7.5775663$\\ \hline
                \end{tabular}
        \end{center}
\end{table}
\begin{table}[htbp]
\caption{\it Fit parameters  as defined 
in Eq.~(\ref{eq:fitqgmu}) for the two functions 
$\tilde{g}^{(1)}_{gq,3,4}$.}\label{tab:qgttXmu}
        \begin{center}\renewcommand{\arraystretch}{1.2}
                \begin{tabular}[h]{|c|*{2}{c|}}
\hline                   &$\tilde{g}^{(1)}_{gq,3}$ & $\tilde{g}^{(1)}_{gq,4}$ \\ \hline
                         $a_{1}$ & $0.016955545$ & $-2.7829949\cdot 10^{-3}$\\ \hline
                         $a_{2}$ & $0.057599714$ & $0.03597281$\\ \hline
                         $a_{3}$ & $0.012017294$ & $1.5756308\cdot 10^{-3}$\\ \hline
                         $a_{4}$ & $-0.061168008$ & $-0.043921612$\\ \hline
                         $a_{5}$ & $0.025443918$ & $0.023078656$\\ \hline
                         $a_{6}$ & $-2.4972922\cdot 10^{-3}$ & $-2.5244529\cdot 10^{-3}$\\ \hline
                         $a_{7}$ & $0.019870969$ & $0.019474128$\\ \hline
                         $a_{8}$ & $5.0330984\cdot 10^{-3}$ & $4.949093\cdot 10^{-3}$\\ \hline
                         $a_{9}$ & $2.216736\cdot 10^{-3}$ & $7.0180624\cdot 10^{-3}$\\ \hline
                         $a_{10}$ & $7.7028477\cdot 10^{-3}$ & $-1.6586156\cdot 10^{-3}$\\ \hline
                         $a_{11}$ & $-1.8013666\cdot 10^{-4}$ & $1.1083146\cdot 10^{-4}$\\ \hline
                         $a_{12}$ & $3.666881\cdot 10^{-5}$ & $-3.7800956\cdot 10^{-5}$\\ \hline
                         $a_{13}$ & $1.7202878\cdot 10^{-3}$ & $1.3536538\cdot 10^{-3}$\\ \hline
                         $a_{14}$ & $-4.2292913\cdot 10^{-6}$ & $2.2933896\cdot 10^{-6}$\\ \hline
                         $a_{15}$ & $-7.5513607\cdot 10^{-3}$ & $1.7188112\cdot 10^{-3}$\\ \hline
                         $a_{16}$ & $8.349152\cdot 10^{-7}$ & $8.6710055\cdot 10^{-7}$\\ \hline
                         $a_{17}$ & $1.6975408\cdot 10^{-4}$ & $-1.1585369\cdot 10^{-4}$\\ \hline
                         $a_{18}$ & $4.1704783\cdot 10^{-6}$ & $-2.2181371\cdot 10^{-6}$\\ \hline
                         $a_{19}$ & $1.645701\cdot 10^{-3}$ & $-1.7941491\cdot 10^{-3}$\\ \hline
                         $a_{20}$ & $-8.3995838\cdot 10^{-7}$ & $-8.6053362\cdot 10^{-7}$\\ \hline
                \end{tabular}               
        \end{center}
\end{table}

\end{document}